%% file: HIG-21-001_temp.tex
\pdfoutput=1
\documentclass[11pt,twoside,a4paper,cmspaper,final,collab]{cms-tdr}
\def\svnVersion{8b9c2f5}\def\svnDate{2023/07/20}\def\cmsCernNoTag{CERN-EP-2022-137}\def\cmsCernDate{\today}\def\cmsMessage{Published in the Journal of High Energy Physics as \href{http://dx.doi.org/10.1007/JHEP07(2023)073}{\doi{10.1007/JHEP07(2023)073}.}}
\begin{document}\cmsNoteHeader{HIG-21-001}

\newcommand{\Phobs}{\ensuremath{\mathrm{\PH(125)}}\xspace}
\newcommand{\ditau}{\ensuremath{\PGt\PGt}\xspace}
\newcommand{\PP}{\ensuremath{\Pp\Pp}\xspace}
\newcommand{\emu}{\ensuremath{\Pe\PGm}\xspace}
\newcommand{\etau}{\ensuremath{\Pe\tauh}\xspace}
\newcommand{\mutau}{\ensuremath{\PGm\tauh}\xspace}
\newcommand{\tautau}{\ensuremath{\tauh\tauh}\xspace}
\newcommand{\qtau}{\ensuremath{\PQq\PGt}\xspace}
\newcommand{\Pphi}{\ensuremath{\phi}\xspace}
\newcommand{\ggA}{\ensuremath{\Pg\Pg\PA}\xspace}
\newcommand{\bbA}{\ensuremath{\PQb\PQb\PA}\xspace}
\newcommand{\ggH}{\ensuremath{\Pg\Pg\PH}\xspace}
\newcommand{\bbH}{\ensuremath{\PQb\PQb\PH}\xspace}
\newcommand{\ggh}{\ensuremath{\Pg\Pg\Ph}\xspace}
\newcommand{\bbh}{\ensuremath{\PQb\PQb\Ph}\xspace}
\newcommand{\Vh}{\ensuremath{\PV\Ph}\xspace}
\newcommand{\ggPhi}{\ensuremath{\Pg\Pg\phi}\xspace}
\newcommand{\bbPhi}{\ensuremath{\PQb\PQb\phi}\xspace}
\newcommand{\mphi}{\ensuremath{m_{\phi}}\xspace}
\newcommand{\ptphi}{\ensuremath{\pt^{\phi}}\xspace}
\newcommand{\SUTwoL}{\ensuremath{\mathrm{SU}(2)_{\mathrm{L}}}\xspace}
\newcommand{\SUTwo}{\ensuremath{\mathrm{SU}(2)}\xspace}
\newcommand{\Phiu}{\ensuremath{\Phi_{\mathrm{u}}}\xspace}
\newcommand{\Phid}{\ensuremath{\Phi_{\mathrm{d}}}\xspace}
\newcommand{\PA}{\ensuremath{\mathrm{A}}\xspace}
\newcommand{\mH}{\ensuremath{m_{\PH}}\xspace}
\newcommand{\mh}{\ensuremath{m_{\Ph}}\xspace}
\newcommand{\mA}{\ensuremath{m_{\PA}}\xspace}
\newcommand{\mSUSY}{\ensuremath{m_{\text{SUSY}}}\xspace}
\newcommand{\mt}{\ensuremath{m_{\PQt}}\xspace}
\newcommand{\Uone}{\ensuremath{\mathrm{U}_1}\xspace}
\newcommand{\SUThreeC}{\ensuremath{\mathrm{SU}(3)_{\mathrm{C}}}\xspace}
\newcommand{\UOneY}{\ensuremath{\text{U}(1)_{\mathrm{Y}}}\xspace}
\newcommand{\gU}{\ensuremath{g_{\text{U}}}\xspace}
\newcommand{\betaL}{\ensuremath{\beta_{\mathrm{L}}}\xspace}
\newcommand{\betaR}{\ensuremath{\beta_{\mathrm{R}}}\xspace}
\newcommand{\mU}{\ensuremath{m_{\text{U}}}\xspace}
\newcommand{\Irelem}{\ensuremath{I_{\text{rel}}^{\Pe(\mu)}}\xspace}
\newcommand{\yDT}{\ensuremath{y^{\text{DT}}}\xspace}
\newcommand{\Dj}{\ensuremath{D_{\text{jet}}}\xspace}
\newcommand{\De}{\ensuremath{D_{\Pe}}\xspace}
\newcommand{\Dm}{\ensuremath{D_{\PGm}}\xspace}
\newcommand{\ZEE}{\mbox{\ensuremath{\PZ/\PGg^{*}\to\Pe\Pe}}\xspace}
\newcommand{\ZMM}{\mbox{\ensuremath{\PZ/\PGg^{*}\to\PGm\PGm}}\xspace}
\newcommand{\ZTT}{\mbox{\ensuremath{\PZ/\PGg^{*}\to\PGt\PGt}}\xspace}
\newcommand{\ZLL}{\mbox{\ensuremath{\PZ/\PGg^{*}\to\Pell\Pell}}\xspace}
\newcommand{\Zgamma}{\ensuremath{\PZ/\PGg^{*}}\xspace}
\newcommand{\ATT}{\ensuremath{\PA\to\PGt\PGt}\xspace}
\newcommand{\HTT}{\ensuremath{\PH\to\PGt\PGt}\xspace}
\newcommand{\mTem}{\ensuremath{m_{\mathrm{T}}^{\Pe(\mu)}}\xspace}
\newcommand{\Dzeta}{\ensuremath{D_{\zeta}}\xspace}
\newcommand{\zetahat}{\ensuremath{\hat{\zeta}}\xspace}
\newcommand{\ptvectauone}{\ensuremath{\ptvec^{\kern1pt\PGt_{1}}}\xspace}
\newcommand{\ptvectautwo}{\ensuremath{\ptvec^{\kern1pt\PGt_{2}}}\xspace}
\newcommand{\ptvece}{\ensuremath{\ptvec^{\kern1pt\Pe}}\xspace}
\newcommand{\ptvecm}{\ensuremath{\ptvec^{\kern1pt\PGm}}\xspace}
\newcommand{\ptvecem}{\ensuremath{\ptvec^{\kern1pt\Pe(\PGm)}}\xspace}
\newcommand{\pzetamiss}{\ensuremath{p_{\zeta}^{\text{miss}}}\xspace}
\newcommand{\pzetavis}{\ensuremath{p_{\zeta}^{\text{vis}}}\xspace}
\newcommand{\Wjets}{\ensuremath{\PW}{+}\,\text{jets}\xspace}
\newcommand{\Zjets}{\ensuremath{\PZ}{+}\,\text{jets}\xspace}
\newcommand{\mTm}{\ensuremath{m_{\mathrm{T}}^{\mu}}\xspace}
\newcommand{\mTtot}{\ensuremath{m_{\mathrm{T}}^{\text{tot}}}\xspace}
\newcommand{\pttt}{\ensuremath{\pt^{\ditau}}\xspace}
\newcommand{\mumu}{\ensuremath{\PGm\PGm}\xspace}
\newcommand{\ptem}{\ensuremath{\pt^{\Pe(\PGm)}}\xspace}
\newcommand{\pte}{\ensuremath{\pt^{\Pe}}\xspace}
\newcommand{\ptm}{\ensuremath{\pt^{\PGm}}\xspace}
\newcommand{\ptell}{\ensuremath{\pt^{\Pell}}\xspace}
\newcommand{\ptth}{\ensuremath{\pt^{\tauh}}\xspace}
\newcommand{\ptjet}{\ensuremath{\pt^{\text{jet}}}\xspace}
\newcommand{\FF}{\ensuremath{F_{\mathrm{F}}}\xspace}
\newcommand{\FFi}{\ensuremath{\FF^{i}}\xspace}
\newcommand{\FFttbar}{\ensuremath{\FF^{\ttbar}}\xspace}
\newcommand{\DRi}{\ensuremath{\mathrm{DR}^{i}}\xspace}
\newcommand{\NSR}{\ensuremath{N_{\text{SR}}}\xspace}
\newcommand{\NAR}{\ensuremath{N_{\text{AR}}}\xspace}
\newcommand{\jettau}{\ensuremath{\text{jet}\to\tauh}\xspace}
\newcommand{\jettoL}{\ensuremath{\text{jet}\to\Pell}\xspace}
\newcommand{\mutoe}{\ensuremath{\PGm\to\Pe}\xspace}
\newcommand{\eormutotau}{\ensuremath{\Pe(\PGm)\to\tauh}\xspace}
\newcommand{\LplusX}{\ensuremath{\hphantom{\PGt}\Pell+\mathrm{X}}\xspace}
\newcommand{\Zh}{\ensuremath{\PZ\Ph}\xspace}
\newcommand{\Wh}{\ensuremath{\PW\Ph}\xspace}
\newcommand{\mtt}{\ensuremath{m_{\PGt\PGt}}\xspace}
\newcommand{\TF}{\ensuremath{F_{\mathrm{T}}}\xspace}
\newcommand{\Irele}{\ensuremath{I_{\text{rel}}^{\Pe}}\xspace}
\newcommand{\Irelm}{\ensuremath{I_{\text{rel}}^{\PGm}}\xspace}
\newcommand{\Irelell}{\ensuremath{I_{\text{rel}}^{\Pell}}\xspace}
\newcommand{\Njet}{\ensuremath{N_{\text{Jets}}}\xspace}
\newcommand{\Nbjet}{\ensuremath{N_{\PQb\text{-jets}}}\xspace}
\newcommand{\mur}{\ensuremath{\mu_{\mathrm{R}}}\xspace}
\newcommand{\muf}{\ensuremath{\mu_{\mathrm{F}}}\xspace}
\newcommand{\hdamp}{\ensuremath{h_{\text{damp}}}\xspace}
\newcommand{\mhBMPone}{\ensuremath{M_{\mathrm{h}}^{125}}\xspace}
\newcommand{\mhBMPtwo}{\ensuremath{M_{\mathrm{h},\,\text{EFT}}^{125}}\xspace}
\newcommand{\betaRbtau}{\ensuremath{\betaR^{\mathrm{b}\tau}}\xspace}
\newcommand{\betaLbtau}{\ensuremath{\betaL^{\PQb\PGt}}\xspace}
\newcommand{\betaLstau}{\ensuremath{\betaL^{\mathrm{s}\tau}}\xspace}
\newcommand{\betaLdtau}{\ensuremath{\betaL^{\mathrm{d}\tau}}\xspace}
\newcommand{\betaLsmu}{\ensuremath{\betaL^{\mathrm{s}\mu}}\xspace}
\newcommand{\betaLbmu}{\ensuremath{\betaL^{\mathrm{b}\mu}}\xspace}
\newcommand{\Qmatch}{\ensuremath{Q_{\text{match}}}\xspace}
\newcommand{\dR}{\ensuremath{{\Delta}R}\xspace}
\newcommand{\order}{\ensuremath{\mathcal{O}}}
\renewcommand{\labelenumi}{\theenumi}
\renewcommand{\theenumi}{\roman{enumi})}
\newlength\cmsTabSkip\setlength{\cmsTabSkip}{2ex}
\providecommand{\cmsTable}[1]{\resizebox{\textwidth}{!}{#1}}

\cmsNoteHeader{HIG-21-001}
\title{Searches for additional Higgs bosons and for vector leptoquarks in \texorpdfstring{$\ditau$}{tau tau} final states in proton-proton collisions at \texorpdfstring{$\sqrt{s}=13\TeV$}{sqrt(s)=13 TeV}}

\date{\today}

\abstract{
Three searches are presented for signatures of physics beyond the standard model (SM) in \ditau final states in proton-proton collisions at the LHC, using a data sample collected with the CMS detector at $\sqrt{s} = 13\TeV$, corresponding to an integrated luminosity of 138\fbinv. Upper limits at 95\% confidence level (CL) are set on the products of the branching fraction for the decay into \PGt leptons and the cross sections for the production of a new boson \Pphi, in addition to the \Phobs boson, via gluon fusion (\ggPhi) or in association with \PQb quarks, ranging from \order(10\unit{pb}) for a mass of 60\GeV to 0.3\unit{fb} for a mass of 3.5\TeV each. The data reveal two excesses for \ggPhi production with local $p$-values equivalent to about three standard deviations at $\mphi=0.1$ and 1.2\TeV. In a search for $t$-channel exchange of a vector leptoquark \Uone, 95\% CL upper limits are set on the dimensionless \Uone leptoquark coupling to quarks and \PGt leptons ranging from 1 for a mass of 1\TeV to 6 for a mass of 5\TeV, depending on the scenario. In the interpretations of the \mhBMPone and \mhBMPtwo minimal supersymmetric SM benchmark scenarios, additional Higgs bosons with masses below 350\GeV are excluded at 95\% CL. 
}

\hypersetup{
pdfauthor={CMS Collaboration},
pdftitle={Searches for additional Higgs bosons and for vector leptoquarks in tau tau final states in proton-proton collisions at sqrt(s)=13 TeV},
pdfsubject={CMS},
pdfkeywords={CMS,Higgs, BSM, MSSM, leptoquarks}} 

\maketitle 

\section{Introduction}
\label{sec:introduction}

The discovery of a Higgs boson with a mass of around 125\GeV, \Phobs, at the  
LHC in 2012~\cite{Aad:2012tfa,Chatrchyan:2012xdj,Chatrchyan:2013lba} has turned 
the standard model (SM) of particle physics into a theory that could be valid up 
to the Planck scale. In the SM, \Phobs emerges from the spontaneous breaking of 
the electroweak \SUTwoL symmetry. While the nature of the underlying mechanism 
leading to this symmetry breaking and the exact form of the required 
symmetry-breaking potential are still to be explored, the measured couplings of 
\Phobs to fermions and gauge bosons, with 5--20\% experimental 
precision~\cite{Khachatryan:2016vau,Sirunyan:2018koj,Aad:2019mbh,Sirunyan:2019twz}, 
are in good agreement with the expectation for an SM Higgs boson with a mass of 
$125.38\pm0.14\GeV$~\cite{Sirunyan:2020xwk}. The SM still leaves several fundamental 
questions related to particle physics unaddressed, including the presence of 
dark matter and the observed baryon asymmetry in nature. Many extensions of the 
SM that address these questions require a more complex structure of the part of 
the theory that is related to \SUTwoL breaking, often referred to as the Higgs 
sector. Such models usually predict additional spin-0 states and modified properties 
of \Phobs with respect to the SM expectation. Models incorporating supersymmetry 
(SUSY)~\cite{Golfand:1971iw,Wess:1974tw} are prominent examples. In the minimal 
extension of the SM, the minimal supersymmetric SM (MSSM)~\cite{Fayet:1974pd,
Fayet:1977yc}, the model predicts three neutral and two charged Higgs bosons.

Searches for additional heavy neutral Higgs bosons in the context of the MSSM
were carried out in electron-positron collisions at the LEP collider at CERN~\cite{Schael:2006cr} 
and in proton-antiproton collisions at the Fermilab Tevatron~\cite{Aaltonen:2009vf,
Abazov:2010ci,Abazov:2011jh,Aaltonen:2011nh}. At the LHC such searches have been 
carried out by the ATLAS and CMS Collaborations in the \PQb 
quark~\cite{Chatrchyan:2013qga,Khachatryan:2015tra,Sirunyan:2018taj,ATLAS:2019tpq}, 
dimuon~\cite{Aad:2012cfr,CMS:2015ooa,CMS:2019mij,ATLAS:2019odt}, and 
\ditau~\cite{Aad:2012cfr,Aad:2014vgg,Aaboud:2016cre,Aaboud:2017sjh,Chatrchyan:2011nx,
Chatrchyan:2012vp,Khachatryan:2014wca,Sirunyan:2018zut,ATLAS:2020zms} final 
states. The \ditau final state has a leading role in these searches, since \PGt 
leptons can be identified with higher purity than \PQb quarks and backgrounds from 
genuine \ditau events can be estimated with higher accuracy, while the branching 
fractions for the decay into \PGt leptons are typically larger than those for the 
decay into muons because of the larger \PGt lepton mass. 
There are several other examples of extended Higgs sectors, which are summarized 
in Ref.~\cite{Steggemann:2020egv}, that could give appreciable resonant \ditau 
production rates in addition to the known SM processes at the LHC. 
Furthermore, models that include additional coloured states carrying both 
baryon and lepton quantum numbers, known as leptoquarks~\cite{Diaz:2017lit,
Schmaltz:2018nls}, can lead to an enhancement in the nonresonant production 
rates of \ditau pairs with large invariant masses via the leptoquark $t$-channel 
exchange. Searches for resonant and nonresonant \ditau signatures are thus 
complementary in the exploration of physics beyond the SM (BSM) at the LHC. 
Recent searches for single- and pair-production of third-generation leptoquarks 
at the LHC are reported in Refs.~\cite{CMS:2017xcw,CMS:2018txo,CMS:2018iye,
ATLAS:2019qpq,ATLAS:2020dsf,CMS:2020wzx,ATLAS:2021jyv,ATLAS:2021yij,ATLAS:2021oiz}.

In this paper the results of three searches for both resonant and nonresonant 
\ditau signatures are presented:
\begin{enumerate}
\item
The first search, which is meant to be as model independent as possible, targets 
the production of a single narrow spin-0 resonance \Pphi, in addition to \Phobs, 
via gluon fusion (\ggPhi) or in association with \PQb quarks (\bbPhi). Assumptions 
that have been made for this search are that the width of \Pphi is small compared 
with the experimental resolution, and that the \Pphi transverse momentum (\pt) 
spectrum for \ggPhi production as well as the relative contributions of \PQt- and 
\PQb-quarks to \ggPhi production are as expected for an SM Higgs boson at the 
tested mass value.
\item
The second search targets the $t$-channel exchange of a vector 
leptoquark \Uone.
\item
The third search exploits selected benchmark scenarios of the MSSM that rely 
on the signal from three neutral Higgs bosons, one of which is associated with 
\Phobs.
\end{enumerate}

The results are based on the proton-proton (\PP) collision data collected at 
the LHC during the years 2016--2018, at $\sqrt{s} = 13\TeV$, by the CMS experiment. 
The data correspond to an integrated luminosity of 138\fbinv. The analysis is 
performed in four \ditau final states: \emu, \etau, \mutau, and \tautau, where 
\Pe, \PGm, and \tauh indicate \PGt decays into electrons, muons, and hadrons, 
respectively. For this analysis the most significant backgrounds are estimated 
from data, which includes all SM processes with two genuine \PGt leptons in the 
final state, and processes where quark- or gluon-induced jets are misidentified 
as \tauh, denoted as \jettau. 

The paper is organized as follows. Section~\ref{sec:phenomenology} gives an 
overview of the phenomenology of the BSM physics scenarios under consideration. 
Section~\ref{sec:detector} describes the CMS detector, and Section~\ref{sec:reconstruction} 
describes the event reconstruction. Section~\ref{sec:selection} summarizes the 
event selection and categorization used for  the extraction of the signal. The 
data model and systematic uncertainties are described in Sections~\ref{sec:data-model} 
and~\ref{sec:systematic-uncertainties}. Section~\ref{sec:results} contains the 
results of the analysis. Section~\ref{sec:summary} briefly summarizes the paper.
A complete set of tabulated results of this search for all tested mass hypotheses 
is available in the HEPData database~\cite{hepdata}.

\section{Signal models}
\label{sec:phenomenology}

\begin{figure}[b]
  \centering
  \includegraphics[width=0.32\textwidth]{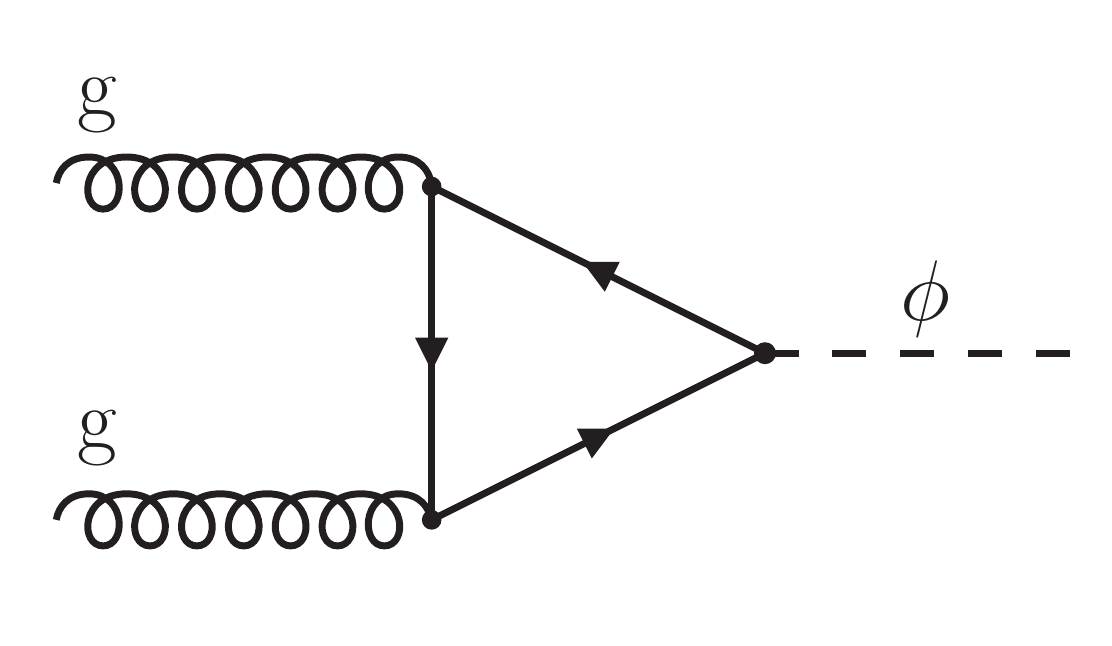}
  \includegraphics[width=0.32\textwidth]{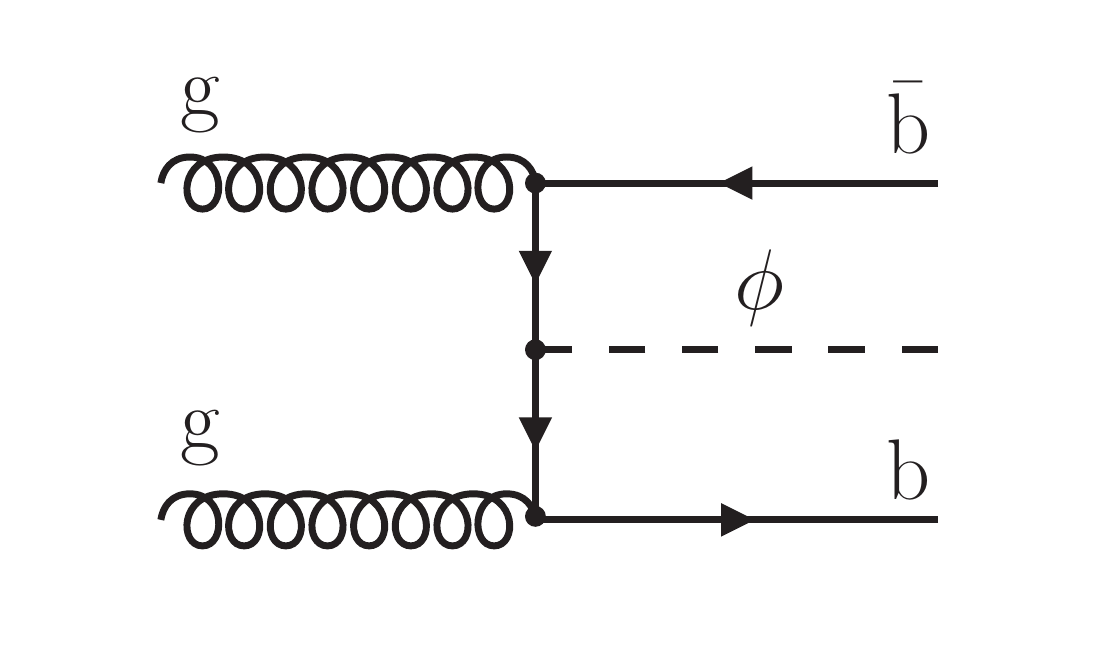}
  \includegraphics[width=0.32\textwidth]{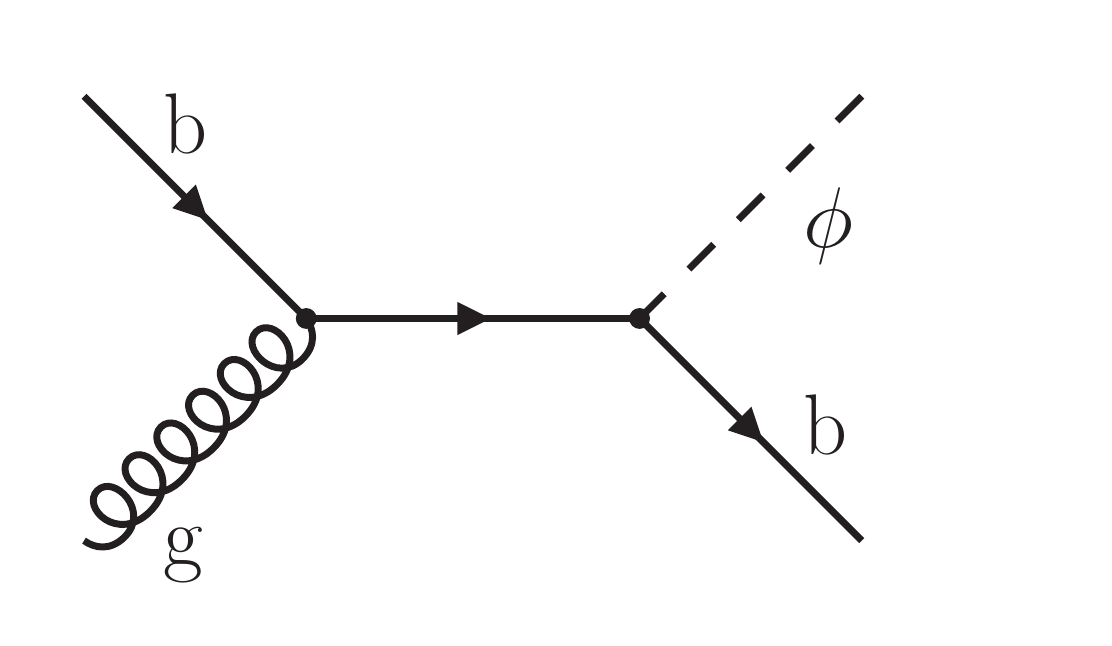}
  \caption {
    Diagrams for the production of neutral Higgs bosons \Pphi (left) via gluon 
    fusion, labelled as \ggPhi, and (middle and right) in association with \PQb 
    quarks, labelled as \bbPhi in the text. In the middle diagram, a pair of \PQb 
    quarks is produced from the fusion of two gluons, one from each proton. In 
    the right diagram, a \PQb quark from one proton scatters from a gluon from 
    the other proton. In both cases \Pphi is radiated off one of the \PQb 
    quarks.
  }
  \label{fig:production-diagrams}
\end{figure}

Neutral (pseudo)scalar bosons \Pphi appear in many extensions of the SM. They 
may have different couplings to the upper and lower components of the \SUTwoL 
fermion fields (associated with up- and down-type fermions) and gauge bosons. 
In several models, like the MSSM models discussed in Section~\ref{sec:MSSM}, 
the \Pphi couplings to down-type fermions are enhanced with respect to the 
expectation for an SM Higgs boson of the same mass, while the couplings to 
up-type fermions and vector bosons are suppressed. This makes down-type fermion 
final states, such as \ditau, particularly interesting for searches for neutral 
Higgs bosons in addition to \Phobs. An enhancement in the couplings to down-type 
fermions also increases the \bbPhi production cross section relative to \ggPhi, 
which is another characteristic signature of these models and motivates the 
search for enhanced production cross sections in this production mode with 
respect to the SM expectation. 
In a first interpretation of the data, which is meant to be as model independent 
as possible, we search for \Pphi production via the \ggPhi and \bbPhi processes 
in a range of $60\leq\mphi\leq3500\GeV$, where \mphi denotes the hypothesized \Pphi 
mass. Diagrams for these processes are shown in Fig.~\ref{fig:production-diagrams}. 
In a second, more specific interpretation of the data, we search for nonresonant 
\ditau production in a model with vector leptoquarks. Finally, in a third 
interpretation of the data, we survey the parameter space of two indicative 
benchmark scenarios of the MSSM, which predict multiresonance signatures, one of 
which is associated with \Phobs. The most important characteristics of the vector 
leptoquark model and the MSSM are described in the following.

\subsection{Vector leptoquarks}

Leptoquarks are hypothetical particles that carry both baryon and lepton 
numbers~\cite{Buchmuller:1986zs}, and are predicted by various BSM theories, such 
as grand unified theories~\cite{Pati:1973uk, Pati:1974yy, GeorgiGlashow, Fritzsch:1974nn},
technicolour models~\cite{Dimopoulos:1979es,Dimopoulos:1979sp,Technicolor, Lane:1991qh},
compositeness scenarios~\cite{LightLeptoquarks,Gripaios:2009dq}, and $R$-parity 
violating SUSY~\cite{Farrar:1978xj,Ramond:1971gb,Golfand:1971iw,Neveu:1971rx,
Volkov:1972jx,Wess:1973kz,Wess:1974tw,Fayet:1974pd,Nilles:1983ge,Barbier:2004ez}.
In recent years there has been a renewed interest in leptoquark models as a means 
of explaining various anomalies observed by a number of \PQb physics measurements 
performed in different experiments~\cite{Tanaka:2012nw,Barbieri:2015yvd,Faroughy:2016osc,
Bordone:2017bld,DiLuzio:2017vat,Greljo:2018tuh,Angelescu:2021lln,Cornella:2021sby}, 
most notably the apparent violation of lepton flavour universality in 
neutral-current~\cite{LHCb:2021trn} and charged-current~\cite{BaBar:2012obs,BaBar:2013mob,
Belle:2015qfa,LHCb:2015gmp,Belle:2016dyj,LHCb:2017rln,LHCb:2017smo} B meson decays. 
Models that contain a \TeVns-scale vector leptoquark (\Uone), characterized by 
its quantum numbers $(\SUThreeC, \SUTwoL, \UOneY) = (\textbf{3}, \textbf{1}, 2/3)$, 
are particularly appealing because they can explain both neutral- and charged-current 
anomalies at the same time~\cite{Barbieri:2015yvd,Faroughy:2016osc,Bordone:2017bld,
DiLuzio:2017vat,Greljo:2018tuh,Angelescu:2021lln,Cornella:2021sby}.

The Lagrangian for the \Uone coupling to SM fermions is given 
by~\cite{Cornella:2021sby}
\begin{linenomath}
  \begin{equation}
    \mathcal{L_{\mathrm{U}}} = \frac{\gU}{\sqrt{2}}\mathrm{U}^\mu \left[
    \betaL^{i\alpha}(\overline{q}_{\mathrm{L}}^{i}\gamma_\mu l_{\mathrm{L}}^\alpha) 
    +\betaR^{i\alpha}(\overline{d}_{\mathrm{R}}^{i}\gamma_\mu e_{\mathrm{R}}^{\alpha})
    \right]+\mathrm{h.c.},
  \end{equation}
\end{linenomath}
with the coupling constant \gU, where $q_{\mathrm{L}}$ and $d_{\mathrm{R}}$ ($l_{
\mathrm{L}}$ and $e_{\mathrm{R}}$) denote the left- and right-handed quark (lepton) 
doublets, and \betaL and \betaR are left- and right-handed coupling matrices, which 
are assumed to have the structures:
\begin{linenomath}
  \begin{equation}
    \betaL = \begin{pmatrix}
      0 & 0 & \betaLdtau \\
      0 & \betaLsmu & \betaLstau \\
      0 & \betaLbmu & \betaLbtau 
    \end{pmatrix},
    \quad
    \betaR = \begin{pmatrix}
      0 & 0 & 0 \\
      0 & 0 & 0 \\
      0 & 0 & \betaRbtau 
    \end{pmatrix}.
  \label{eqn:beta_couplings}
  \end{equation}
\end{linenomath}
The motivations for the assumed structures of these matrices are given in 
Ref.~\cite{Cornella:2021sby}. The normalization of \gU is chosen to give 
$\betaLbtau=1$. Two benchmark scenarios are considered, with different assumptions 
made about the value of \betaRbtau. In the first benchmark scenario (``VLQ BM 1''), 
\betaRbtau is assumed to be zero. In the second benchmark scenario (``VLQ BM 2''), 
\betaRbtau is assumed to be $-1$, which corresponds to a 
Pati--Salam-like~\cite{Pati:1974yy,Bordone:2017bld} \Uone leptoquark. The 
\betaLstau couplings are set to their preferred values from global fits to the 
low-energy observables presented in Ref.~\cite{Cornella:2021sby}, as summarized 
in Table~\ref{tab:betal_values}. The \betaLdtau, \betaLsmu, and \betaLbmu 
couplings are small and have negligible influence on the \ditau signature, and 
therefore have been set to zero. 

\begin{figure}[t]
  \centering
  \includegraphics[width=0.49\textwidth]{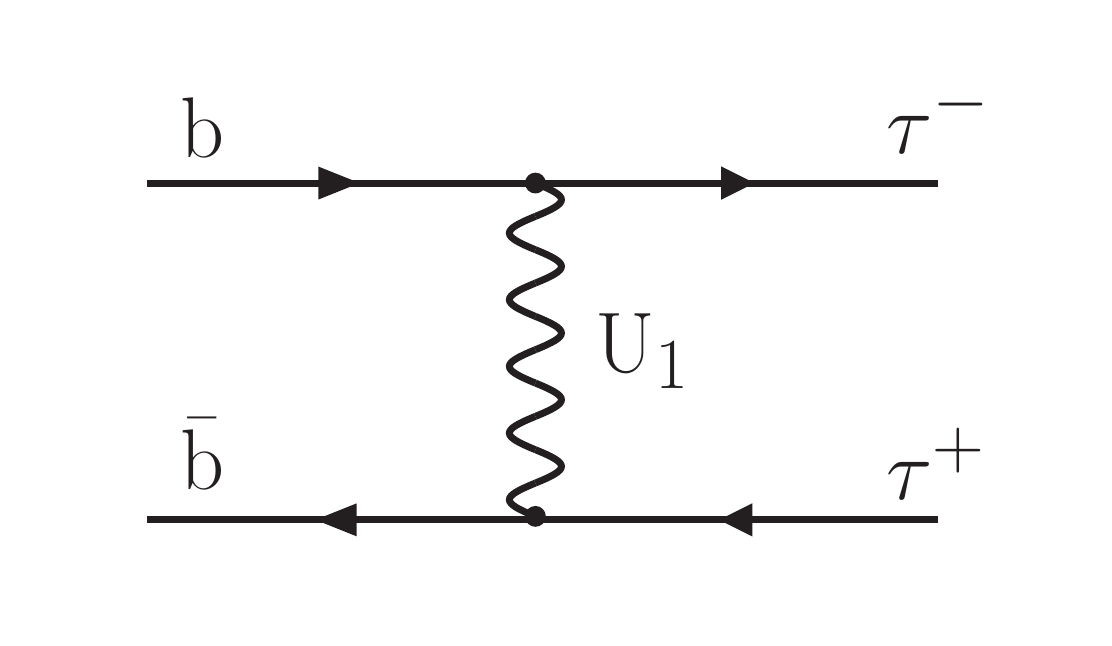}
  \caption {
    Diagram for the production of a pair of \PGt leptons via the $t$-channel
    exchange of a vector leptoquark \Uone.
  }
  \label{fig:production-diagrams-vlq}
\end{figure}

If the \Uone leptoquark mass (\mU) is sufficiently small, the \Uone particle will 
contribute to the \ditau spectrum via pair production with each \Uone subsequently 
decaying to a \qtau pair. For larger \mU, the pair production cross section is 
suppressed because of the decreasing probability that the initial-state partons possess 
sufficiently large momentum fractions of the corresponding protons to produce 
on-shell \Uone pairs. In this case the dominant contribution to the \ditau spectrum 
is via \Uone $t$-channel exchange in the $\PQb\PAQb$ initial-state as 
illustrated in Fig.~\ref{fig:production-diagrams-vlq}, with subdominant contributions 
from the equivalent $\PQb\PAQs$, $\PQs\PAQb$, and $\PQs\PAQs$ 
initiated processes. In our analysis we target the kinematic region of $\mU\gtrsim
1\TeV$, motivated by the experimental exclusion limits on \mU by direct searches, 
\eg in Ref.~\cite{CMS:2020wzx}. The contribution to the \ditau spectrum from 
\Uone pair production is negligible in this case, and we therefore consider only 
production through the $t$-channel exchange. 

\begin{table}[b]
  \topcaption{
    Summary of the preferred values and uncertainties of \betaLstau in the two 
    considered \Uone benchmark scenarios from Ref.~\cite{Cornella:2021sby}. 
  }
  \renewcommand{\arraystretch}{1.3}       
  \label{tab:betal_values}
  \centering
  \begin{tabular}{cc}
    Benchmark & \betaLstau \\ 
    \hline
    VLQ BM 1 & $0.19 {}^{+0.06}_{-0.09}$ \\
    VLQ BM 2 & $0.21 {}^{+0.05}_{-0.09}$ \\
  \end{tabular}
\end{table}

\subsection{The MSSM}
\label{sec:MSSM}

In the MSSM, which is a concrete example of the more general class of two Higgs 
doublet models (2HDMs)~\cite{Lee:1973iz,Branco:2011iw}, the Higgs sector requires 
two \SUTwo doublets, \Phiu and \Phid, to provide masses for up- and down-type 
fermions. In $CP$-conserving 2HDMs, this leads to the prediction of two charged 
($\PH^{\pm}$) and three neutral \Pphi bosons (\Ph, \PH, and \PA), where \Ph and 
\PH (with masses $\mh<\mH$) are scalars, and \PA (with mass \mA) is a pseudoscalar. 
The physical states \Ph and \PH arise as mixtures of the pure gauge fields with 
a mixing angle $\alpha$.

At tree level in the MSSM, the masses of these five Higgs bosons and $\alpha$ 
can be expressed in terms of the known gauge boson masses and two additional 
parameters, which can be chosen as \mA and the ratio of the vacuum expectation 
values of the neutral components of \Phiu and \Phid,
\begin{linenomath}
  \begin{equation}
    \tanb = \frac{\langle \Phiu^{0}\rangle}{\langle \Phid^{0}\rangle}.
  \end{equation}
\end{linenomath}
Dependencies on additional parameters of the soft SUSY breaking mechanism enter 
via higher-order corrections in perturbation theory. In the exploration of the 
MSSM Higgs sector these additional parameters are usually set to fixed values in 
the form of indicative benchmark scenarios to illustrate certain properties of 
the theory. The most recent set of MSSM benchmark scenarios provided by the LHC 
Higgs Working Group has been introduced in Refs.~\cite{Bahl:2018zmf,Bahl:2020kwe,
Bahl:2019ago} and summarized in Ref.~\cite{Bagnaschi:2791954}. The corresponding 
predictions of masses, cross sections, and branching fractions can be obtained 
from Ref.~\cite{MSSM_benchmark}. With one exception (the $M_{H}^{125}$ scenario), 
in these scenarios \Ph takes the role of \Phobs, and \PH and \PA are nearly 
degenerate in mass ($\mH\approx\mA$) in a large fraction of the provided parameter 
space.

For values of \mA much larger than the mass of the \PZ boson, the coupling of \PH 
and \PA to down-type fermions is enhanced by \tanb with respect to the expectation 
for an SM Higgs boson of the same mass, while the coupling to vector bosons and 
up-type fermions is suppressed. For increasing values of \tanb, \bbPhi (with 
$\Pphi=\PH,\,\PA$) is enhanced relative to \ggPhi production. The larger 
contribution of \PQb quarks to the loop in Fig.~\ref{fig:production-diagrams} 
(left) in addition leads to softer spectra of the \PH and \PA transverse momentum. 
Extra SUSY particles influence the production and decay via higher-order 
contributions to the interaction vertices that belong to \PQb quark lines. They 
also contribute directly to the loop in Fig.~\ref{fig:production-diagrams} (left). 

\section{The CMS detector}
\label{sec:detector}

The central feature of the CMS apparatus is a superconducting solenoid of 6\unit{m}
internal diameter, providing a magnetic field of 3.8\unit{T}. Within the
solenoid volume are a silicon pixel and strip tracker, a lead tungstate crystal
electromagnetic calorimeter (ECAL), and a brass and scintillator hadron calorimeter,
each composed of a barrel and two endcap sections. Forward calorimeters extend
the pseudorapidity ($\eta$) coverage provided by the barrel and endcap detectors.
Muons are measured in gas-ionization detectors embedded in the steel flux-return
yoke outside the solenoid. Events of interest are selected using a two-tiered 
trigger system. The first level (L1), composed of custom hardware processors, uses 
information from the calorimeters and muon detectors to select events at a rate of 
around 100\unit{kHz} within a fixed latency of about 4\mus~\cite{Sirunyan:2020zal}. 
The second level, known as the high-level trigger (HLT), consists of a farm of 
processors running a version of the full event reconstruction software optimized 
for fast processing, and reduces the event rate to around 1\unit{kHz} before data 
storage~\cite{Khachatryan:2016bia}. A more detailed description of the CMS detector,
together with a definition of the coordinate system used and the relevant kinematic
variables, can be found in Ref.~\cite{Chatrchyan:2008zzk}.

\section{Event reconstruction}
\label{sec:reconstruction}

The reconstruction of the \PP collision products is based on the particle-flow 
(PF) algorithm~\cite{Sirunyan:2017ulk}, which combines the information from all 
CMS subdetectors to reconstruct a set of particle candidates (PF candidates), 
identified as charged and neutral hadrons, electrons, photons, and muons. In the 
2016 (2017--2018) data sets the average number of interactions per bunch crossing 
was 23 (32). The primary vertex (PV) is taken to be the vertex corresponding to 
the hardest scattering in the event, evaluated using tracking information alone, 
as described in Ref.~\cite{CMS-TDR-15-02}. Secondary vertices, which are displaced 
from the PV, might be associated with decays of long-lived particles emerging 
from the PV. Any other collision vertices in the event are associated with 
additional, mostly soft, inelastic \PP collisions, referred to as pileup (PU).

Electrons are reconstructed using tracks from hits in the tracking system and 
energy deposits in the ECAL~\cite{Khachatryan:2015hwa,CMS:2020uim}. To 
increase their purity, reconstructed electrons are required to pass a multivariate 
electron identification discriminant, which combines information on track quality, 
shower shape, and kinematic quantities. For this analysis, a working point with 
an identification efficiency of 90\% is used, for a rate of jets misidentified as 
electrons of ${\approx}1\%$. Muons in the event are reconstructed by combining 
the information from the tracker and the muon detectors~\cite{Sirunyan:2018fpa}.
The presence of hits in the muon detectors already leads to a strong suppression 
of particles misidentified as muons. Additional identification requirements on 
the track fit quality and the compatibility of individual track segments with the 
fitted track can reduce the misidentification rate further. For this analysis, 
muon identification requirements with an efficiency of ${\approx}99\%$ are chosen, 
with a misidentification rate below 0.2\% for pions. 

The contributions from backgrounds to the electron and muon selections are further 
reduced by requiring the corresponding lepton to be isolated from any hadronic 
activity in the detector. This property is quantified by an isolation variable
\begin{linenomath}
  \begin{equation}
    \Irelem=\frac{1}{\ptem}\left(\sum\pt^{\text{charged}} + \max\left(0, 
    \sum\et^{\text{neutral}}+\sum\et^{\PGg}-\pt^{\text{PU}}\right)\right),
  \end{equation}
\end{linenomath}
where \ptem corresponds to the electron (muon) \pt and $\sum\pt^{\text{charged}}$, 
$\sum\et^{\text{neutral}}$, and $\sum\et^{\PGg}$ to the \pt (or transverse energy 
\et) sum of all charged particles, neutral hadrons, and photons, in a predefined 
cone of radius $\dR = \sqrt{\smash[b]{\left(\Delta\eta\right)^{2}+\left(\Delta
\varphi\right)^{2}}}$ around the lepton direction at the PV, where $\Delta\eta$ 
and $\Delta\varphi$ (measured in radians) correspond to the angular distances of 
the particle to the lepton in the $\eta$ and azimuthal $\varphi$ directions. The 
chosen cone size is $\dR=0.3\,(0.4)$ for electrons (muons). The lepton itself is 
excluded from the calculation. To mitigate any distortions from PU, only those 
charged particles whose tracks are associated with the PV are included. 
Since an unambiguous association with the PV is not possible for neutral hadrons 
and photons, an estimate of the contribution from PU ($\pt^{\text{PU}}$) is 
subtracted from the sum of $\sum\et^{\text{neutral}}$ and $\sum\et^{\PGg}$. 
This estimate is obtained from tracks not associated with the PV in the case of 
\Irelm and the mean energy flow per area unit in the case of \Irele. For negative 
values, the \Irelem is set to zero.

For further characterization of the event, all reconstructed PF candidates are 
clustered into jets using the anti-\kt algorithm with a distance parameter of 
0.4, as implemented in the \FASTJET software package~\cite{Cacciari:2008gp,
Cacciari:2011ma}. To identify jets resulting from the hadronization of \PQb 
quarks (\PQb jets) the \textsc{DeepJet} algorithm is used, as described in 
Refs.~\cite{Sirunyan:2017ezt,Bols:2020bkb}. In this analysis a working point 
of this algorithm is chosen that corresponds to a \PQb jet identification 
efficiency of ${\approx}80\%$ for a misidentification rate for jets originating
from light-flavour quarks or gluons of \order(1\%)~\cite{CMS-DP-2018-058}. Jets 
with $\pt>30\GeV$ and $\abs{\eta}<4.7$ and \PQb jets with $\pt>20\GeV$ and 
$\abs{\eta}<2.4$ are used in 2016. From 2017 onwards, after the upgrade of the 
silicon pixel detector, the \PQb jet $\eta$ range is extended to $\abs{\eta}<
2.5$. 

Jets are also used as seeds for the reconstruction of \tauh candidates. This 
is done by exploiting the substructure of the jets using the ``hadrons-plus-strips'' 
algorithm, as described in Refs.~\cite{Sirunyan:2018pgf,CMS:2022prd}. Decays into 
one or three charged hadrons with up to two neutral pions with $\pt>2.5\GeV$ are 
used. Neutral pions are reconstructed as strips with dynamic size in $\eta$-$\varphi$ 
from reconstructed photons and electrons contained in the seeding jet, where the 
latter originate from photon conversions. The strip size varies as a function of 
the \pt of the electron or photon candidates. The \tauh decay mode is then obtained 
by combining the charged hadrons with the strips. To distinguish \tauh candidates 
from jets originating from the hadronization of quarks or gluons, and from electrons 
or muons, the \textsc{DeepTau} (DT) algorithm is used, as described in 
Ref.~\cite{CMS:2022prd}. This algorithm exploits the information of the reconstructed 
event record (comprising tracking, impact parameter, and calorimeter cluster 
information), the kinematic and object identification properties of the PF candidates 
in the vicinity of the \tauh candidate and those of the \tauh candidate itself, 
and quantities that estimate the PU density of the event. It results in a 
multiclassification output $\yDT_{\alpha}\,(\alpha=\PGt,\,\Pe,\,\PGm,\,\text{jet})$ 
equivalent to a Bayesian probability of the \tauh candidate to originate from a 
genuine \PGt lepton, the hadronization of a quark or gluon, an electron, or a muon. 
From this output three discriminants are built according to 
\begin{linenomath}
  \begin{equation}
    D_{\alpha} = \frac{\yDT_{\PGt}}{\yDT_{\PGt}+\yDT_{\alpha}}, \quad
    \alpha=\,\Pe,\,\PGm,\text{ jet}.
  \end{equation}
\end{linenomath}
For the analysis presented here, predefined working points of \De, \Dm, and 
\Dj~\cite{CMS:2022prd} are chosen depending on the \ditau final state, for which 
the \tauh selection efficiencies and misidentification rates are given in 
Table~\ref{tab:dt-working-points}. Since the \jettau misidentification rate 
strongly depends on the \pt and initiating parton type of the misidentified jet, 
it should be viewed as approximate.    

The missing transverse momentum vector \ptvecmiss is also used for further 
categorization of the events. It is calculated as the negative vector \pt sum 
of all PF candidates, weighted by their probability to originate from the 
PV~\cite{Sirunyan:2019kia}, and exploits the pileup-per-particle identification 
algorithm~\cite{Bertolini:2014bba} to reduce the PU dependence. With \ptmiss we 
refer to the magnitude of this quantity.

\begin{table}[t]
  \centering
  \topcaption{
    Efficiencies for the identification of \tauh decays and corresponding 
    misidentification rates (given in parentheses) for the working points of \De, \Dm, 
    and \Dj, chosen for the \ditau selection, depending on the \ditau final state. 
    The numbers are given as percentages.
  }
  \begin{tabular}{cccc}
    & \De (\%) & \Dm (\%) & \Dj (\%) \\
    \hline
    \etau   & 54 (0.05) & 71.1 (0.13) &  \\
    \mutau  & \multirow{2}{*}{70 (2.60)} & 70.3 (0.03) & 49 (0.43) \\
    \tautau & & 71.1 (0.13) &  \\
  \end{tabular}
  \label{tab:dt-working-points}
\end{table}

\section{Event selection and categorization}
\label{sec:selection}

\subsection{Selection of \texorpdfstring{\ditau}{tau tau} 
candidates}

Depending on the final state, the online selection in the HLT step is based either 
on the presence of a single electron, muon, or \tauh candidate, or an \emu, \etau, 
\mutau, or \tautau pair in an event. In the offline selection further requirements 
on \pt, $\eta$, and \Irelem are applied in addition to the object identification 
requirements described in Section~\ref{sec:reconstruction}, as summarized in 
Table~\ref{tab:selection_kin}.

In the \emu final state an electron and a muon with $\pt>15\GeV$ and $\abs{\eta
}<2.4$ are required. Depending on the trigger path that has led to the online 
selection of an event, a stricter requirement of $\pt>24\GeV$ is imposed on one 
of the two leptons to ensure a sufficiently high trigger efficiency of the HLT 
selection. Both leptons are required to be isolated from any hadronic activity 
in the detector according to $\Irelem<0.15\,(0.2)$.

In the \etau (\mutau) final state, an electron (muon) with $\pt>25\,(20)\GeV$ 
is required if the event was selected by a trigger based on the presence of the 
\etau (\mutau) pair in the event. From 2017 onwards, the threshold on the muon is 
raised to 21\GeV. If the event was selected by a single-electron trigger, the \pt 
requirement on the electron is increased to 26, 28, or 33\GeV for the years 2016, 
2017, or 2018, respectively. For muons, the \pt requirement is increased to 23 
(25)\GeV for 2016 (2017--2018), if selected by a single-muon trigger. The electron 
(muon) is required to be contained in the central part of the detector with $\abs{
\eta}<2.1$, and to be isolated according to $\Irelem<0.15$. The \tauh candidate 
is required to have $\abs{\eta}<2.3$ and $\pt>35\,(32)\GeV$ if selected by an \etau 
(\mutau) pair trigger, or $\pt>30\GeV$ if selected by a single-electron (single-muon) 
trigger. In the \tautau final state, both \tauh candidates are required to have 
$\abs{\eta}<2.1$ and $\pt>40\GeV$. For events only selected by a single \tauh 
trigger, the \tauh candidate that has been identified with the triggering object 
is required to have $\pt>120\,(180)\GeV$ for events recorded in 2016 (2017--2018). 

The selected \PGt lepton decay candidates are required to be of opposite charge 
and to be separated by more than $\dR=0.3$ in the $\eta$-$\varphi$ plane in the 
\emu final state and by more than 0.5 otherwise. The closest distance of their 
tracks to the PV is required to be $d_{z}<0.2\cm$ along the beam axis. For electrons 
and muons, an additional requirement of $d_{xy}<0.045\cm$ in the transverse plane 
is applied. In rare cases, where more than the required number of \tauh candidates 
fulfilling all selection requirements is found, the candidate with the highest 
\Dj score is chosen. For electrons and muons, the most isolated candidate is chosen.

To avoid the assignment of single events to more than one final state, events with 
additional electrons or muons, fulfilling looser selection requirements than those 
given for each corresponding \ditau final state above, are rejected from the 
selection. These requirements also help with the suppression of background processes, 
such as \ZEE or \ZMM. 

\begin{table}[t]
  \centering
  \topcaption
  {
    Offline selection requirements applied to the electron, muon, and \tauh 
    candidates used for the selection of the \PGt pair. The expressions first 
    and second lepton refer to the label of the final state in the first column. 
    The \pt requirements are given in {\GeVns}. For the \emu final state two lepton 
    pair trigger paths, with a stronger requirement on the \pt of the electron 
    (muon), are used for the online selection of the event. For the \etau, 
    \mutau, and \tautau final states, the values (in parentheses) correspond to 
    the lepton pair (single lepton) trigger paths that have been used in the 
    online selection. A detailed discussion is given in the text. 
  }
  \begin{tabular}{lccccccc}
    Final state & Obs. 
    & \multicolumn{3}{c}{First lepton} 
    & \multicolumn{3}{c}{Second lepton} \\ 
    & & 2016 & 2017 & 2018 & 2016 & 2017 & 2018 \\
    \hline
    \emu 
    & \pt & \multicolumn{3}{c}{$>15\,(24)$} & \multicolumn{3}{c}{$>24\,(15)$} \\ 
    & $\abs{\eta}$ & \multicolumn{3}{c}{$<2.4$} & \multicolumn{3}{c}{$<2.4$} \\ 
    & \Irele & \multicolumn{3}{c}{$<0.15$} & \multicolumn{3}{c}{$<0.20$} \\ [\cmsTabSkip]
    \etau 
    & \pt & $>25\,(26)$ & $>25\,(28)$ & $>25\,(33)$ 
    & $\hphantom{>20\,(21)}$ & $>35\,(30)$ & $\hphantom{>20\,(21)}$ \\ 
    & $\abs{\eta}$ 
    & \multicolumn{3}{c}{$<2.1$}  & \multicolumn{3}{c}{$<2.3$} \\ 
    & \Irele 
    & \multicolumn{3}{c}{$<0.15$} & \multicolumn{3}{c}{--} \\ [\cmsTabSkip]
    \mutau 
    & \pt & $>20\,(23)$ & $>21\,(25)$ & $>21\,(25)$ 
    & \multicolumn{3}{c}{$>32\,(30)$} \\
    & $\abs{\eta}$ & \multicolumn{3}{c}{$<2.1$} & \multicolumn{3}{c}{$<2.3$} \\ 
    & \Irelm & \multicolumn{3}{c}{$<0.15$} & \multicolumn{3}{c}{--} \\ [\cmsTabSkip]
    \tautau 
    & \pt & $>40\,(120)$ & \multicolumn{2}{c}{$>40\,(180)$} & \multicolumn{3}{c}{$>40$} \\
    & $\abs{\eta}$ & \multicolumn{3}{c}{$<2.1$} & \multicolumn{3}{c}{$<2.1$} \\ 
  \end{tabular}
  \label{tab:selection_kin}
\end{table}

\subsection{Event categorization}
\label{sec:event-categories}

\subsubsection{Standard categories and signal extraction}
\label{sec:standard-categorisation}

To increase the sensitivity of the searches, all selected events are further split 
into categories. Events with at least one \PQb jet, according to the selection 
requirements given in Section~\ref{sec:reconstruction}, are combined into a global 
``\PQb tag'' category, used to target \bbPhi production and to control the 
background from top quark pair (\ttbar) production. All other events are subsumed 
into a global ``no \PQb tag'' category. The events in the \tautau final state are 
not further categorized beyond that point. In the \etau and \mutau final states, 
more categories are introduced in the global ``\PQb tag'' and ``no \PQb tag'' 
categories, based on the transverse mass of the \Pe (\PGm)-\ptvecmiss system 
defined as 
\begin{linenomath}
  \begin{equation}
    \label{eqn:mt_definition}
    \mTem =\mT(\ptvecem,\ptvecmiss),\quad\text{with} \quad
    \mT(\ptvec^{\kern1pt{i}}, \ptvec^{\kern1pt{j}}) = \sqrt{2\,\pt^{\kern1pt{i}}
      \,\pt^{\kern1pt{j}}\left(1-\cos\Delta\varphi
    \right)},
  \end{equation} 
\end{linenomath}
where $\Delta\varphi$ refers to the azimuthal angular difference between $\ptvec^{
\kern1pt{i}}$ and $\ptvec^{\kern1pt{j}}$. Events are divided into a tight-\mT 
($\mTem<40\GeV$) and a loose-\mT ($40<\mTem<70\GeV$) category. The \Pphi signal 
is expected to be concentrated in the tight-\mT category. However, the loose-\mT 
category increases the acceptance for $\mphi\gtrsim700\GeV$. 

\begin{figure}[t]
  \centering
  \includegraphics[width=0.90\textwidth]{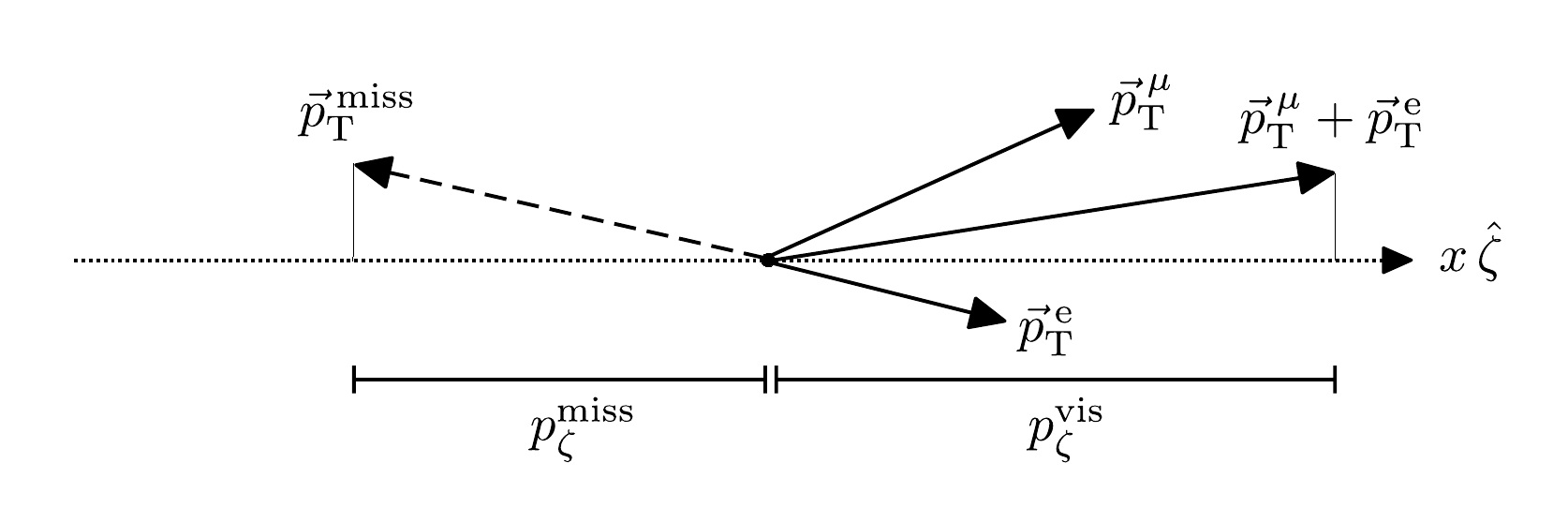}
  \caption{
    Inputs to the reconstruction of the event observable \Dzeta in the \emu final 
    state, as described in the text.
  }
  \label{fig:dzeta}
\end{figure}

In the \emu final state, events are categorized based on the observable 
\Dzeta~\cite{Abulencia:2005kq} defined as
\begin{linenomath}
  \begin{equation}
    \label{eqn:Dzeta}
      \Dzeta = \pzetamiss - 0.85\,\pzetavis ; \qquad
      \pzetamiss = \ptvecmiss\cdot\zetahat ; \qquad
      \pzetavis = \left(\ptvece+\ptvecm\right)\cdot\zetahat, \\
  \end{equation}
\end{linenomath}
where \zetahat corresponds to the bisectional direction between \ptvece and 
\ptvecm. The scalar products \pzetamiss and \pzetavis can take positive or 
negative values. Their linear combination has been optimized to maximize the 
sensitivity of the search. For events originating from \PW boson production in 
association with jets (\Wjets) or \ttbar production, the \ptvece, \ptvecm, and 
\ptvecmiss directions are more isotropically distributed leading to nonpeaking 
distributions in \Dzeta. For \ditau events from resonant decays, \ptvecmiss is 
expected to roughly coincide with \zetahat, and a stronger correlation between 
\pzetamiss and \pzetavis is expected to lead to a peaking distribution about 
$\Dzeta\approx0\GeV$. The inputs to the reconstruction of \Dzeta are illustrated 
in Fig.~\ref{fig:dzeta}. Three further categories are introduced as high-\Dzeta 
($\Dzeta>30\GeV$), medium-\Dzeta ($-10<\Dzeta<30\GeV$), and low-\Dzeta ($-35<
\Dzeta<-10\GeV$). A \Pphi signal is expected to be concentrated in the 
medium-\Dzeta category. However, the low- and high-\Dzeta categories still 
contribute to an increase of the sensitivity of the model-independent \Pphi 
search in the \emu final state by ${\approx}10\%$. A control category in the 
\emu final state with at least one \PQb jet and $\Dzeta<-35\GeV$ is used to 
constrain the normalization of \ttbar events in the fit used for signal 
extraction. 

\begin{figure}[t]
  \centering
  \includegraphics[width=0.48\textwidth]{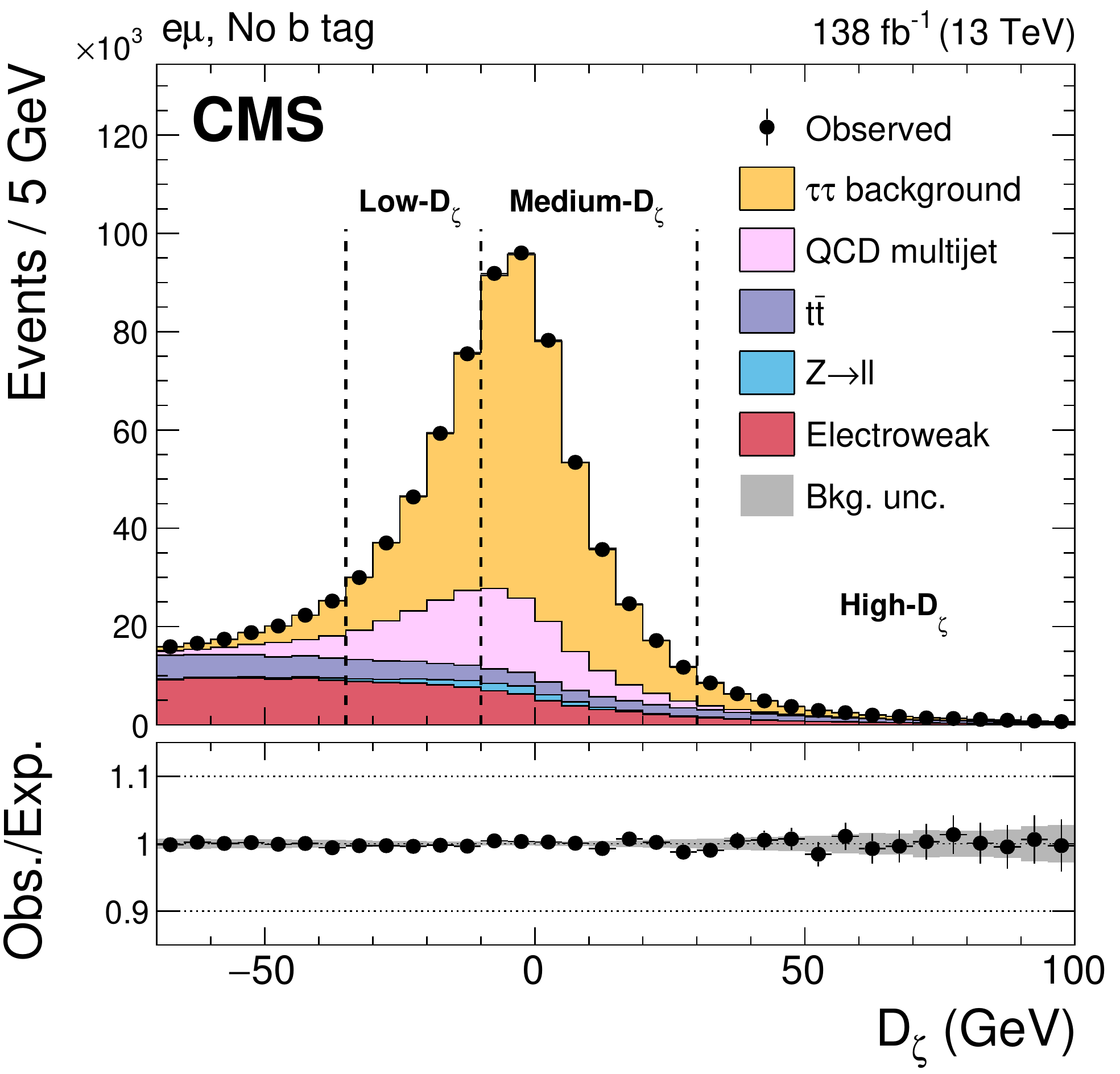}
  \includegraphics[width=0.48\textwidth]{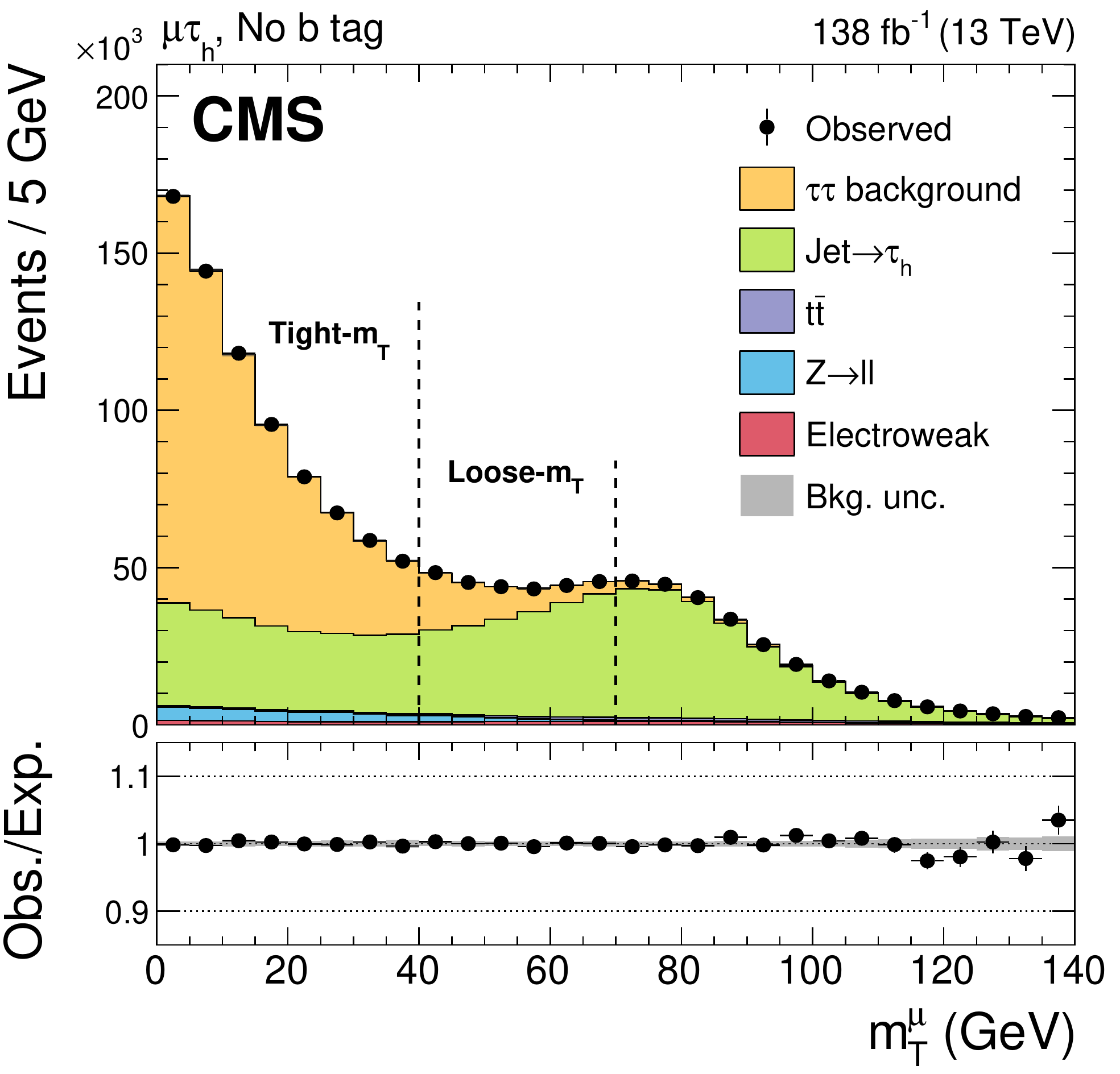}
  \caption {
    Observed and expected distributions of (left) \Dzeta in the \emu final state 
    and (right) \mTm in the \mutau final state. The distributions are shown in 
    the global ``no \PQb tag'' category before any further event categorization 
    and after an individual background-only fit to the data in each corresponding 
    variable. The grey shaded band represents the complete set of uncertainties 
    used for signal extraction, after the fit. A detailed discussion of the data 
    modelling is given in Section~\ref{sec:data-model}. The vertical dashed lines 
    indicate the category definitions in each of the final states, as described 
    in the text. In the lower panels of each figure the ratio of the observed 
    numbers of events per bin to the background expectation is shown.
  }
  \label{fig:sub-categories}
\end{figure}

\begin{figure}[!b]
  \centering
  \includegraphics[width=0.98\textwidth]{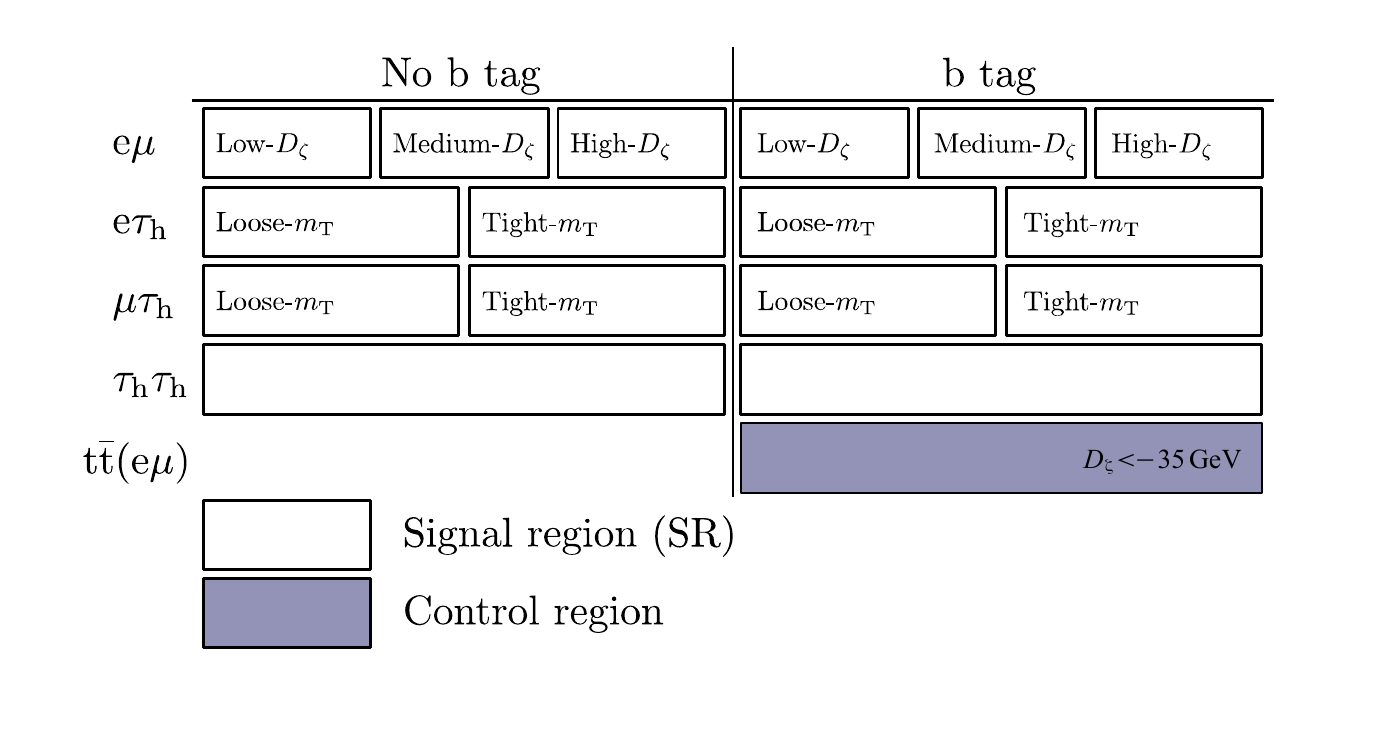}
  \caption {
    Overview of the categories used for the extraction of the signal for the 
    model-independent \Pphi search for hypothesized values of $\mphi\geq250\GeV$, 
    the vector leptoquark search, and the interpretation of the data in 
    MSSM benchmark scenarios. 
  }
  \label{fig:categories}
\end{figure}

In summary, this leads to 17 event categories per data-taking year. 
Figure~\ref{fig:sub-categories} shows the \Dzeta and \mTm distributions in the 
\emu and \mutau final states, before splitting the events into the categories 
described above. The category definitions are indicated by the vertical dashed 
lines in the figures. An overview of the categories described above is given in 
Fig.~\ref{fig:categories}. 

In all cases the signal is extracted from distributions of the total transverse 
mass~\cite{Aad:2014vgg} defined as
\begin{linenomath}
  \begin{equation}
      \mTtot = \sqrt{\mT^{2}(\ptvectauone,\ptvectautwo)+\mT^{2}(\ptvectauone,
      \ptvecmiss)+\mT^{2}(\ptvectautwo,\ptvecmiss)},
      \label{eq:mttot}
  \end{equation}
\end{linenomath}
where $\tau_{1(2)}$ refers to the first (second) \PGt final state indicated in 
the \emu, \etau, \mutau, and \tautau final state labels, and \mT between two 
objects with transverse momenta \ptvectauone and \ptvectautwo is defined in 
Eq.~(\ref{eqn:mt_definition}). This quantity is expected to provide superior 
discriminating power between resonant signals with $\mphi\gtrsim250\GeV$ and 
nonpeaking backgrounds, such as \Wjets or \ttbar production in the high-mass 
tails of the distribution.

This strategy is used for the model-independent \Pphi search, to extract the 
expected signal for hypothesized values of $\mphi\geq250\GeV$. It is also used 
for the extraction of the \PA and \PH signal (for $\mA,\,\mH\gtrsim250\GeV$), 
when interpreting the data in MSSM benchmark scenarios, and for the vector 
leptoquark search, which is most sensitive to an excess over the background 
expectation for $\mTtot\gtrsim250\GeV$ as will be discussed in 
Section~\ref{sec:simulation-signal}.

To increase the sensitivity of the analyses for the model-independent \Pphi 
search for hypothesized values of $\mphi<250\GeV$ and the low-mass resonance \Ph 
for the interpretation of the data in MSSM benchmark scenarios, this signal 
extraction strategy is modified as discussed in the following sections. 

\begin{figure}[hb]
  \centering
  \includegraphics[width=0.95\textwidth]{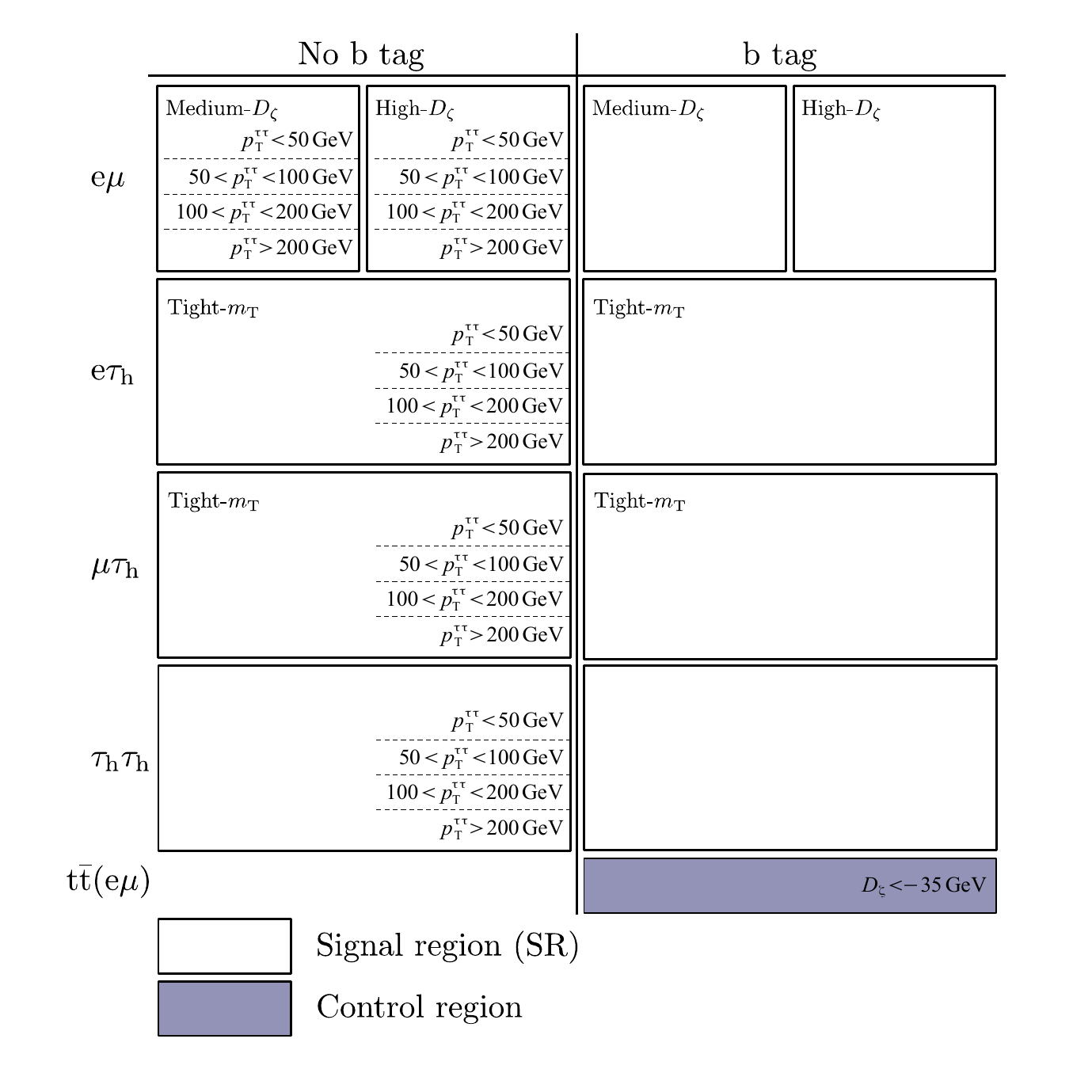}
  \caption{ 
    Overview of the categories used for the extraction of the signal for the 
    model-independent \Pphi search for $60\leq\mphi<250\GeV$.
  }
  \label{fig:categories_lowmass}
\end{figure}

\subsubsection{Modifications for the low-mass model-independent 
\texorpdfstring{\Pphi}{phi} search}
\label{sec:low-mass-categorisation}

For hypothesized values of $\mphi<250\GeV$, the background from \ZTT production,  
which features a peaking mass distribution in a region close to the signal mass, 
starts to exceed the nonpeaking backgrounds. The \mTtot distribution loses 
discrimination power and some of the categories that were introduced to increase 
the acceptance for high-mass signals are not useful anymore. Therefore, the 
signal extraction strategy is modified in the following way. The low-\Dzeta and 
loose-\mT categories are removed. The remaining ``no \PQb tag'' categories are 
further split by \pttt, obtained from the vectorial sum \pt of the visible \PGt 
decay products and \ptvecmiss, according to $\pttt<50\GeV$, $50<\pttt<100\GeV$, 
$100<\pttt<200\GeV$, and $\pttt<200\GeV$, where \pttt is used as an estimate for 
the \Pphi \pt (\ptphi) in data. No further splitting based on \pttt is applied 
to the ``\PQb tag'' categories because of the lower event populations in these 
categories. In summary, this leads to 26 event categories per data-taking year. 
An overview of this modified set of categories is given in 
Fig.~\ref{fig:categories_lowmass}. 

In these categories, the signal is extracted from a likelihood-based fit of the 
invariant mass of the \ditau system, \mtt, before the decay of the \PGt 
leptons~\cite{Bianchini:2014vza}. This estimate combines the measurement of 
\ptvecmiss and its covariance matrix with the measurements of the visible \ditau 
decay products, utilizing the matrix elements for unpolarized \PGt 
decays~\cite{Bullock:1992yt} for the decay into leptons and the two-body phase 
space~\cite{ParticleDataGroup:2020ssz} for the decay into hadrons. On average 
the resolution of \mtt amounts to about 10--25\%, depending on the kinematic 
properties of the \ditau system and the \ditau final states, where the latter 
is related to the number of neutrinos that escape detection. This approach 
exploits the better \mphi resolution of \mtt compared to \mTtot, together with 
the usually harder \ptphi, compared to the \ZTT \pt spectrum. 

\begin{table}[b]
  \centering
  \topcaption{
    Event categories and discriminants used for the extraction of the signals, 
    for the searches described in this paper. We note that \mphi refers to the 
    hypothesized mass of the model-independent \Pphi search, while \mtt refers 
    to the reconstructed mass of the \ditau system before the decays of the \PGt 
    leptons, and thus to an estimate of \mphi in data. The variable $y_{l}$ 
    refers to the output functions of the NNs used for signal extraction in 
    Ref.~\cite{CMS:2022kdi}. 
  }
  \begin{tabular}{llllc}
    \multicolumn{2}{l}{Search} & Categories & Additional & Discr. \\
    & & & selections & variable \\ 
    \hline
    \multirow{2}{*}{Model-independent (\Pphi)}  
    & $\mphi<250\GeV$ 
    & Fig.~\ref{fig:categories_lowmass} 
    & \NA
    & \mtt \\
    & $\mphi\geq 250\GeV$ 
    & Fig.~\ref{fig:categories}
    & \NA
    & \mTtot \\[\cmsTabSkip] 
    \multicolumn{2}{l}{Vector leptoquark (\Uone)}  
    & Fig.~\ref{fig:categories}
    & \NA
    & \mTtot \\[\cmsTabSkip] 
    \multicolumn{2}{l}{\multirow{3}{*}{MSSM benchmark scenarios (\PA, \PH, \Ph)}}
    & \multirow{2}{*}{NN-analysis}
    & $\mtt<250\GeV$,
    & \multirow{2}{*}{$y_{l}$} \\
    &&& $\Dzeta>-35\GeV$ (in \emu) &\\ [1ex]
    & & Fig.~\ref{fig:categories}
    & $\mtt>250\GeV$
    & \mTtot \\
  \end{tabular}
  \label{tab:categories-vs-analyses}
\end{table}

\subsubsection{Modifications for the MSSM interpretation}

The MSSM predicts three neutral Higgs bosons \Pphi, one of which is identified 
with \Phobs. Each benchmark scenario has to match the observed \Phobs properties. 
To exploit the best possible experimental knowledge about \Phobs all events in 
the global ``no \PQb tag'' category are split by \mtt. For events with $\mtt>250
\GeV$, the categories described in Section~\ref{sec:standard-categorisation} are 
used. For events with $\mtt<250\GeV$, the neural-network-based (NN) analysis, 
which was used for the stage-0 simplified template cross section measurements of 
Ref.~\cite{CMS:2022kdi}, is used to obtain the most precise estimates from data 
for \Phobs production via gluon fusion (\ggh), vector boson fusion (VBF), and 
vector boson associated production (\Vh). Although the NN is trained specifically 
to target events with an SM-like \Pphi with $\mphi=125\GeV$, signal events for 
the additional Higgs bosons can also enter the NN categories for $\mphi\lesssim 
250\GeV$, and the $y_{l}$ discriminators contribute to the separation of such 
events from the background.

This modification adds 18 background and 8 signal categories from the NN-analysis 
per data-taking year. We will refer to these as the "NN categories" throughout 
this paper. In these categories, the \Phobs signal is extracted from distributions 
of the NN output functions $y_{l}$ in each signal and background category $l$. 
For the NN-analysis in Ref.~\cite{CMS:2022kdi}, $\mT^{\emu}$ calculated from 
$\ptvece+\ptvecm$ and \ptvecmiss is required to be less than 60\GeV in the \emu 
final state, to prevent event overlap with analyses of other \Phobs decay modes 
in the SM interpretation. For the analysis presented here, this requirement is 
replaced by $\Dzeta>-35\GeV$. 

A summary of the categories and discriminating variables used for signal 
extraction for each of the analyses presented in this paper is given in 
Table~\ref{tab:categories-vs-analyses}.  

\begin{table}[t]
  \topcaption{
    Background processes contributing to the event selection, as discussed in 
    Section~\ref{sec:selection}. The symbol $\Pell$ corresponds to an electron or 
    muon. The second column refers to the experimental signature in the analysis, 
    the last four columns indicate the estimation methods used to model each 
    corresponding signature, as described in 
    Sections~\ref{sec:tau-embedding}--\ref{sec:simulation}. Diboson and single 
    \PQt production are part of the process group iv) discussed in 
    Section~\ref{sec:data-model}. QCD(\emu) refers to QCD multijet production 
    with an \emu pair in the final state.
  }
  \label{tab:bg-processes}
  \centering
  \begin{tabular}{lr@{$\to$}lcccc}
    & \multicolumn{2}{c}{} &\multicolumn{4}{c}{Estimation method} \\
    Background process & \multicolumn{2}{c}{Final-state signature} & \PGt-emb. & \FF & SS & Sim. \\
    \hline
    \multirow{3}{*}{\Zgamma} & \multicolumn{2}{c}{\ditau} & $\checkmark$ & \NA & \NA & \NA \\
    & $\hspace{1.1cm}$Jet & \tauh  & \NA & $\checkmark$ & \NA & \NA \\
    & \multicolumn{2}{c}{$\Pell\Pell$} & \NA & \NA & \NA & $\checkmark$ \\ [\cmsTabSkip]
    \multirow{3}{*}{\ttbar} & \multicolumn{2}{c}{\ditau} & $\checkmark$ & \NA & \NA & \NA \\
    & Jet & \tauh  & \NA & $\checkmark$ & \NA & \NA \\ 
    & \multicolumn{2}{c}{\LplusX}   & \NA & \NA & \NA & $\checkmark$ \\ [\cmsTabSkip]
    \multirow{3}{*}{Diboson+single \PQt} & \multicolumn{2}{c}{\ditau} & $\checkmark$ & \NA & \NA & \NA \\
    & Jet & \tauh  & \NA & $\checkmark$ & \NA & \NA \\
    & \multicolumn{2}{c}{\LplusX}   & \NA & \NA & \NA & $\checkmark$ \\ [\cmsTabSkip]
    \multirow{2}{*}{\Wjets}   & Jet & \tauh & \NA & $\checkmark$ & \NA & \NA \\
    & Jet & $\Pell$ & \NA & \NA & \NA & $\checkmark$ \\[\cmsTabSkip]
    \multirow{2}{*}{QCD multijet} & Jet & \tauh & \NA & $\checkmark$ & \NA & \NA \\ 
    & \multicolumn{2}{c}{QCD(\emu)} & \NA & \NA & $\checkmark$ & \NA \\[\cmsTabSkip]
    \Phobs & \multicolumn{2}{c}{\ditau} & \NA & \NA & \NA & $\checkmark$ \\
  \end{tabular}
\end{table}

\section{Background and signal modelling}
\label{sec:data-model}

All SM background sources that are relevant after the event selection 
described in Section~\ref{sec:selection} are listed in Table~\ref{tab:bg-processes}. 
The expected background composition depends on the \ditau final state, event 
category, and the tested signal mass hypothesis. The most abundant source of 
background in the ``\PQb tag'' categories is \ttbar production. In the 
``no \PQb tag'' categories \ZTT forms the largest fraction of background 
processes, followed by \Wjets production and events containing purely quantum 
chromodynamics (QCD) induced gluon and light-flavour quark jets, referred to 
as QCD multijet production. These backgrounds are grouped according to their 
experimental signatures into: 

\begin{enumerate}
\item events containing genuine \PGt lepton pairs (\ditau); 
\item events with quark- or gluon-induced jets misidentified as \tauh candidates 
(\jettau) or light leptons (\jettoL) in the \emu final state; 
\item \ttbar events where an intermediate \PW boson decays into an electron, muon, 
or \PGt lepton, which do not fall into the previous groups (labelled as ``\ttbar'' 
in later figures); 
\item remaining background processes that are of minor importance for the analysis 
and not yet included in any of the previous event groups (labelled as ``others'' 
in later figures). 
\end{enumerate}

Event group (i) mostly contains $\Zgamma\to\PGt\PGt$ events, with smaller 
contributions from \ttbar, diboson, and single~\PQt quark production. These 
events are modelled using the \PGt-embedding method~\cite{Sirunyan:2019drn}, 
labelled "\Pgt-emb." in Table~\ref{tab:bg-processes}, and discussed in 
Section~\ref{sec:tau-embedding}. Event group (ii) contains events from QCD 
multijet, \Wjets, \Zjets, \ttbar, diboson and single~\PQt quark production 
with \jettau misidentification, and QCD multijet production with \jettoL 
misidentification in the \emu final state. The events with \jettau 
misidentification are estimated from the ``fake factor'' (\FF) method, labelled 
\mbox{"\FF"} in Table~\ref{tab:bg-processes}, and discussed in 
Section~\ref{sec:FF-method}. The events with \jettoL misidentification in the 
\emu final state are estimated from the ``same-sign'' (SS) method, labelled "SS" 
in Table~\ref{tab:bg-processes}, and discussed in Section~\ref{sec:em-background}. 
Event group (iv) comprises diboson and single~\PQt quark production (labelled as 
``electroweak'' in Fig.~\ref{fig:sub-categories} left), H(125) production, \ZMM 
and \ZEE events, and \Wjets events with \jettoL misidentification. Events from 
event groups iii) and iv), and all signal processes are estimated from full event 
simulation, labelled "Sim." in Table~\ref{tab:bg-processes}, and discussed in 
Section~\ref{sec:simulation}. 

\subsection{\texorpdfstring{Backgrounds with genuine \PGt lepton pairs (\ditau)}
{Backgrounds with genuine tau lepton pairs (tau tau)}}
\label{sec:tau-embedding}

For all events where the decay of a \PZ boson results in two genuine \PGt 
leptons, the \PGt-embedding method, as described in Ref.~\cite{Sirunyan:2019drn}, 
is used. For this purpose, \mumu events are selected in data. All energy deposits 
of the muons are removed from the event record and replaced by simulated \PGt 
lepton decays with the same kinematic properties as the selected muons. In this 
way the method relies only on the simulation of the well-understood \PGt lepton 
decay and its energy deposits in the detector, while all other parts of the event, 
such as the identification and reconstruction of jets, including \PQb jets, or 
the non-\PGt related parts of \ptvecmiss are obtained from data. This results 
in an improved modelling of the data compared with the simulation of the full 
process. In turn, several simulation-to-data corrections, as detailed in 
Section~\ref{sec:corrections}, are not needed. The selected muons predominantly 
originate from \ZMM events. However, contributions from other processes resulting 
in two genuine \PGt leptons, like \ttbar or diboson production, are also covered 
by this model. For a selection with no (at least one) \PQb jet in the event, as 
described in Section~\ref{sec:selection}, 97\% (84\%) of the \mumu events selected 
for the \PGt-embedding method are expected to originate from \ZMM and ${<}1\%$ 
(14\%) from \ttbar production. A detailed discussion of the selection of the 
original \mumu events, the exact procedure itself, its range of validity, and 
related uncertainties is reported in Ref.~\cite{Sirunyan:2019drn}. 

\subsection{\texorpdfstring{Backgrounds with jets misidentified as hadronically decaying \PGt leptons (\jettau)}{Backgrounds with jets misidentified 
as hadronically decaying tau leptons (jet to tau)}}
\label{sec:FF-method}

The main processes contributing to \jettau events in the \etau, \mutau, and 
\tautau final states are QCD multijet, \Wjets, and \ttbar production. These 
events are estimated using the \FF method described in Refs.~\cite{Sirunyan:2018qio, 
Sirunyan:2018zut}, and adapted to the analyses described in this paper. 

For this purpose, the signal region (SR), defined by the event selection given 
in Section~\ref{sec:selection}, is complemented by three additional regions: the 
application region (AR) and two determination regions \DRi, where $i$ stands for 
QCD or \Wjets. For the AR a looser working point for the identification of the 
\tauh candidate is chosen and the events from the SR are excluded, which is the 
only selection difference with respect to the SR. In this way the AR forms an 
orthogonal, though still adjacent, sideband to the SR that is enriched in \jettau 
events. The events in the AR are then multiplied with a transfer function, which 
is obtained from each corresponding \DRi or simulation, to estimate the contribution 
of \jettau events in the SR. The background processes in the AR and each 
corresponding \DRi that are not targeted by this method are estimated either 
from simulation or the \PGt-embedding method and subtracted from the data. 

In the \tautau final state, where QCD multijet production contributes 
${\gtrsim}95\%$ of the events in the AR, the transfer function is determined 
from DR$^{\text{QCD}}$ only, for which the charges of the two selected \tauh 
candidates are required to be of same sign. This function is assumed to be 
applicable also for the small fraction of \Wjets and \ttbar events in the AR. 
In this final state, both \tauh candidates usually originate from \jettau 
misidentification. We require only the \tauh candidate with the larger \pt to 
fulfil the AR requirements, which provides an estimate for events where only 
this \tauh candidate is misidentified. Events in which the \tauh candidate with 
the larger \pt is a genuine \PGt lepton and the one with the lower \pt is 
misidentified, which constitute ${\approx}2\%$ of the total \jettau background, 
are modelled from simulation. 

In the \etau (\mutau) final state, where the sharing of processes contributing 
to the AR is more equal, separate contributions to the transfer function \FFi 
are used, where the index $i$ runs over the processes of QCD multijet, \Wjets, 
and \ttbar production. For QCD multijet and \Wjets production each \FFi is derived 
in its corresponding \DRi. For DR$^{\text{QCD}}$ we require $0.05<\Irelem<0.15$ 
and the charges of the selected $\Pe (\PGm)$ and the \tauh candidate to be of same 
sign. For DR$^{\Wjets}$ we require $\mTem>70\GeV$ and the absence of \PQb jets. 
The estimate of \FFttbar is obtained from simulation. Each \FFi is then used to 
estimate the yield \NSR and kinematic properties of the combination of the main 
contributing backgrounds $i$ in the SR from the number of events \NAR in the AR 
according to 
\begin{linenomath}
  \begin{equation}
    \label{eq:FF}
    \NSR = \left(\sum\limits_{i}w_{i}\FFi\right)\NAR,
    \qquad i=\text{QCD, }\PW\text{+jets, }\ttbar.
  \end{equation}
\end{linenomath}
Each \FFi is combined into a weighted sum, using the simulation-based estimate 
of the fractions $w_{i}$ of each process in the AR. A template fit to the data 
in the AR yields a similar result for the $w_{i}$.

Each \FFi is computed on an event-by-event basis. It mainly depends on the \pt 
of the \tauh candidate with the larger \pt, \ptth, the ratio $\ptjet/\ptth$ 
where \ptjet corresponds to the \pt of the jet seeding the \tauh reconstruction, 
and the jet multiplicity \Njet. Each \FFi is further subject to a number of residual 
corrections derived from both control regions in data and simulation to take 
subleading dependencies of the \FFi into account. Depending on the transfer 
function \FFi and the \ditau final state these are dependencies on \ptell, 
the invariant mass of the visible decay products of the \ditau system, \Irelell, 
or \ptth of the second-leading \tauh candidate.  

\subsection{\texorpdfstring{Backgrounds with jets misidentified as electron-muon pairs (QCD(\emu))}{Backgrounds with jets misidentified as electron-muon pairs (QCD(emu))}}
\label{sec:em-background}

The background from QCD multijet production where two quark- or gluon-induced 
jets are misidentified as an \emu pair is estimated using the SS method. In this 
case, an AR is distinguished from the SR by requiring the charges of the electron 
and muon to have the same sign. A sideband region DR is defined requiring the 
muon to be nonisolated ($0.2<\Irelm<0.5$). From this DR an SS to opposite-sign 
(OS) transfer function \TF is obtained to extrapolate the number \NAR of events 
in the AR to the number \NSR of events in the SR according to
\begin{linenomath}
  \begin{equation}
    \label{eq:TF}
    \NSR = \TF\,\NAR.
  \end{equation}
\end{linenomath} 

The function \TF primarily depends on the distance $\dR(\Pe, \PGm)$ between the 
\Pe and \PGm trajectories in $\eta$-$\varphi$ and \Njet. Additional dependencies 
on the electron and muon \pt enter via a bias correction, ranging from 0.85--0.9. To 
validate the method, a second transfer function $\TF^{\prime}$ is calculated from 
a modified DR$^{\prime}$ with an isolated muon ($\Irelm<0.2$) and nonisolated 
electron ($0.15<\Irele<0.5$), which is applied to the SS selection of the DR. The 
resulting event yield and shapes of the \mTtot and \mtt distributions are compared 
to the OS selection of the DR. This test reveals a consistent result within the 
statistical uncertainties of the estimate, for events with $\Nbjet=0$. For events 
with $\Nbjet\geq1$, a global correction factor $r_{\PQb}$ is required, with a value of 
0.71--0.75 depending on the year of data-taking. 

A potential bias from requiring the muon to be nonisolated in the definition of 
DR is checked from a third definition of the transfer function $\TF^{\prime
\prime}$, in a DR$^{\prime\prime}$ with a nonisolated muon ($0.2<\Irelm<0.5$) 
and electron ($0.15<\Irele<0.5$). This test reveals another correction of 
0.94--0.95, depending on the year of data-taking, to correct for the fact that 
$r_{\PQb}$, with an isolated muon, is systematically smaller by ${\approx}5\%$ 
than in the case of a nonisolated muon.    

\subsection{Simulated backgrounds and signal}
\label{sec:simulation}

In the \tautau final state, the \PGt-embedding and \FF methods cover 97\% of all 
expected background events. The fractions of expected background events described 
by these two methods are 83\% in the \etau and 90\% in the \mutau final states. In the 
\emu final state, 53\% of all events are obtained by either the \PGt-embedding or 
SS method. All remaining events originate from processes such as \PZ boson, \ttbar, 
or diboson production, where at least one decay of a vector boson into an electron 
or muon is not covered by any of the previously discussed methods. These backgrounds 
and the signal processes are modelled using the simulation of the full processes. 

\subsubsection{Background processes}

The \Wjets and \ZLL processes are simulated at leading order (LO) accuracy in the 
strong coupling \alpS, using the \MGvATNLO 2.2.2 (2.4.2) event generator~\cite{Alwall:2011uj,
Alwall:2014hca} for the simulation of the data taken in 2016 (2017--2018). To 
increase the number of simulated events in regions of high signal purity, 
supplementary samples are generated with up to four outgoing partons in the hard 
interaction. For diboson production, \MGvATNLO is used at next-to-LO (NLO) precision 
in \alpS. In each case, the FxFx~\cite{Frederix:2012ps} (MLM~\cite{Alwall:2007fs}) 
prescription is used to match the NLO (LO) matrix element calculation with the 
parton shower model. For \ttbar~\cite{Alioli:2011as} and ($t$-channel) single~\PQt 
quark production~\cite{Frederix:2012dh}, samples are generated at NLO precision in 
\alpS using \POWHEG 2.0~\cite{Nason:2004rx,Frixione:2007vw,Alioli:2010xd,Jezo:2015aia}. 
The \POWHEG version 1.0 at NLO precision is used for single~\PQt quark production 
in association with a \PW boson ($\PQt\PW$ channel)~\cite{Re:2010bp}. 

When compared with data, \Wjets, \ZLL, \ttbar, and single~\PQt quark events in the 
$\PQt\PW$ channel are normalized to their cross sections at next-to-NLO (NNLO) 
precision in \alpS~\cite{Melnikov:2006kv,Czakon:2011xx,Kidonakis:2013zqa}. 
Single~\PQt quark ($t$-channel) and diboson events are normalized to their cross 
sections at NLO precision in \alpS or higher~\cite{Kidonakis:2013zqa,
Campbell:2011bn,Gehrmann:2014fva}.

\subsubsection{Signal processes}
\label{sec:simulation-signal}

The kinematic properties of single \Ph production are simulated at NLO precision 
in \alpS using \POWHEG 2.0 separately for the production via \ggh~\cite{Bagnaschi:2011tu}, 
VBF~\cite{Nason:2009ai}, or in association with a \PZ (\Zh) or \PW (\Wh) 
boson~\cite{Luisoni:2013cuh,Granata:2017iod}. For \ggh production, the distributions 
of the \Ph boson \pt and the jet multiplicity in the simulation are tuned to match 
the NNLO accuracy obtained from full phase space calculations with the NNLOPS event 
generator~\cite{Hamilton:2013fea,Hamilton:2015nsa}. For this purpose, \Ph is 
assumed to behave as expected from the SM. This applies to the modelling of \Phobs 
as part of the background for the model-independent \Pphi search, as well as for 
the SM and the MSSM hypotheses for the interpretation of the data in MSSM benchmark 
scenarios, where \Ph is associated with \Phobs with properties as expected from 
the SM. 

The production of \Pphi, \PH, and \PA bosons via gluon fusion is simulated at NLO 
precision in \alpS using the 2HDM implementation of \POWHEG 2.0~\cite{Bagnaschi:2011tu}. 
To account for the multiscale nature of the process in the NLO plus parton shower 
prediction, the \pt spectra corresponding to the contributions from the \PQt quark 
only, \PQb quark only, and $\PQt\PQb$-interference are each calculated separately. 
The \POWHEG damping factor \hdamp, which controls the matching between the matrix 
element calculation and the parton shower, is set specifically for each contribution 
as proposed in Refs.~\cite{Harlander:2014uea,Bagnaschi:2015bop,Bagnaschi:2015qta}. 

\begin{figure}[t]
  \centering
  \includegraphics[width=0.48\textwidth]{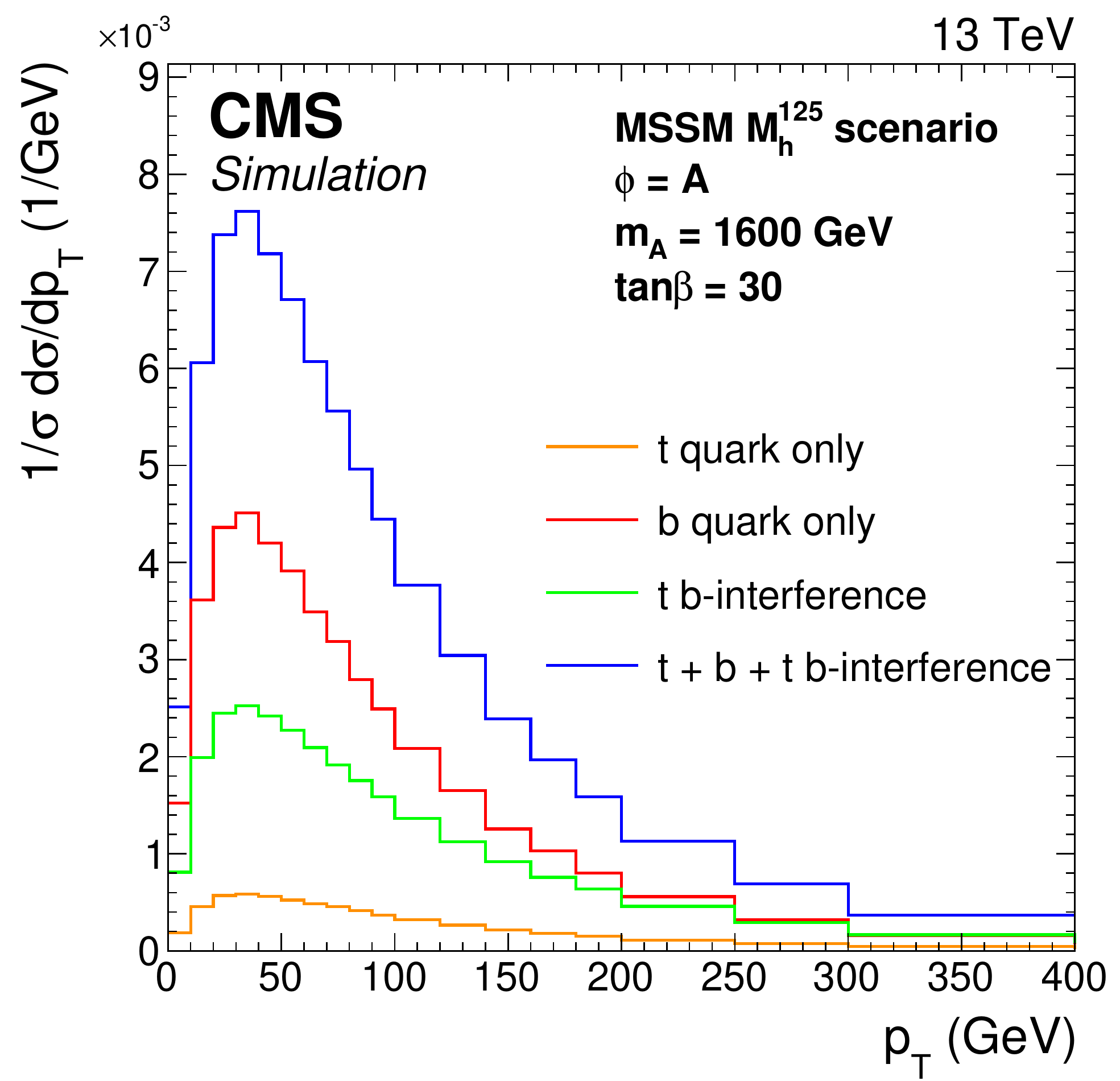}
  \includegraphics[width=0.48\textwidth]{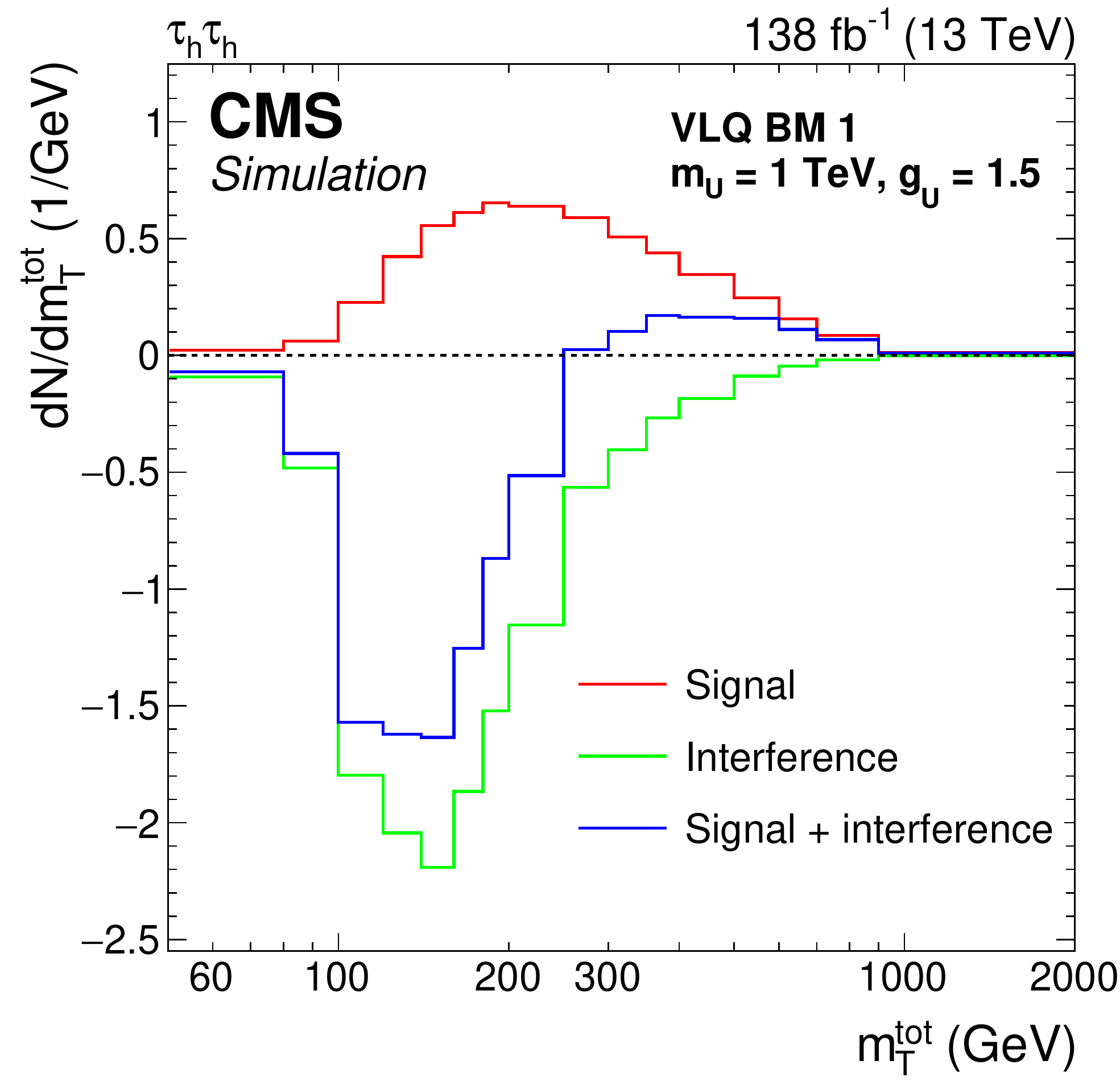}
  \caption{
    Composition of the signal for the MSSM interpretation of the data and the 
    vector leptoquark search. The left figure shows the generator level \PA 
    boson \pt density for the MSSM \mhBMPone scenario for $\mA=1.6\TeV$ and $\tanb
      =30$, split by the contributions from the \PQt quark only, the $\PQb$ quark 
    only, and the $\PQt\PQb$-interference term. The right figure shows the 
    distribution of \mTtot at reconstruction level in the \tautau final state 
    for \Uone $t$-channel exchange with $\mU=1\TeV$ and $\gU=1.5$, for the signal 
    with and without the interference term for the VLQ BM 1 scenario. The 
    \tautau final state is shown, since it is the most sensitive one for this 
    search. The bins of the distributions are divided by their width and the 
    distribution is normalized to the expected signal yield for 138\fbinv. 
  }
  \label{fig:signal-templates}
\end{figure}

For the model-independent \Pphi search, the individual distributions are combined 
according to their contribution to the total cross section as expected for an SM-like 
Higgs boson with given mass. For the tests of MSSM benchmark scenarios, where the 
contributions of the individual distributions also depend on the model parameters, 
these distributions are scaled using the effective Yukawa couplings as predicted 
by the corresponding benchmark model~\cite{MSSM_benchmark}, before combining them 
into one single prediction. In this context, the \tanb-enhanced SUSY corrections 
to the $\Pphi\PQb\PQb$ couplings are also included via the corresponding 
effective Yukawa couplings, where appropriate. Other SUSY contributions have been 
checked to amount to less than a few percent and are neglected. An example of the 
\PA boson \pt spectrum for $\mA=1.6\TeV$ and $\tanb=30$ is shown in 
Fig.~\ref{fig:signal-templates} (left). The \bbPhi production is simulated at NLO 
precision in \alpS using the corresponding \POWHEG 2.0 
implementation~\cite{Jager:2015hka} in the four-flavour scheme (4FS).

The signal process of the \Uone $t$-channel exchange is simulated in the 
five-flavour scheme (5FS) at LO precision in \alpS using the \MGvATNLO event 
generator, v2.6.5~\cite{Baker:2019sli}. Events are generated with up to one 
additional outgoing parton from the matrix element calculation and matched 
following the MLM prescription, with the matching scale \Qmatch set to 40\GeV. 
The contribution from on-shell $\Uone\to\PQq\PGt$ production and decay is excluded 
during the event generation. Samples are produced with $\gU=1$, for several 
values of \mU between 1 and 5\TeV. We observe no large dependence, neither of the 
templates used for signal extraction nor of the overall cross section, on the 
assumed \Uone decay width $\Gamma$, even after variations of factors of 0.5 and 
2 and therefore, for each considered value of \mU, we choose $\Gamma$ to 
approximately match the value predicted for \Uone production with couplings as 
obtained from the global fit presented in Ref.~\cite{Cornella:2021sby}. 

We expect a sizeable effect of destructive interference between the \Uone signal 
and \ZTT production, where the relative sizes of the interference and 
noninterference contributions depend on \gU. To include this dependence  
we generate separate samples for each contribution to form signal templates, 
which are negative in case of the interference contribution. These are scaled by 
$\gU^{4}$ (for the noninterference contribution) and $\gU^{2}$ (for the interference 
contribution), respectively, and combined to form the overall signal distributions 
for any value of \gU. Finally, the resulting signal event yields are normalized 
to the cross sections for the inclusive \Uone mediated $\PP\to\PGt\PGt$ process, 
computed at LO precision in \alpS. The contribution of the \Uone $t$-channel 
exchange to the \mTtot distribution in the \tautau final state for $\mU=1\TeV$ 
and $\gU=1.5$ for the VLQ BM 1 scenario is shown in Fig.~\ref{fig:signal-templates} 
(right). As visible from the figure, a complex contribution of the signal to the 
overall \ditau event yield in \mTtot is expected, with a reduction for $\mTtot
\lesssim250\GeV$ and an enhancement otherwise. Both features may contribute to 
the signal inference, while the sensitivity of the analysis relies on the yield 
enhancement for $\mTtot\gtrsim250\GeV$, as will be discussed in more detail in 
Section~\ref{sec:results_VLQ}. We note that for the \Pphi searches presented in 
this paper interference effects with \ditau backgrounds, \eg, from \ZTT production 
are not an issue due to the different spin configurations of the \ditau final 
states.  

\subsubsection{Common processing}

The PDF4LHC15~\cite{Butterworth:2015oua} (NNPDF3.1~\cite{Ball:2017nwa}) parton 
distribution functions (PDFs) are used for the simulation of the \ggPhi and \bbPhi 
(\Uone) signal processes. For all other processes, the NNPDF3.0~\cite{Ball:2014uwa} 
(NNPDF3.1) PDFs are used for the simulation of the data taken in 2016 (2017--2018). 
The description of the underlying event is parameterized according to the 
CUETP8M1~\cite{Khachatryan:2015pea} and CP5~\cite{Sirunyan:2019dfx} tunes for the 
simulation of the data taken in 2016 and 2017--2018, respectively. 

Parton showering and hadronization, as well as the \PGt lepton decays, are 
modelled using the \PYTHIA event generator~\cite{Sjostrand:2014zea}, where 
versions 8.212 and 8.226 are used for the simulation of the data taken in 2016, 
and version 8.230 is used for the data taken in 2017--2018. For all simulated 
events, additional inclusive inelastic \PP collisions generated with \PYTHIA 
are added according to the expected PU profile in data. All events generated are 
passed through a \GEANTfour-based~\cite{Agostinelli:2002hh} simulation of the CMS 
detector and reconstructed using the same version of the CMS event reconstruction 
software  used for the data.

\subsection{Corrections to the model}
\label{sec:corrections}

The capability of the model to describe the data is monitored in various control
regions orthogonal to the signal and background classes, and corrections and
corresponding uncertainties are derived where necessary. All corrections that 
have been applied to the model are described in the following. Their uncertainties 
are discussed in Section~\ref{sec:systematic-uncertainties}.

The following corrections apply equally to simulated and \PGt-embedded events, 
where the \PGt decay is also simulated. Since the simulation part of \PGt-embedded 
events happens under detector conditions that are different from the case of 
fully simulated events, corrections and related uncertainties may differ, as 
detailed in Ref.~\cite{Sirunyan:2019drn}. Corrections are derived for residual 
differences in the efficiencies of the selected triggers, differences in the 
electron and muon tracking efficiencies, and in the efficiencies of the identification 
and isolation requirements for electrons and muons. These corrections are obtained 
in bins of \pt and $\eta$ of the corresponding lepton, using the ``tag-and-probe'' 
method, as described in Ref.~\cite{Khachatryan:2010xn}, with \ZEE and \ZMM events. 
They usually do not amount to more than a few percent. The electron energy scale 
is adjusted to the scale measured in data using the \PZ boson mass peak in \ZEE 
events.

In a similar way, corrections are obtained for the efficiency of triggering on 
the \tauh decay signature and for the \tauh identification efficiency. 
The trigger efficiency corrections are obtained from parametric fits to the trigger 
efficiency, as a function of \pt, derived for simulated events and data. The 
identification efficiency corrections are also derived as a function of the \pt of 
the \tauh candidate. For $\ptth>40\GeV$, a correction is moreover derived for each 
\tauh decay mode individually, which is used only in the \tautau final state. For 
each data-taking year and each \tauh decay mode, corrections to the energy scale 
of the \tauh candidates and of electrons misidentified as \tauh candidates are 
derived from likelihood scans of discriminating observables, such as the 
reconstructed \tauh candidate mass. For muons misidentified as \tauh candidates, 
the energy scale correction has been checked to be negligible. 

Corrections are applied to the magnitude and resolution of \ptvecmiss in 
\PGt-embedded events to account for rare cases of an incomplete removal of the 
energy deposits from the muons that are replaced by simulated \PGt decays during 
the embedding procedure. These corrections are derived by comparing \ptmiss in 
\PGt-embedded events with fully simulated events. 

The following corrections only apply to fully simulated events. During the 2016 
and 2017 data taking, a gradual shift in the timing of the inputs of the ECAL L1 
trigger in the region at $\abs{\eta}>2.0$ caused a specific trigger 
inefficiency~\cite{Sirunyan:2020zal}. For events containing an electron (a jet) 
with \pt larger than ${\approx}50\,({\approx}100)\GeV$, in the region of $2.5<
\abs{\eta}<3.0$ the efficiency loss is 10--20\%, depending on \pt, $\eta$, and 
time. Corresponding corrections have been derived from data and applied to the 
simulation, where this effect is not present. 

The energies of jets are corrected to the expected response of the jet at the 
stable hadron level, using corrections measured in bins of the jet \pt and $\eta$. 
These corrections are usually less than 10--15\%. Residual data-to-simulation 
corrections are applied to the simulated event samples. They usually range from 
subpercent level at high jet \pt in the central part of the detector to a few 
percent in the forward region. The energy resolution of simulated jets is also 
adjusted to match the resolution in data. A correction is applied to the direction 
and magnitude of \ptvecmiss based on differences between estimates of the hadronic 
recoil in \ZMM events in data and simulation. This correction is applied to the 
simulated \ZLL, \Wjets, \Ph, and \Pphi signal events, where a hadronic recoil against 
a single particle is well defined. The efficiencies for genuine and misidentified 
\PQb jets to pass the working points of the \PQb jet identification discriminant, 
as given in Section~\ref{sec:selection}, are determined from data, using \ttbar 
events for genuine \PQb jets and jet-associated \PZ boson production 
for jets originating from light-flavour quarks. Data-to-simulation corrections 
are obtained for these efficiencies and used to correct the number of \PQb jets 
in the simulation. 

Data-to-simulation corrections are further applied to simulated events where 
an electron (muon) is reconstructed as a \tauh candidate, to account for residual 
differences in the \eormutotau misidentification rate between data and simulation. 
In a similar way, a correction is applied to account for residual differences in 
the \mutoe misidentification rate between data and simulation.

The dilepton mass and \pt spectra in simulated \ZLL events are corrected to better 
match the data. To do this, the dilepton mass and \pt are measured in data and 
simulation in \mumu events, and the simulated events are corrected to match the 
spectra in data. In addition, all simulated \ttbar events are weighted to better 
match the top quark \pt distribution observed in data~\cite{Khachatryan:2015oqa}. 
The overall normalization of \ttbar events is constrained using the \ttbar control 
region described in Section~\ref{sec:selection}. 

\section{Systematic uncertainties}
\label{sec:systematic-uncertainties}

The uncertainty model used for the analysis comprises theoretical and experimental 
uncertainties, and uncertainties due to the limited population of template 
distributions available for the background model. The last group of uncertainties 
is incorporated for each bin of each corresponding template individually following 
the approach proposed in Refs.~\cite{Barlow:1993dm,Conway:2011in}. All other 
uncertainties lead to correlated changes across bins either in the form of 
normalization changes or as general nontrivial shape-altering variations. 
Depending on the way they are derived, correlations may also arise across 
data-taking years, samples, or individual uncertainties.

\subsection{Uncertainties related to the \texorpdfstring{\PGt}{tau}-embedding method or the simulation}
\label{sec:uncertainties}

The following uncertainties, related to the reconstruction of electrons, muons, 
and \tauh candidates after selection, apply to simulated and \PGt-embedded events. 
Unless stated otherwise, they are partially correlated across \PGt-embedded and 
simulated events.

\subsubsection{Uncertainties common to signal and background events}

Uncertainties in the identification efficiency of electrons and muons amount to 
2\%, correlated across all years. Since no significant dependence on the \pt or 
$\eta$ of each corresponding lepton is observed, these uncertainties are introduced 
as normalization uncertainties. A similar reasoning applies to uncertainties in 
the electron and muon trigger efficiencies, which also amount to 2\% each. Because 
of differences in the online selections they are treated as uncorrelated 
for single-lepton and dilepton triggers. This may result in shape-altering effects 
in the overall model, since the two trigger types act on different ranges of 
lepton \pt.

For fully simulated events, an uncertainty in the electron energy scale is derived 
from the calibration of ECAL crystals, and applied on an event-by-event basis. 
For \PGt-embedded events, uncertainties of 0.5--1.25\%, determined separately for 
the ECAL barrel and endcap regions, are derived for the corrections described in 
Section~\ref{sec:corrections}. Because of the varying detector conditions, and 
the different ways the uncertainties are determined, they are treated as uncorrelated
across simulated and \PGt-embedded events. They lead to shape-altering variations 
and are treated as correlated across data-taking years. The muon momentum is 
precisely known, and a variation within the expected uncertainties was verified 
to have no effect on the analysis. 

Uncertainties in the \tauh identification efficiency are between 3--9\% in bins 
of \tauh lepton \pt. These are dominated by statistical uncertainties and are, 
therefore, treated as uncorrelated across decay modes, \pt bins, and data-taking 
years. The same is true for the uncertainties in the \tauh energy scale, which 
amount to 0.2--1.1\%, depending on the \tauh lepton \pt and decay mode. For the 
energy scale of electrons misidentified as \tauh candidates, the uncertainties 
are 1--6.5\%. All \tauh energy scale uncertainties are also treated as uncorrelated 
across data-taking years as they are predominantly statistical in nature. The 
uncertainty in the energy scale of muons misidentified as \tauh is 1\%. 
Uncertainties in the \tauh trigger efficiencies are typically \order(10\%), 
depending on the \tauh lepton \pt. They are obtained from parametric fits to data 
and simulation, and are treated as uncorrelated across triggers and data-taking 
years. All uncertainties discussed in this paragraph lead to shape-altering 
variations. 

Four further sources of uncertainty are considered for \PGt-embedded events.
A 4\% normalization uncertainty arises from the efficiency of the \mumu selection 
in data, which is unfolded during the \PGt-embedding procedure. Most of this 
uncertainty originates from the triggers used for selection. Since the trigger 
configurations changed over time, this uncertainty is treated as uncorrelated 
across data-taking years. An additional shape uncertainty is introduced to 
quantify the consistency of the embedding method in a sample of \mumu events. 
For this purpose, dedicated event samples are produced where the muons selected 
in data are replaced by simulated muons instead of \PGt lepton decays. These 
events are compared with the originally selected \mumu events in data and residual 
differences in the \mumu mass and \pt spectra are used as uncertainties. Another 
shape- and normalization-altering uncertainty in the yield of $\ttbar\to\mumu+
\mathrm{X}$ decays, which are part of the \PGt-embedded event samples, ranges 
from subpercent level to 8\%, depending on the event composition of the model. 
For this uncertainty, the number and shape of \ttbar events contained in the 
\PGt-embedded event samples are estimated from simulation, where the corresponding 
decay has been selected at the parton level. This estimate is then varied by 
${\pm}10\%$ to account for the \ttbar cross section and acceptance uncertainties. 
Finally, an uncertainty in the \ptmiss correction for the \PGt-embedded events 
described in Section~\ref{sec:corrections} is applied. Since this correction is 
derived from a comparison with fully simulated events, this uncertainty is related 
to the imperfect \ptvecmiss reconstruction in the simulation.  

For fully simulated events, the following additional uncertainties arise. 
Uncertainties in the \eormutotau misidentification rate are 18--40\% for 
electrons and 7--65\% for muons, depending on the $\eta$ of the \tauh candidate. 
These uncertainties apply only to simulated \ZEE and \ZMM events, which are of 
marginal importance for the analysis. The same is true for the uncertainty in 
the reweighting in the \ZLL dilepton mass and \pt, discussed in 
Section~\ref{sec:corrections}, which is typically smaller than 1\%. A 
normalization uncertainty due to the timing shift of the inputs of the ECAL L1 
trigger described in Section~\ref{sec:corrections} amounts to 2--3\%.

Uncertainties in the energy calibration and resolution of jets are applied with 
different correlations depending on their sources, which arise from the statistical 
limitations of the measurements used for calibration, the time-dependence of 
the energy measurements in data due to detector ageing, and bias corrections 
introduced to cover residual differences between simulation and data. They range 
from subpercent level to \order(10\%), depending on the kinematic properties 
of the jets in the event. Similar uncertainties, with similar ranges, are applied 
for the identification rates for \PQb jets and for the misidentification rates 
for light-flavour quark or gluon jets. 

Depending on the process under consideration, two independent uncertainties in 
\ptmiss are applied. For processes that are subject to recoil corrections, 
\ie \ZLL, \Wjets, \Ph, or \Pphi production, uncertainties in the calibration and 
resolution of the hadronic recoil are applied; they typically result in changes 
to the event yields ranging from 0--5\%. For all other processes, an uncertainty 
in \ptvecmiss is derived from the amount of energy carried by unclustered particle 
candidates, which are not contained in jets, in the event~\cite{Sirunyan:2019kia}. 
This uncertainty typically results in changes to the event yields ranging from 0--10\%.

The integrated luminosities of the 2016, 2017, and 2018 data-taking periods are 
individually known with uncertainties in the 1.2--2.5\% range~\cite{CMS-LUM-17-003,
CMS-PAS-LUM-17-004,CMS-PAS-LUM-18-002}, while the total integrated luminosity for 
the years 2016--2018 has an uncertainty of 1.6\%; the improvement in precision 
reflects the (uncorrelated) time evolution of some systematic effects. 
Uncertainties in the predictions of the normalizations of all simulated processes 
amount to 4\% for \ZLL and \Wjets production~\cite{Melnikov:2006kv}, 6\% for 
\ttbar production~\cite{Czakon:2011xx,Kidonakis:2013zqa}, and 5\% for diboson and 
single~\PQt quark production~\cite{Kidonakis:2013zqa,Campbell:2011bn,Gehrmann:2014fva}, 
where used in the analyses. These uncertainties are correlated across data-taking 
years. A shape-altering uncertainty is derived in the reweighting of the top 
quark \pt described in Section~\ref{sec:corrections} by applying the correction 
twice or not applying it at all. This uncertainty has only a very small effect 
on the final discriminant.

\subsubsection{Uncertainties in the signal modelling}

Theoretical uncertainties in the acceptance of \bbPhi signal events are obtained 
from variations of the renormalization (\mur) and factorization (\muf) scales, 
the \hdamp factor, and the PDFs. The scale uncertainty is obtained from the 
envelope of the six variations of \mur and \muf by factors of 0.5 and 2, omitting 
the variations where one scale is multiplied by 2 and the corresponding other 
scale by 0.5, as recommended in Ref.~\cite{deFlorian:2016spz}. The scale \hdamp 
is varied by factors of $1/\sqrt{2}$ and $\sqrt{2}$. The uncertainty from the 
variation of \mur and \muf, and the uncertainty from the variation of \hdamp are 
added linearly, following the recommendation in Ref.~\cite{deFlorian:2016spz}, 
resulting in an overall uncertainty that ranges from 1--8\% (1--5\%) for the \PQb 
tag (``no \PQb tag'') categories depending on the tested mass. The uncertainties 
due to PDF variations and the uncertainty in \alpS are obtained following the 
PDF4LHC recommendations, taking the root mean square of the variation of the 
results when using different replicas of the default PDF4LHC sets as described, 
\eg, in Ref.~\cite{Butterworth:2015oua}. They range from 1--2\%. 

Uncertainties in the acceptance of the \ggPhi process are also obtained from 
variations of \mur, \muf, and \hdamp. The \mur and \muf scales are varied as 
described above for the \bbPhi process, whereas the \hdamp scale is varied by 
factors of 0.5 and 2 as suggested in Ref.~\cite{Bagnaschi:2015bop}. The 
influence of the former (latter) variation on the signal acceptance amounts to 
20\% (35\%) for the smallest \mphi values. For larger \mphi values, the variation 
is at the subpercent level. In both cases the uncertainties also result in 
shape-altering effects in the overall model. 

For the parameter scan in the MSSM interpretations, theoretical uncertainties 
in the \ggPhi and \bbPhi cross sections are included, as described in 
Ref.~\cite{Bagnaschi:2791954}. This includes uncertainties in the \mur and \muf 
scales, PDFs, and \alpS. The uncertainties are evaluated separately for each 
\mA-\tanb point under consideration. They are typically 5--20\% (10--25\%) for 
\ggPhi (\bbPhi) production.

Several sources of theoretical uncertainty in the \Uone signal prediction are 
included. The uncertainty due to the \mur and \muf scale variations is about 
15\%. The uncertainties due to the PDFs and \alpS variations are about 15 and 
4\%, respectively. The \Qmatch and parton shower uncertainties affect the signal 
acceptances in the ``\PQb tag'' categories, with magnitudes of about 11 and 
1\% respectively, and in the ``no \PQb tag'' categories, with magnitudes of 5 
and 6\% respectively. The uncertainty in the \betaLstau parameter is estimated 
by varying the coupling strength by the uncertainties obtained in the fit 
presented in Ref.~\cite{Cornella:2021sby} and summarized in 
Table~\ref{tab:betal_values}. The resulting uncertainty varies the signal yields 
by 4--12\%. The uncertainty in the signal acceptance due to the choice of flavour 
scheme is estimated by comparing the predictions in the 4FS and 5FS 
calculations, which mainly affect the \Nbjet distribution. The resulting 
uncertainty has a magnitude of 25\% (18\%) for the ``\PQb tag'' 
(``no \PQb tag'') categories.   

For all results shown in the following, the expectation for SM Higgs boson 
production is included in the model used for the statistical inference of the 
signal. Uncertainties due to different choices of \mur and \muf for the 
calculation of the production cross section of the SM Higgs boson amount to 
3.9\% for \ggh, 0.4\% for VBF, 2.8\% for \Zh, and 0.5\% for \Wh 
production~\cite{Alioli:2008tz,Bagnaschi:2011tu,Nason:2009ai,Luisoni:2013cuh,
Hartanto:2015uka}; uncertainties due to different choices for the PDFs and \alpS 
amount to 3.2\%, 2.1\%, 1.6\%, and 1.9\% for these four production modes, respectively. 

\subsection{\texorpdfstring{Uncertainties related to jets misidentified as an electron, muon, or \tauh candidate}{Systematic uncertainties related to jets misidentified as an electron, muon, or tauh candidate}}
\label{sec:uncertainties-FF}

For the \FF method, the following uncertainties apply. The \FFi and their 
corrections are subject to statistical fluctuations in each corresponding \DRi 
and simulation. The corresponding uncertainties are split into a normalization 
and a shape-altering part and propagated into the final discriminant. They are 
typically 1--10\% and are treated as uncorrelated across the kinematic and 
topological bins where they are derived. An additional uncertainty is defined 
by varying the choice of the functional form for the parametric fits.

Uncertainties are also applied to cover residual corrections and extrapolation 
factors, varying from a few percent to \order(10\%), depending on the kinematic 
properties of the \tauh candidate and the topology of the event. These are both 
normalization and shape-altering uncertainties.

An additional source of uncertainty concerns the subtraction of processes other 
than the enriched process in each corresponding \DRi. These are subtracted 
from the data using simulated or \PGt-embedded events. The combined shape of 
the events to be removed is varied by 10\%, and the measurements are repeated. 
The impacts of these variations are then propagated to the final discriminant 
as shape-altering uncertainties. 

An uncertainty in the estimation of the three main background fractions in the 
AR is estimated from a variation of each individual contribution by 10\%, 
increasing or decreasing the remaining fractions such that the sum of all 
contributions remains unchanged. The amount of variation is motivated by the 
uncertainty in the production cross sections and acceptances of the involved 
processes and the constraint on the process composition that can be clearly 
obtained from the AR. The effect of this variation is found to be very small, 
since usually one of the contributions dominates the event composition in the 
AR. 

\begin{table}[!htbp]
  \centering
  \topcaption{
    Summary of systematic uncertainties discussed in the text. The columns 
    indicate the source of uncertainty, the variation, and how it is correlated 
    between data-taking years. A checkmark is given also for partial 
    correlations. More details are given in the text.
  }
    \begin{tabular}{llr@{--}lc}
      \multicolumn{2}{l}{Uncertainty} & \multicolumn{2}{c}{Variation} 
      & Correlated across years \\
      \hline 
      \multirow{4}{*}{\PGt-emb.} & Acceptance 
      & \multicolumn{2}{c}{4\%} \\
      & \mumu closure  
      & \multicolumn{2}{c}{See text} & $\checkmark$ \\
      & \ttbar fraction  
      & 0.1 & 8\% & \NA \\
      & \ptmiss 
      & \multicolumn{2}{c}{See text} & \NA \\[\cmsTabSkip]
      \multirow{2}{*}{\PGm} 
      & Identification      
      & \multicolumn{2}{c}{2\%} & $\checkmark$ \\
      & Trigger 
      & \multicolumn{2}{c}{2\%} & \NA \\[\cmsTabSkip]
      \multirow{3}{*}{\Pe} 
      & Identification
      & \multicolumn{2}{c}{2\%} & $\checkmark$ \\
      & Trigger 
      & \multicolumn{2}{c}{2\%} & \NA \\
      & Energy scale 
      & \multicolumn{2}{c}{See text} & $\checkmark$ \\[\cmsTabSkip]
      \multirow{3}{*}{\tauh} 
      & Identification
      & 3 & 8\% & \NA \\
      & Trigger 
      & 5 & 10\% & \NA \\
      & Energy scale 
      & 0.2 & 1.1\% & \NA \\[\cmsTabSkip]
      \multirow{2}{*}{$\PGm\to\tauh$} & Misidentification 
      & 7 & 67\% & \NA \\
      & Energy scale 
      & \multicolumn{2}{c}{1\%} & \NA \\[\cmsTabSkip]
      \multirow{2}{*}{$\Pe\to\tauh$} & Misidentification 
      & 18 & 41\% & \NA \\
      & Energy scale 
      & 1 & 6.5\% & \NA \\[\cmsTabSkip]
      \multicolumn{2}{l}{$\text{Jet}\to\Pe$ misidentification} 
      & \multicolumn{2}{c}{10\%} & $\checkmark$ \\ 
      \multicolumn{2}{l}{$\text{Jet}\to\PGm$ misidentification} 
      & \multicolumn{2}{c}{10\%} & $\checkmark$ \\ 
      \multicolumn{2}{l}{\mutoe misidentification} 
      & 15 & 45\% & \NA \\ 
      \multicolumn{2}{l}{\Zgamma mass and \pt reweighting} 
      & \multicolumn{2}{c}{${<}1\%$} \\ 
      \multicolumn{2}{l}{Jet energy scale \& resolution} 
      & 0.1 & 10\% & $\checkmark$ \\ 
      \multicolumn{2}{l}{\PQb-jet (mis)identification} 
      & 1 & 10\% & \NA \\ 
      \multicolumn{2}{l}{\ptmiss calibration} 
      & \multicolumn{2}{c}{See text} & $\checkmark$ \\ 
      \multicolumn{2}{l}{ECAL timing shift} 
      & 2 & 3\% & $\checkmark$ \\ 
      \multicolumn{2}{l}{\PQt quark \pt reweighting} 
      & \multicolumn{2}{c}{See text} & $\checkmark$ \\ 
      \multicolumn{2}{l}{Integrated luminosity} 
      & 1.2 & 2.5\% & $\checkmark$ \\ 
      \multicolumn{2}{l}{Background cross sections} 
      & 2 & 5\% & $\checkmark$ \\ 
      \multicolumn{2}{l}{Signal theoretical uncertainties} 
      & \multicolumn{2}{c}{See text} & $\checkmark$ \\ [\cmsTabSkip]
      \multirow{4}{*}{\FF} & Event count 
      & \order(1 & 10\%) & \NA \\
      & Corrections 
      & \multicolumn{2}{c}{\order(10\%)} & \NA \\
      & Non-\FF processes 
      & \multicolumn{2}{c}{10\%} & \NA \\
      & \FF proc. composition 
      & \multicolumn{2}{c}{10\%} & \NA \\[\cmsTabSkip] 
      \multirow{2}{*}{QCD (\emu)} & Event count 
      & 2 & 4\% & \NA \\ 
      & AR to SR extrapolations 
      & \multicolumn{2}{c}{\order(10\%)} & \NA \\ 
    \end{tabular}
    \label{tab:uncertainties}
\end{table}

Since the background from QCD multijet events in the \emu final state is 
also determined from a DR, uncertainties that account for the statistical 
uncertainty in the data and the subtracted backgrounds in this DR are applied in 
a similar way. These uncertainties amount to 2--4\%. In addition, this background 
is subject to uncertainties related to the extrapolations from the DR to the 
corresponding SRs. These uncertainties are \order(10\%) depending on \pte, \ptm, 
and \Nbjet. Because of their mostly statistical nature, all uncertainties related to 
the \FF and SS methods are treated as uncorrelated across data-taking years.

In the \emu final state, the subdominant contribution to the \jettoL and \mutoe 
backgrounds is estimated from simulation. Uncertainties in the simulated 
$\text{jet}\to\Pe$ and $\text{jet}\to\PGm$ misidentification rates are 10 and 
12\%, respectively. They are treated as correlated across data-taking years. The 
uncertainty in the \mutoe misidentification rate is 15--45\%, and is treated as 
uncorrelated across data-taking years since it is mostly statistical in nature. A 
summary of all systematic uncertainties that have been discussed in this section 
is given in Table~\ref{tab:uncertainties}, in which we also state the correlations 
between the data-taking years.   

\section{Results}
\label{sec:results}

The statistical model used to infer the signal from the data is defined by an 
extended binned likelihood of the form
\begin{linenomath}
  \begin{equation}
    \mathcal{L}\left(\{k_{i}\},\{\mu_{s}\},\{\theta_{j}\}\right) =\prod
    \limits_{i}\mathcal{P}\Bigl(k_{i}|\sum\limits_{s}\mu_{s}\,S_{si}(\{\theta_{j}
    \})+\sum\limits_{b}B_{bi}(\{\theta_{j}\})\Bigr)\,
    \prod\limits_{j}\mathcal{C}(\widetilde{\theta}_{j}|\theta_{j}),
    \label{eq:likelihood}
  \end{equation}
\end{linenomath}
where $i$ labels the bins of the discriminating distributions of all categories, 
split by \ditau final state and data-taking year. The function $\mathcal{P}(k_{i}
|\sum\mu_{s}\,S_{si}(\{\theta_{j}\})+\sum B_{bi}(\{\theta_{j}\}))$ corresponds to 
the Poisson probability to observe $k_{i}$ events in bin $i$ for a prediction of 
$\sum \mu_{s}\,S_{si}$ signal and $\sum B_{bi}$ background events. The predictions 
for $S_{si}$ and $B_{bi}$ are obtained from the signal and background models 
discussed in Section~\ref{sec:data-model}. The parameters $\mu_{s}$ act as linear 
scaling parameters of the corresponding signal yields $S_{s}$. Systematic 
uncertainties are incorporated in the form of penalty terms for additional nuisance 
parameters $\{\theta_{j}\}$ in the likelihood, appearing as a product with 
predefined probability density functions $\mathcal{C}(\widetilde{\theta}_{j}|\theta_{j})$, 
where $\widetilde{\theta}_{j}$ corresponds to the nominal value for $\theta_{j}$. The 
predefined uncertainties in the $\widetilde{\theta}_{j}$, as discussed in 
Section~\ref{sec:systematic-uncertainties}, may be constrained by the fit to the 
data. 

The test statistic used for the inference of the signal is the profile likelihood 
ratio, as discussed in Refs.~\cite{CMS-NOTE-2011-005,Chatrchyan:2012tx}:
\begin{linenomath}
  \begin{equation}
    q_{\mu_{s}}=-2\ln
    \left(
    \frac{\mathcal{L}(\left.\{k_{i}\}\right|\sum\limits_{s}\mu_{s}\,S_{si}
    (\{\hat{\theta}_{j,\mu_{s}}\})+\sum\limits_{b}B_{bi}(\{\hat{\theta}_{j,\mu_{s}}\}))}
    {\mathcal{L}(\left.\{k_{i}\}\right|\sum\limits_{s}\hat{\mu}_{s}\,S_{si}(\{\hat{
    \theta}_{j,\hat{\mu}_{s}}\})+\sum\limits_{b}B_{bi}(\{\hat{\theta}_{j,\hat{\mu}_{
    s}}\}))}\right) , \;\; 0\leq \hat{\mu}_{s} \leq \mu_{s},
    \label{eq:profile-likelihood-ratio}
  \end{equation}
\end{linenomath}
where one or more parameters $\mu_{s}$ are the parameters of interest (POIs) and 
$\hat{\mu}_{s}$, $\hat{\theta}_{j,\mu_{s}}$, and $\hat{\theta}_{j,\hat{\mu}_{s}}$ 
are the values of the given parameters that maximize the corresponding likelihood. 
The index of $q_{\mu_{s}}$ indicates that the test statistic is evaluated for a 
fixed value of $\mu_{s}$. In the large number limit, the sampling distribution of 
$q_{\mu_{s}}$ can be approximated by analytic functions, from which the expected 
median and central intervals can be obtained as described in Ref.~\cite{Cowan:2010js}. 
The signal is inferred from the data in three different ways: 

\begin{enumerate}
\item 
the model-independent \Pphi search features a signal model for a single narrow 
resonance \Pphi; 
\item 
for the search for vector leptoquarks, the data are interpreted in terms of 
the nonresonant \Uone $t$-channel exchange; 
\item 
the interpretation of the data in terms of MSSM benchmark scenarios relies on 
three resonances in the \ditau mass spectrum with mass values and rates determined 
by the parameters of the corresponding scenario. 
\end{enumerate}

In all cases the \ttbar control region, as defined in Section~\ref{sec:selection} 
and shown in Figs.~\ref{fig:categories}--\ref{fig:categories_lowmass} is used to 
constrain the normalization of \ttbar events and all \ttbar related uncertainties. 
Detailed descriptions of the specific statistical procedures and the results 
obtained in each case are given in the following sections. 

\begin{figure}[htbp]
  \centering
  \includegraphics[width=.42\textwidth]{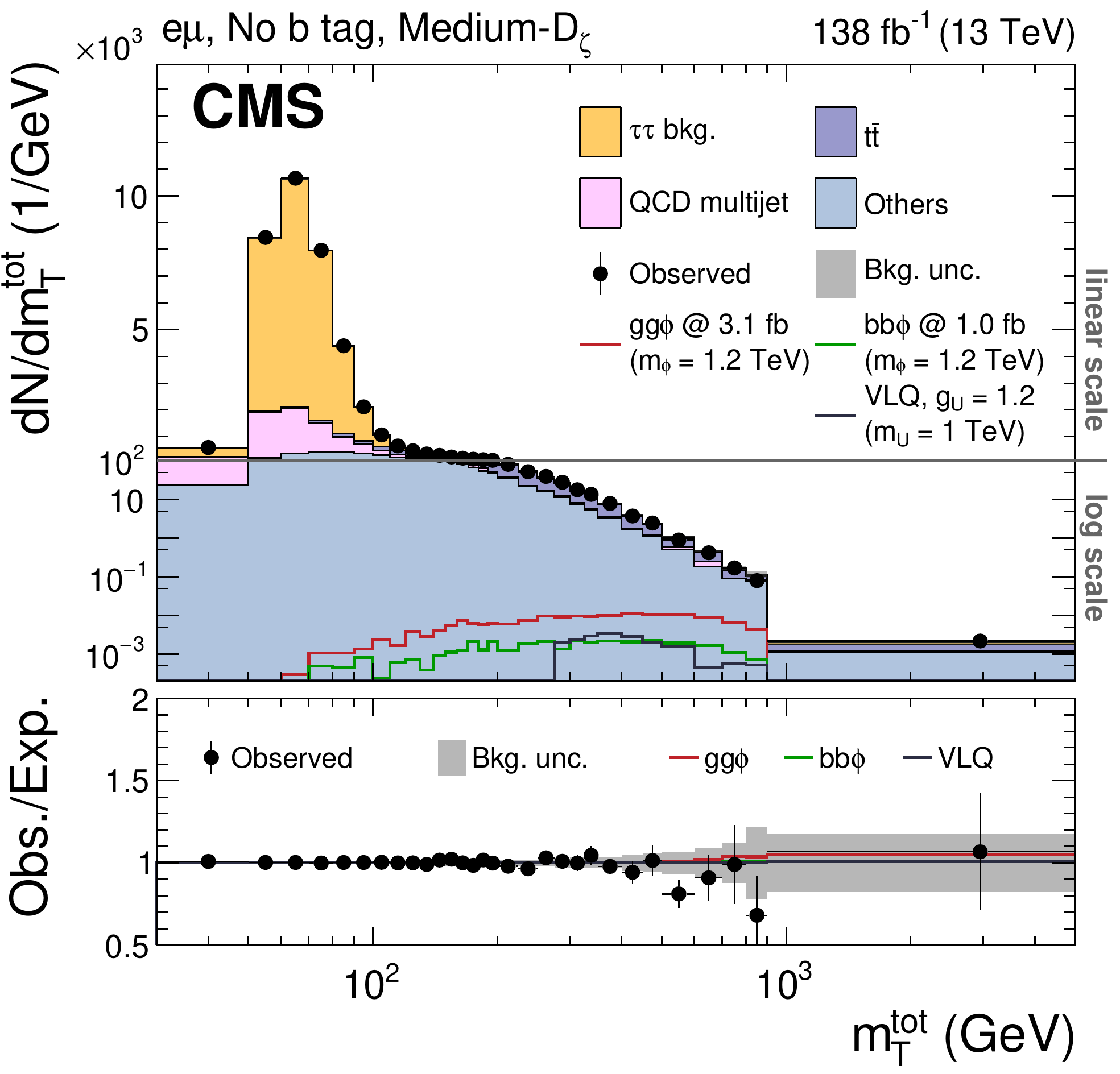}
  \includegraphics[width=.42\textwidth]{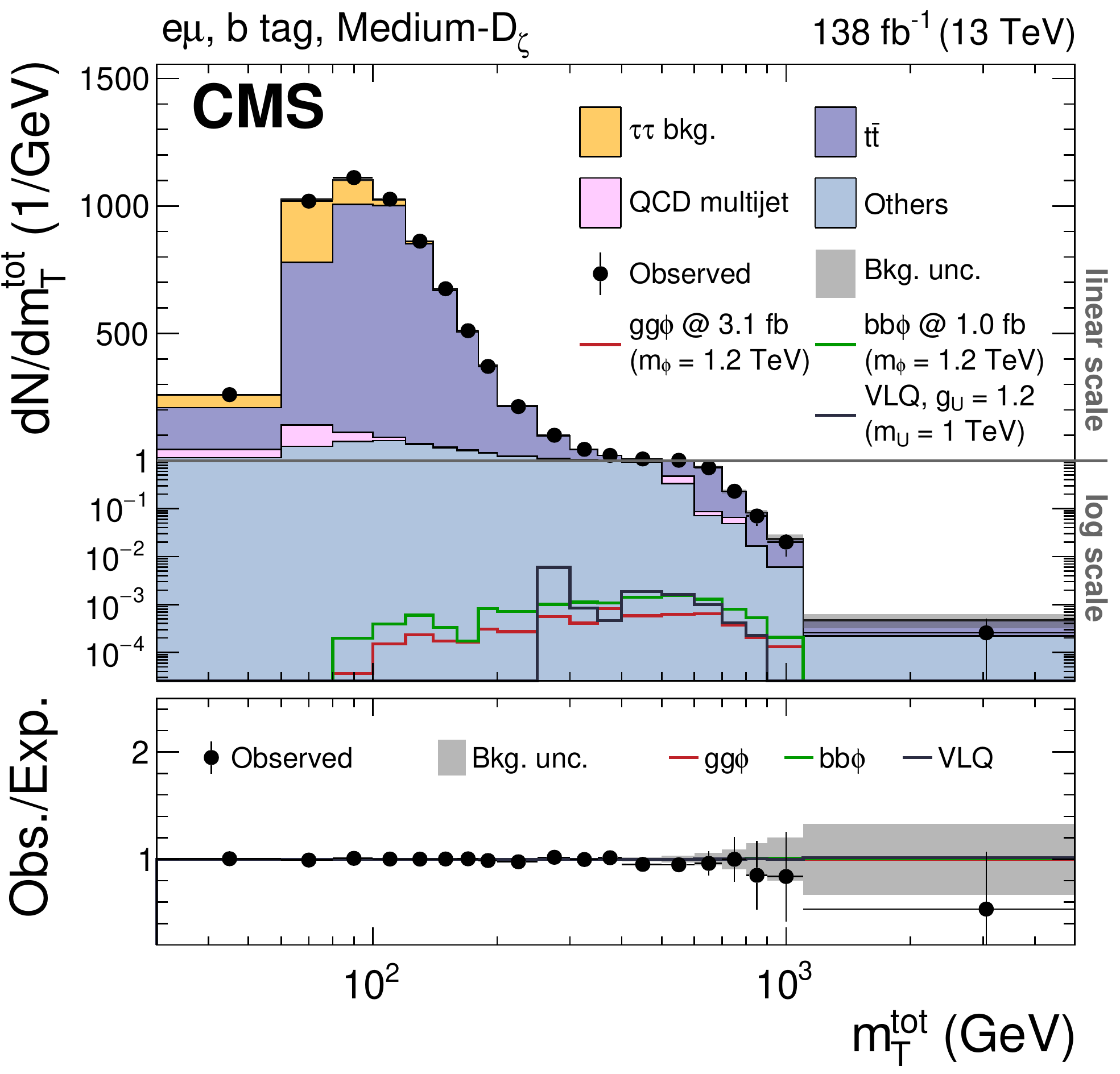}
  \includegraphics[width=.42\textwidth]{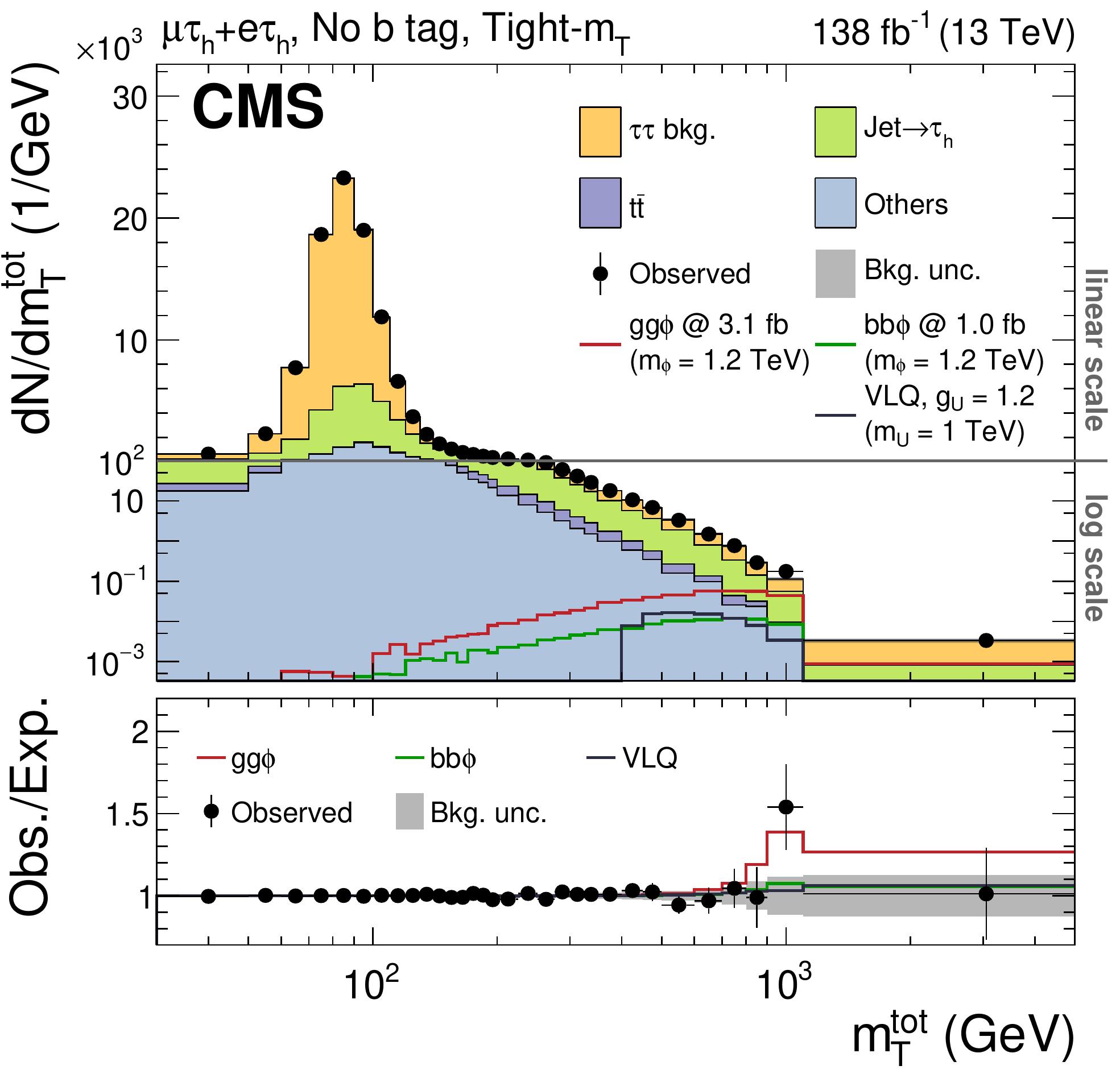}
  \includegraphics[width=.42\textwidth]{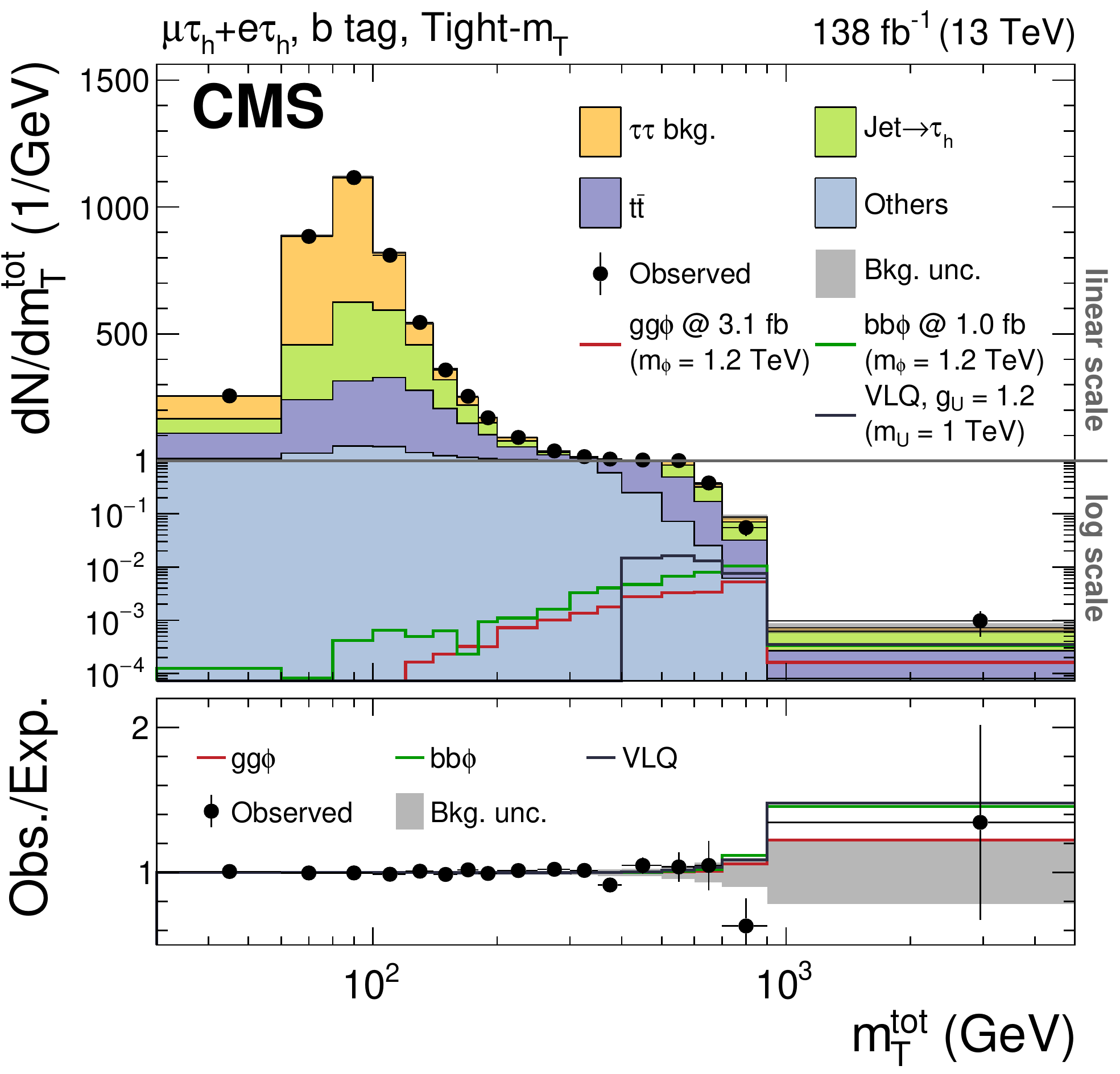}
  \includegraphics[width=.42\textwidth]{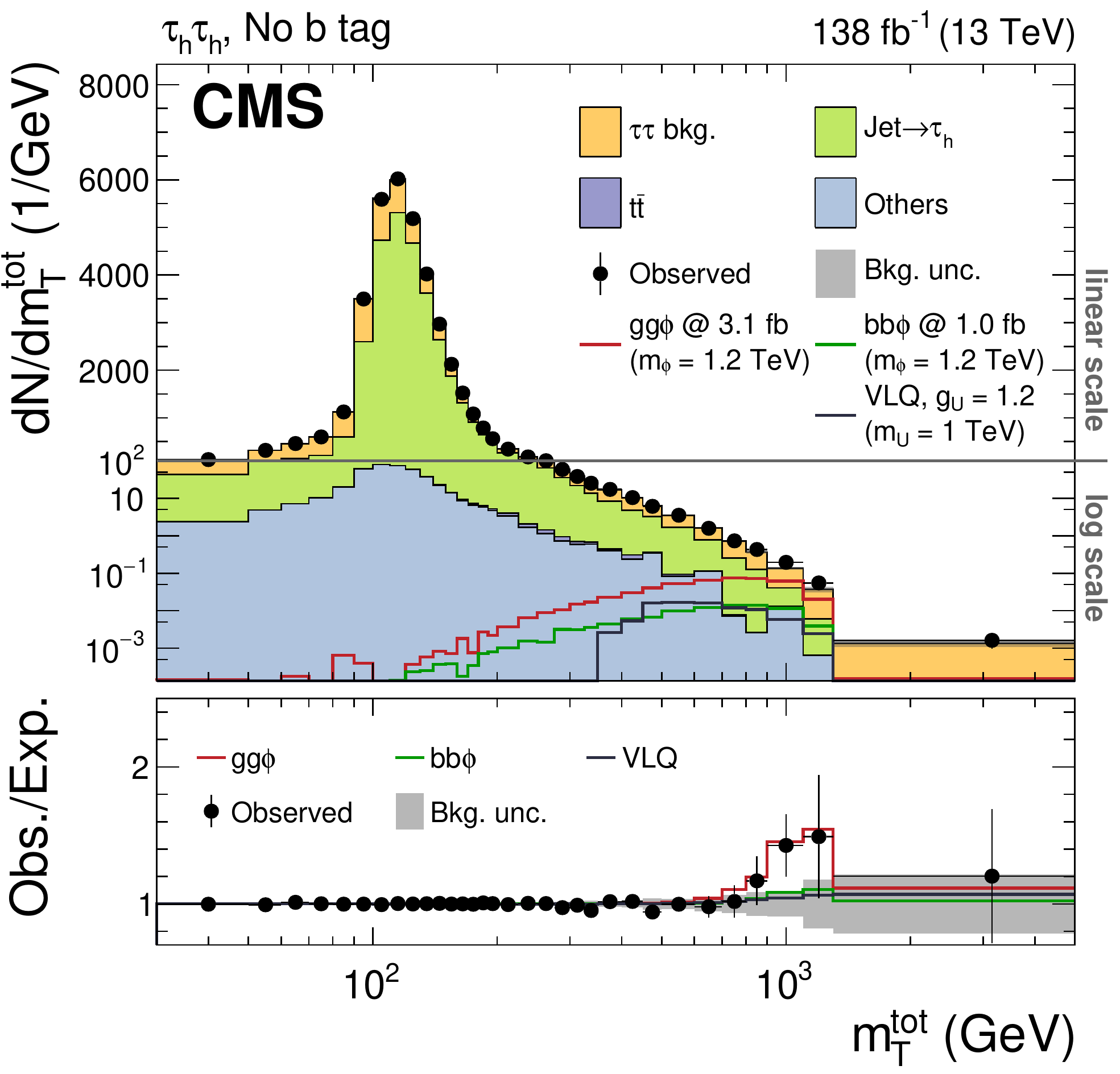}
  \includegraphics[width=.42\textwidth]{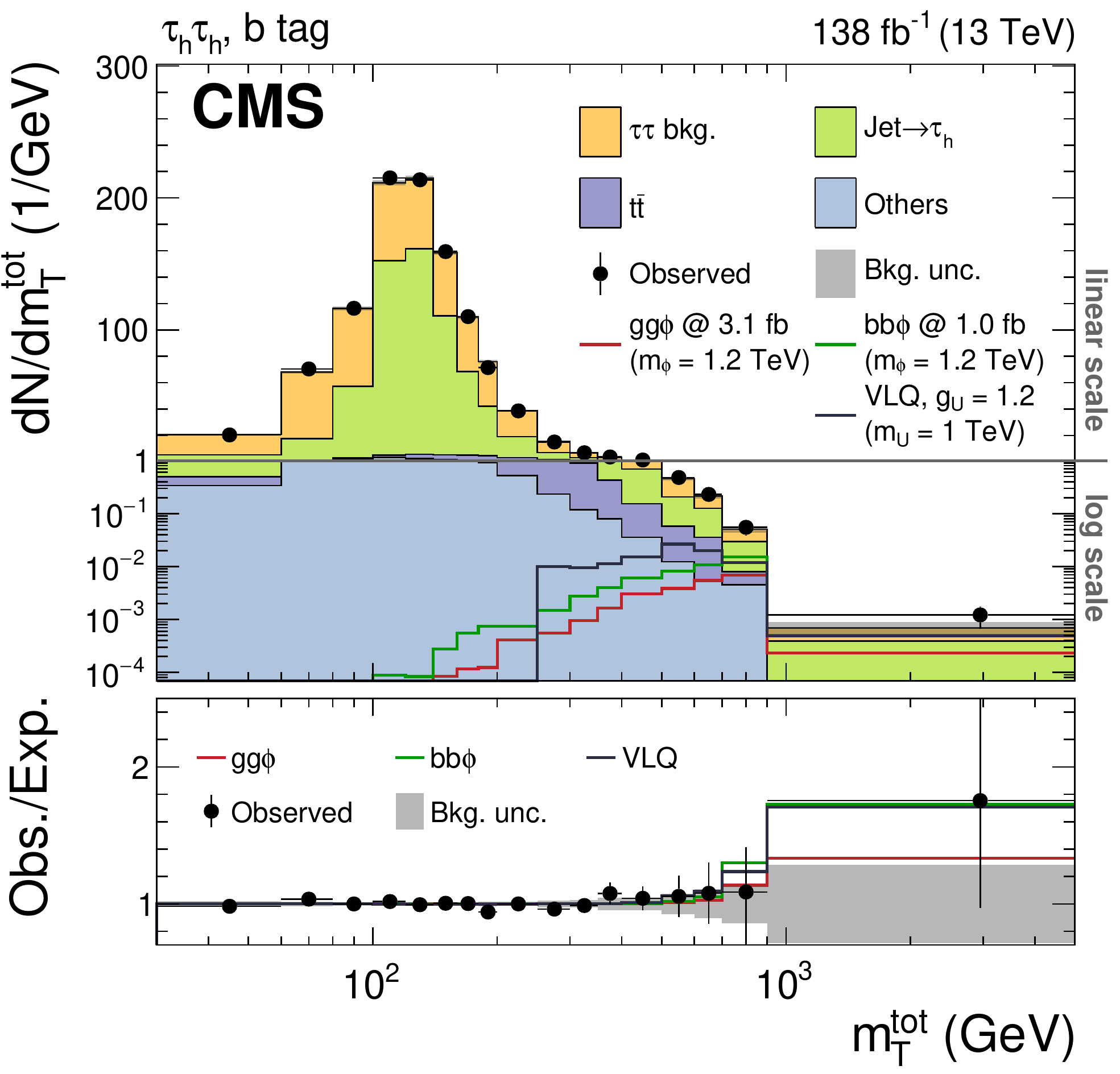}
  \caption {
    Distributions of \mTtot in the global (left) ``no \PQb tag'' and (right) 
    ``\PQb tag'' categories in the (upper) \emu, (middle) \etau and \mutau, and 
    (lower) \tautau final states. For the \emu final state, the medium-\Dzeta 
    category is displayed; for the \etau and \mutau final states the tight-\mT 
    categories are shown. The solid histograms show the stacked background 
    predictions after a background-only fit to the data. The best fit \ggPhi 
    signal for $\mphi=1.2\TeV$ is shown by the red line. The \bbPhi and \Uone 
    signals are also shown for illustrative purposes. For all histograms, the bin 
    contents show the event yields divided by the bin widths. The lower panel 
    shows the ratio of the data to the background expectation after the 
    background-only fit to the data. 
  }
  \label{fig:mTtot-distributions}
\end{figure}

\begin{figure}[htbp]
  \centering
  \includegraphics[width=.42\textwidth]{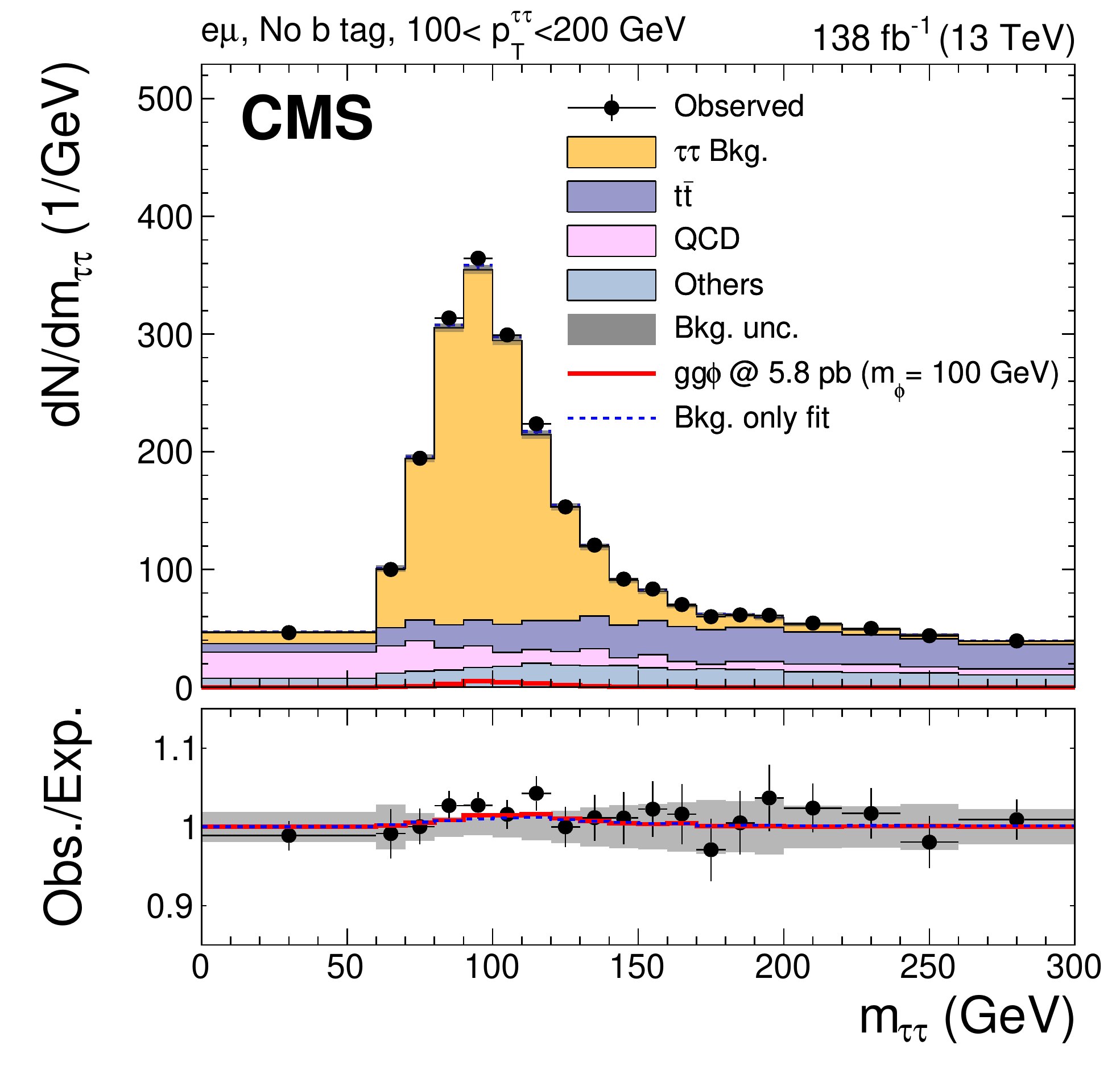}
  \includegraphics[width=.42\textwidth]{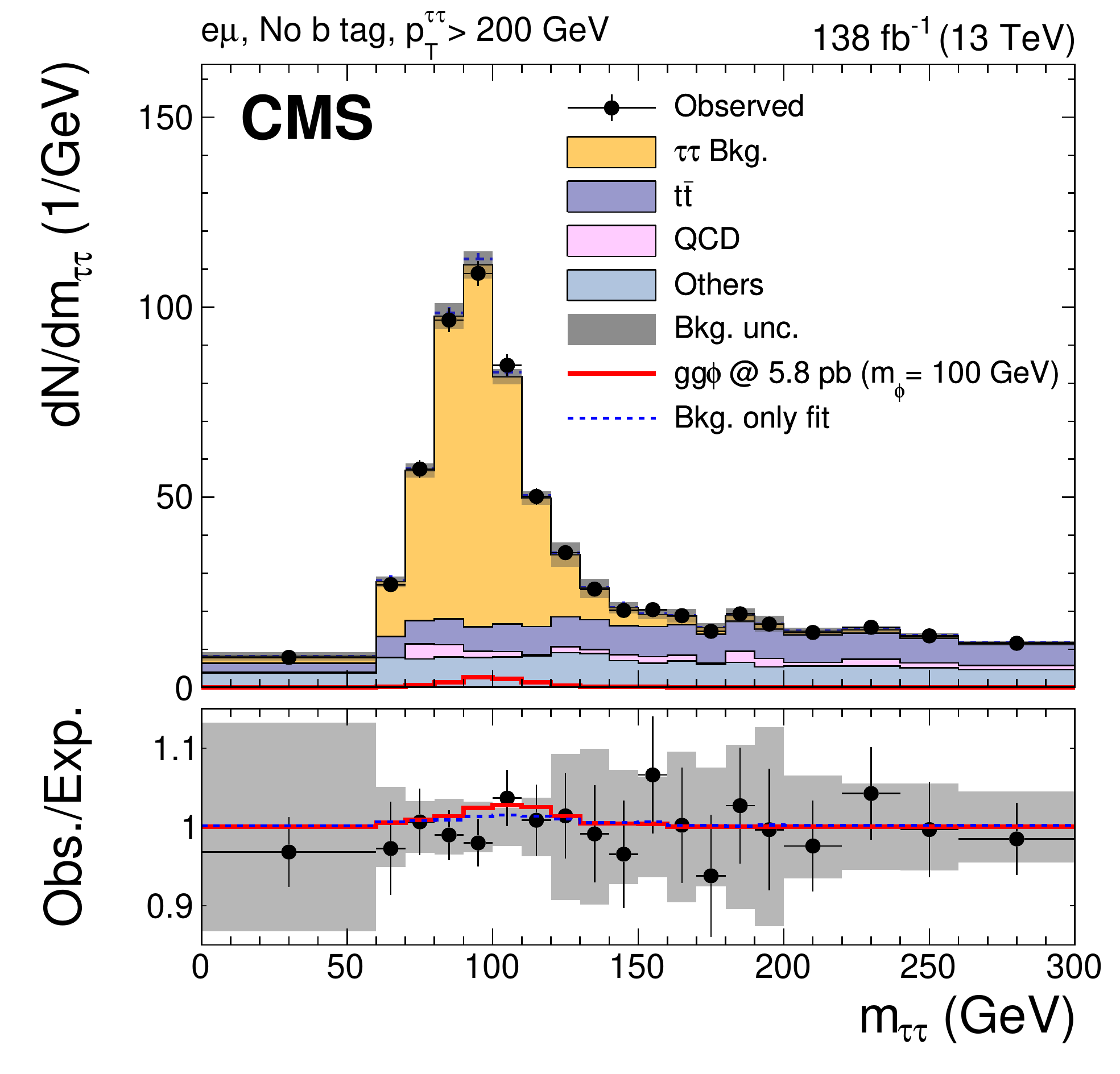} \
  \includegraphics[width=.42\textwidth]{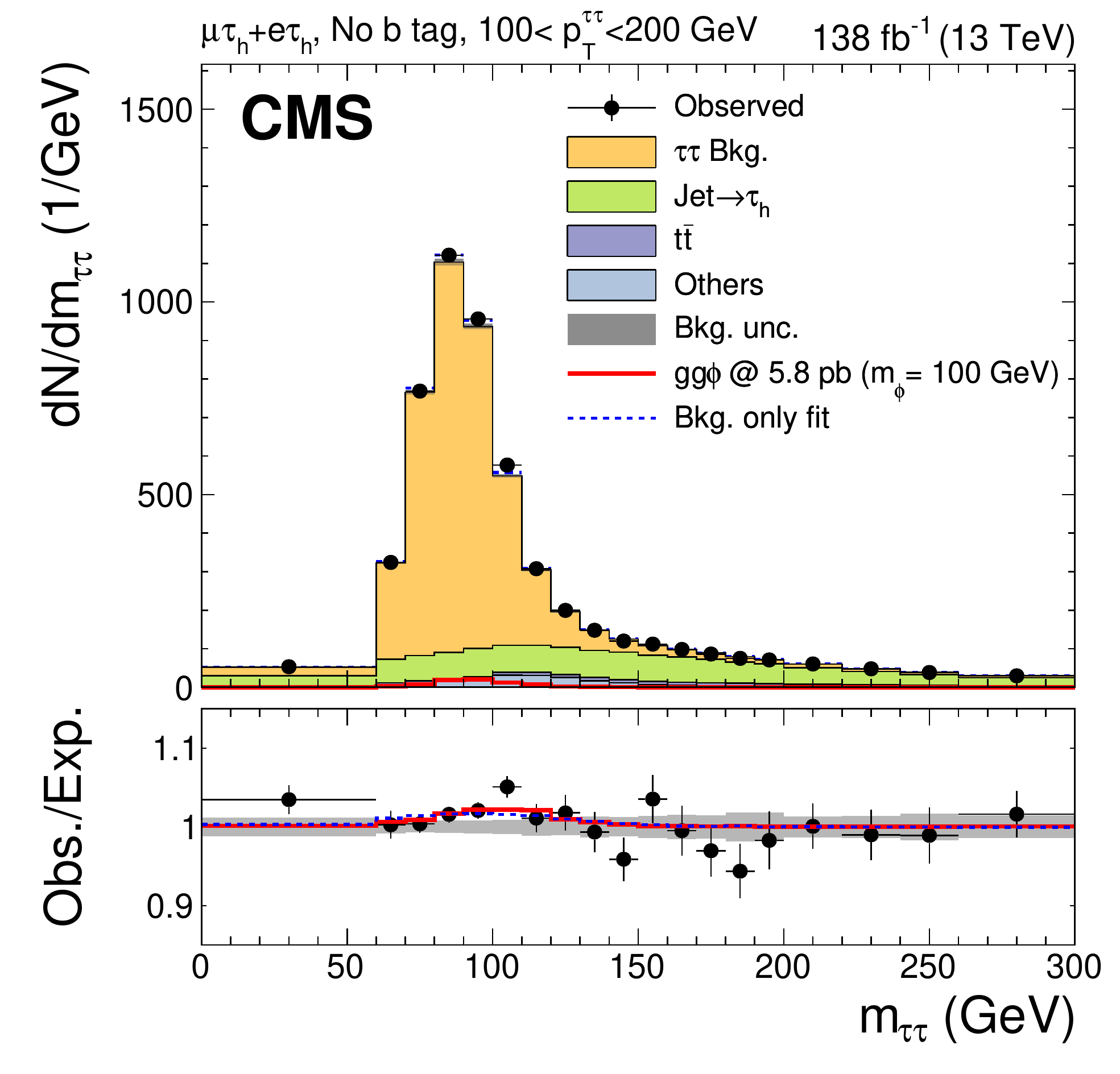}
  \includegraphics[width=.42\textwidth]{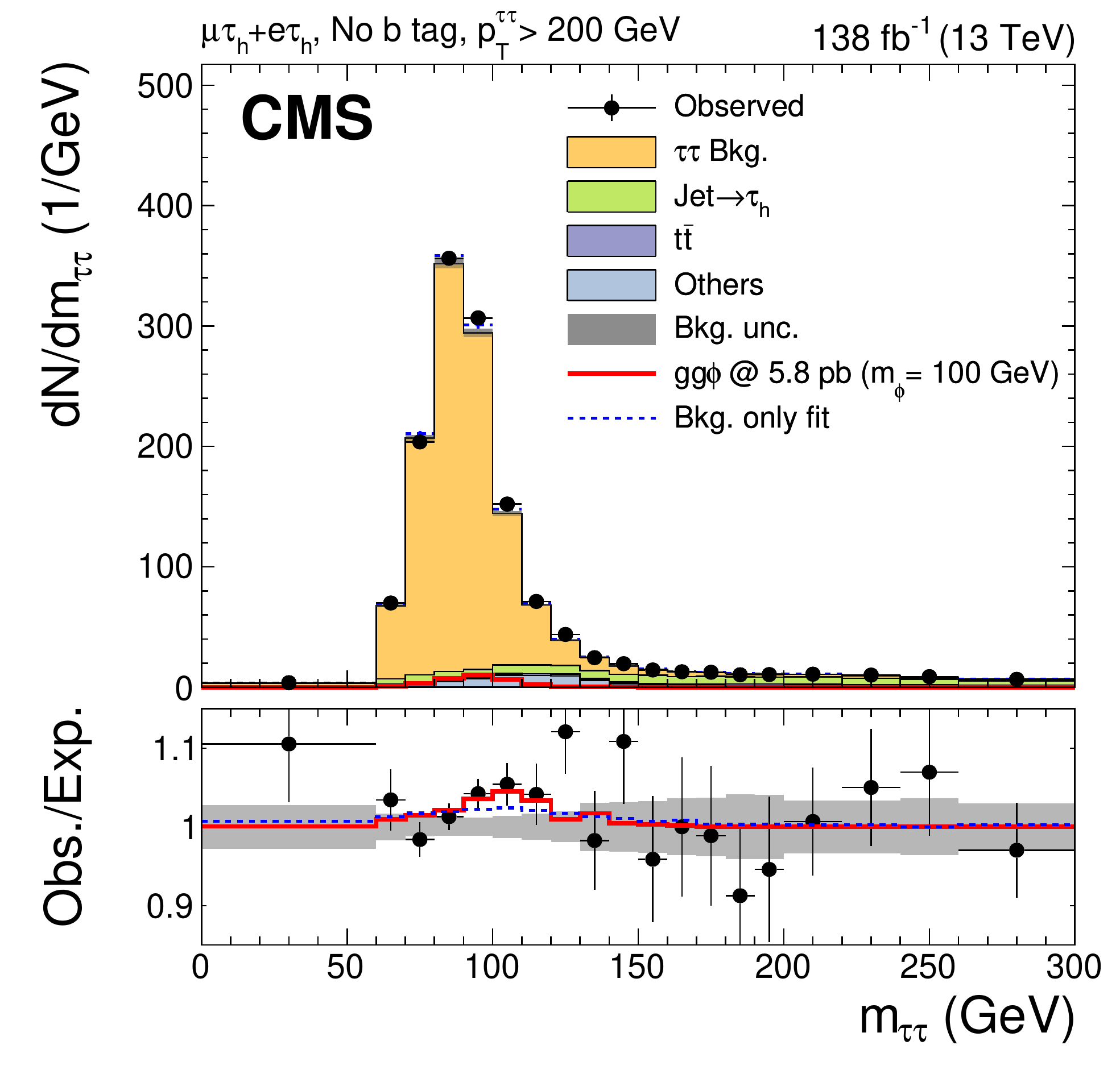} \
  \includegraphics[width=.42\textwidth]{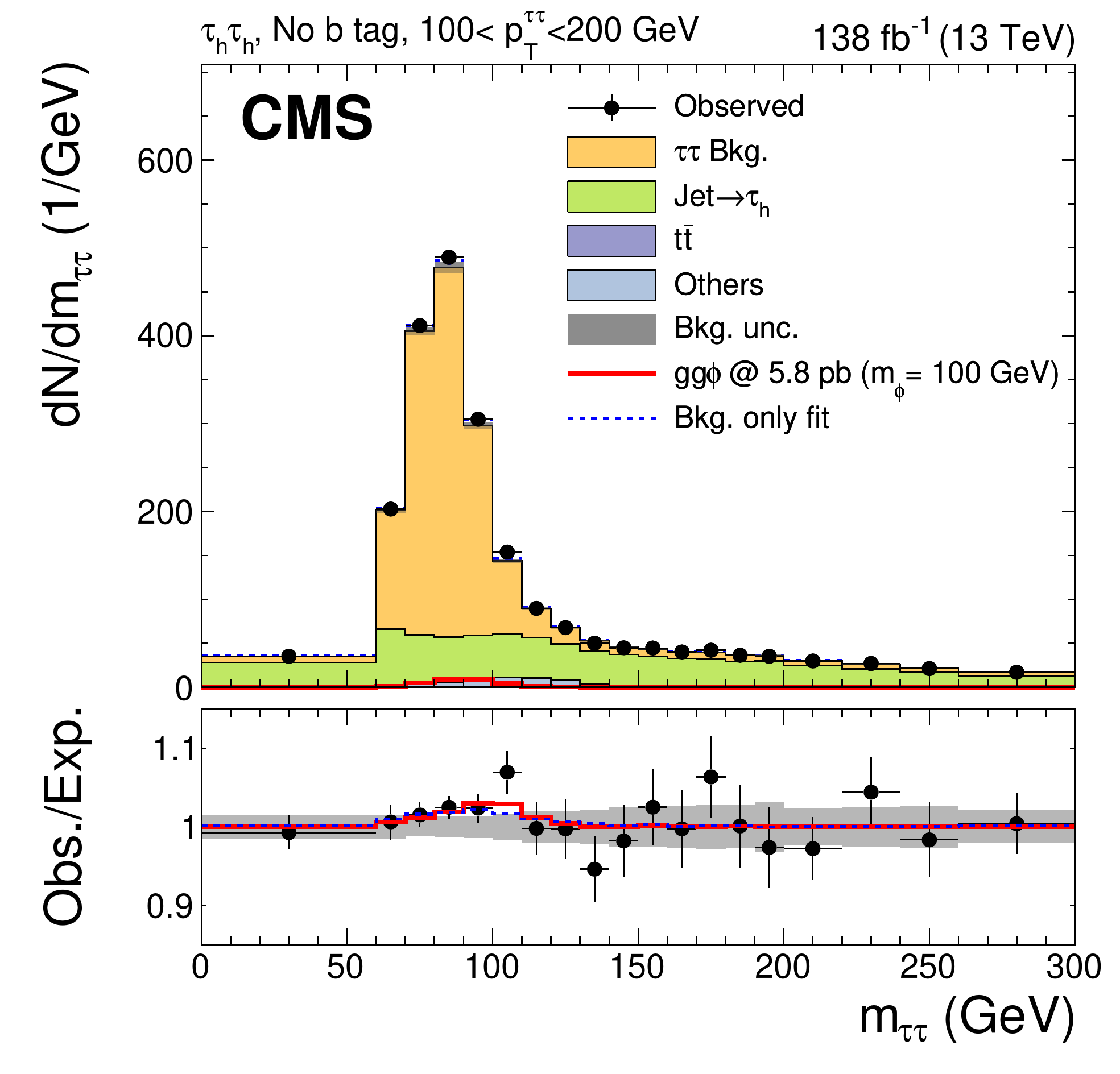}
  \includegraphics[width=.42\textwidth]{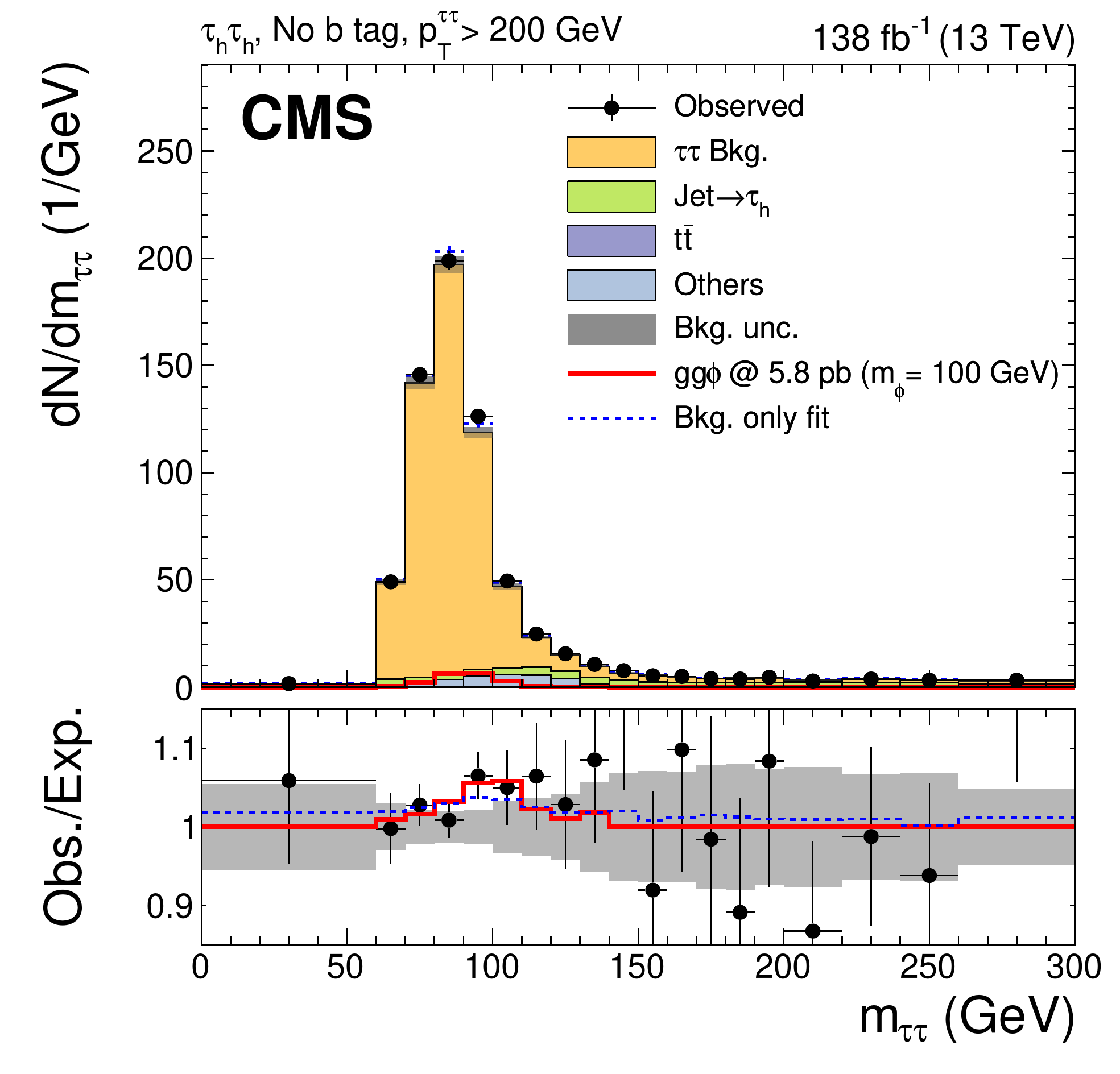}
  \caption {
    Distributions of \mtt in the (left) $100<\pttt<200\GeV$ and (right) $\pttt
      >200\GeV$ ``no \PQb tag'' categories for the (upper) \emu, (middle) \etau 
    and \mutau, and (lower) \tautau final states. The solid histograms show the 
    stacked background predictions after a signal-plus-background fit to the data 
    for $\mphi=100\GeV$. The best fit \ggPhi signal is shown by the red line. The 
    total background prediction as estimated from a background-only fit to the 
    data is shown by the dashed blue line for comparison. For all histograms, the 
    bin contents show the event yields divided by the bin widths. The lower panel 
    shows the ratio of the data to the background expectation after the 
    signal-plus-background fit to the data. The signal-plus-background and 
    background-only fit predictions are shown by the solid red and dashed blue 
    lines, respectively, which are also shown relative to the background expectation 
    obtained from the signal-plus-background fit to the data. 
  }
  \label{fig:mtt-distributions}
\end{figure}

\subsection{\texorpdfstring{Model-independent \Pphi search}
{Model-independent phi search}}

For the model-independent \Pphi search, we investigate \ggPhi and \bbPhi production 
corresponding to two independent POIs $\mu_{\ggPhi}$ and $\mu_{\bbPhi}$ in the 
likelihood of Eq.~(\ref{eq:likelihood}). In the model, \Phobs is treated as 
background assuming the production cross sections and branching fraction for the decay 
into \PGt leptons as expected from the SM. For $\mphi\geq250\GeV$, the signal 
extraction is based on binned template distributions of \mTtot in the 17 categories 
per data-taking year shown in Fig.~\ref{fig:categories}, resulting in a total of 
51 input distributions for signal extraction. For $60\leq\mphi<250\GeV$, binned 
template distributions of \mtt are used in the 26 categories shown in 
Fig.~\ref{fig:categories_lowmass}, resulting in 78 input distributions for signal 
extraction. A few examples of these input distributions in a subset of the most 
sensitive categories per final state are shown in Figs.~\ref{fig:mTtot-distributions} 
and~\ref{fig:mtt-distributions}. In each figure the expected background 
distributions are represented by the stack of filled histograms in the upper 
panel of each subfigure, where each filled histogram corresponds to a process as 
discussed in Section~\ref{sec:data-model}. The grey shaded band associated with 
the sum of filled histograms corresponds to the combination of all uncertainties 
discussed in Section~\ref{sec:systematic-uncertainties}, including all correlations 
as obtained from the fit of the background model to the data. The lower panel 
of each subfigure shows the ratio of the data points to the expectation from the 
background model in each bin. The statistical uncertainty in the data is 
represented by the error bars and the uncertainty in the sum of all background 
processes, after the fit to the data, by the shaded band. The expected \mTtot 
(\mtt) distributions for a \ggPhi or \bbPhi signal with $\mphi=1200\,(100)\GeV$ 
are also shown.

\begin{figure}[b]
  \centering
  \includegraphics[width=0.48\textwidth]{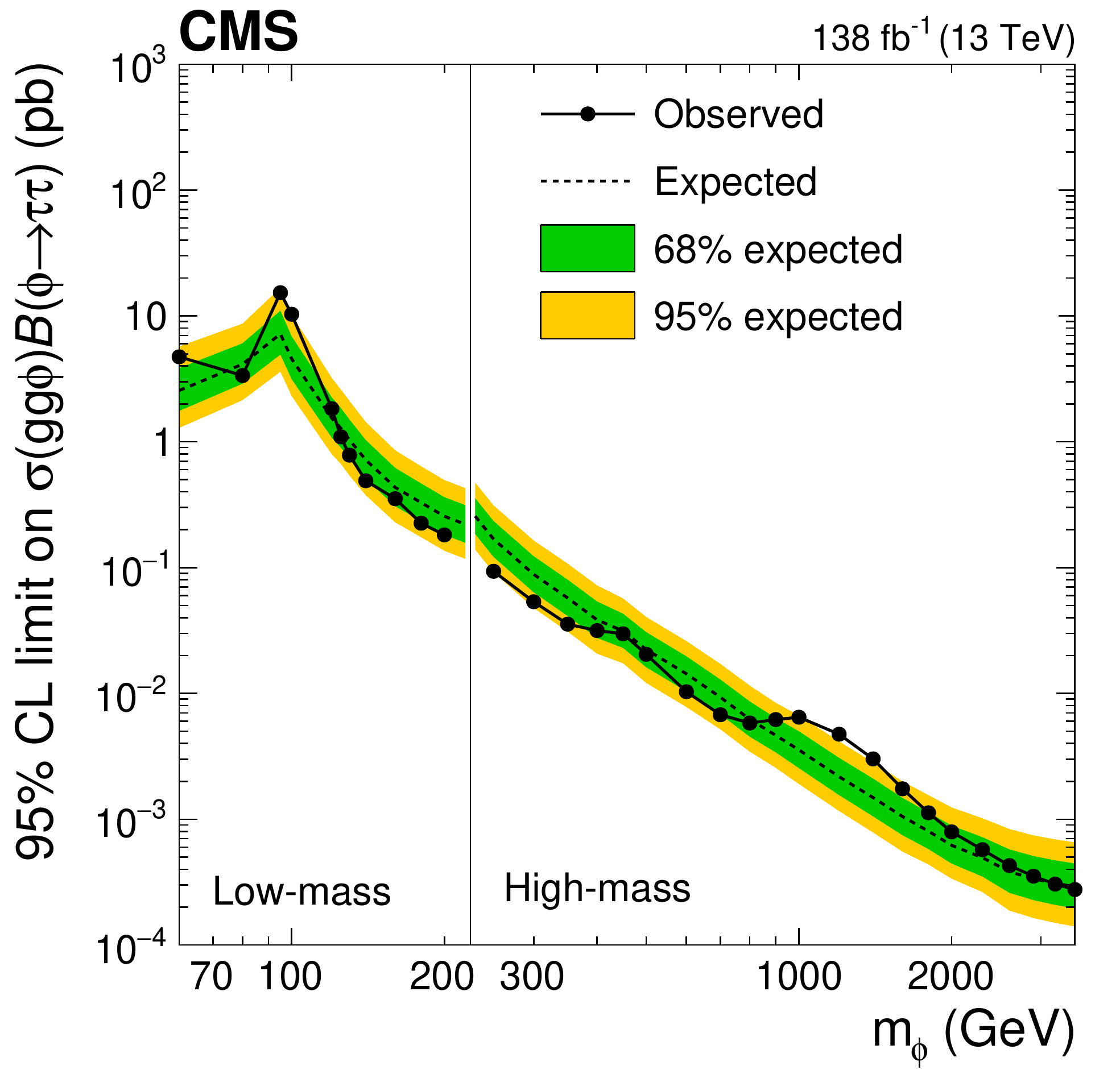}
  \includegraphics[width=0.48\textwidth]{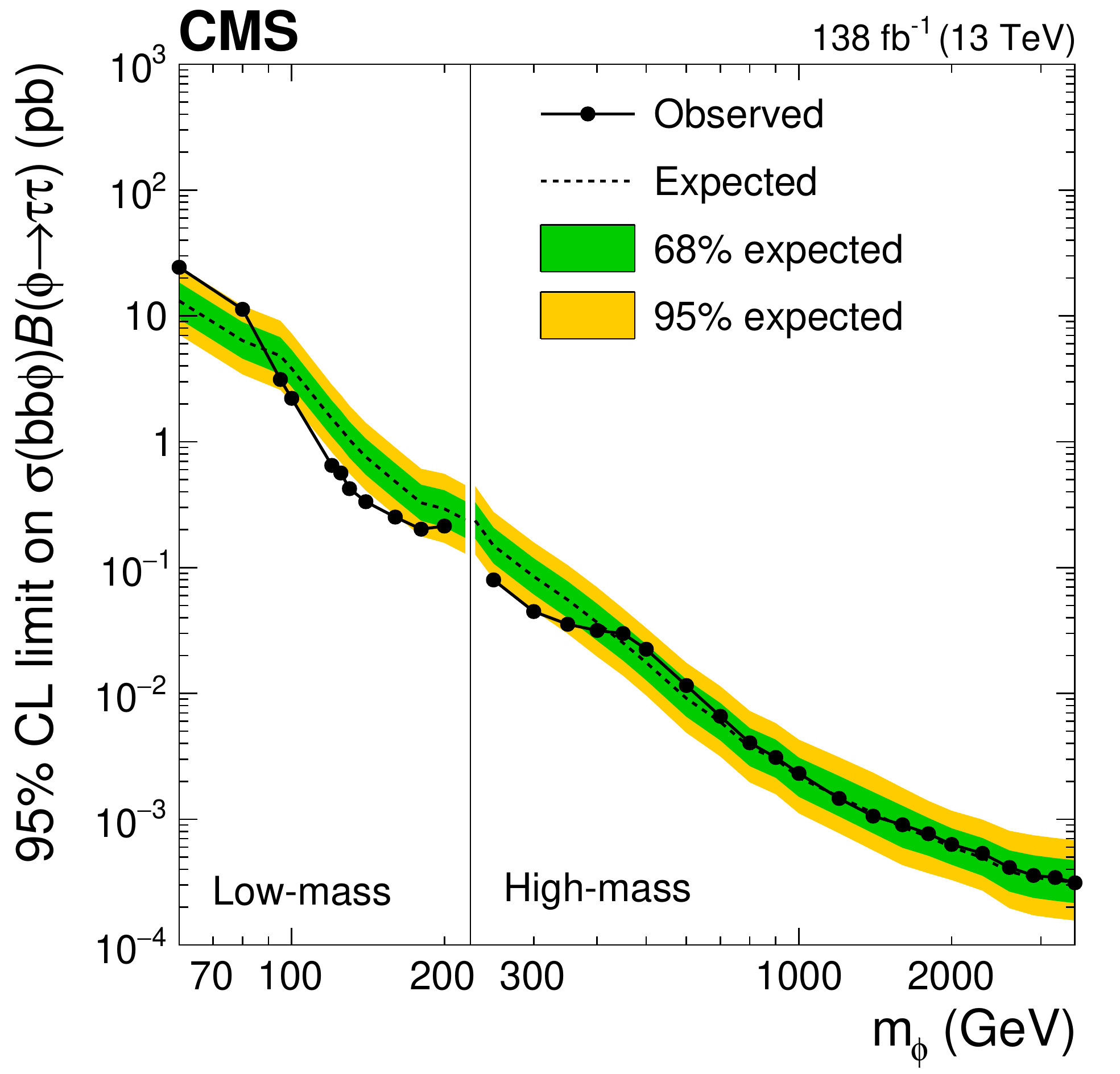}
  \caption{
    Expected and observed 95\% \CL upper limits on the product of the cross 
    sections and branching fraction for the decay into \PGt leptons for (left) 
    \ggPhi and (right) \bbPhi production in a mass range of $60\leq\mphi\leq
    3500\GeV$, in addition to \Phobs. The expected median of the exclusion limit 
    in the absence of signal is shown by the dashed line. The dark green and 
    bright yellow bands indicate the central 68\% and 95\% intervals for the 
    expected exclusion limit. The black dots correspond to the observed limits. 
    The peak in the expected \ggPhi limit emerges from the loss of sensitivity 
    around 90\GeV due to the background from \ZTT events. 
  }
  \label{fig:results_modelindep_cmb}
\end{figure}

\begin{figure}[t]
  \centering
  \includegraphics[width=0.3\textwidth]{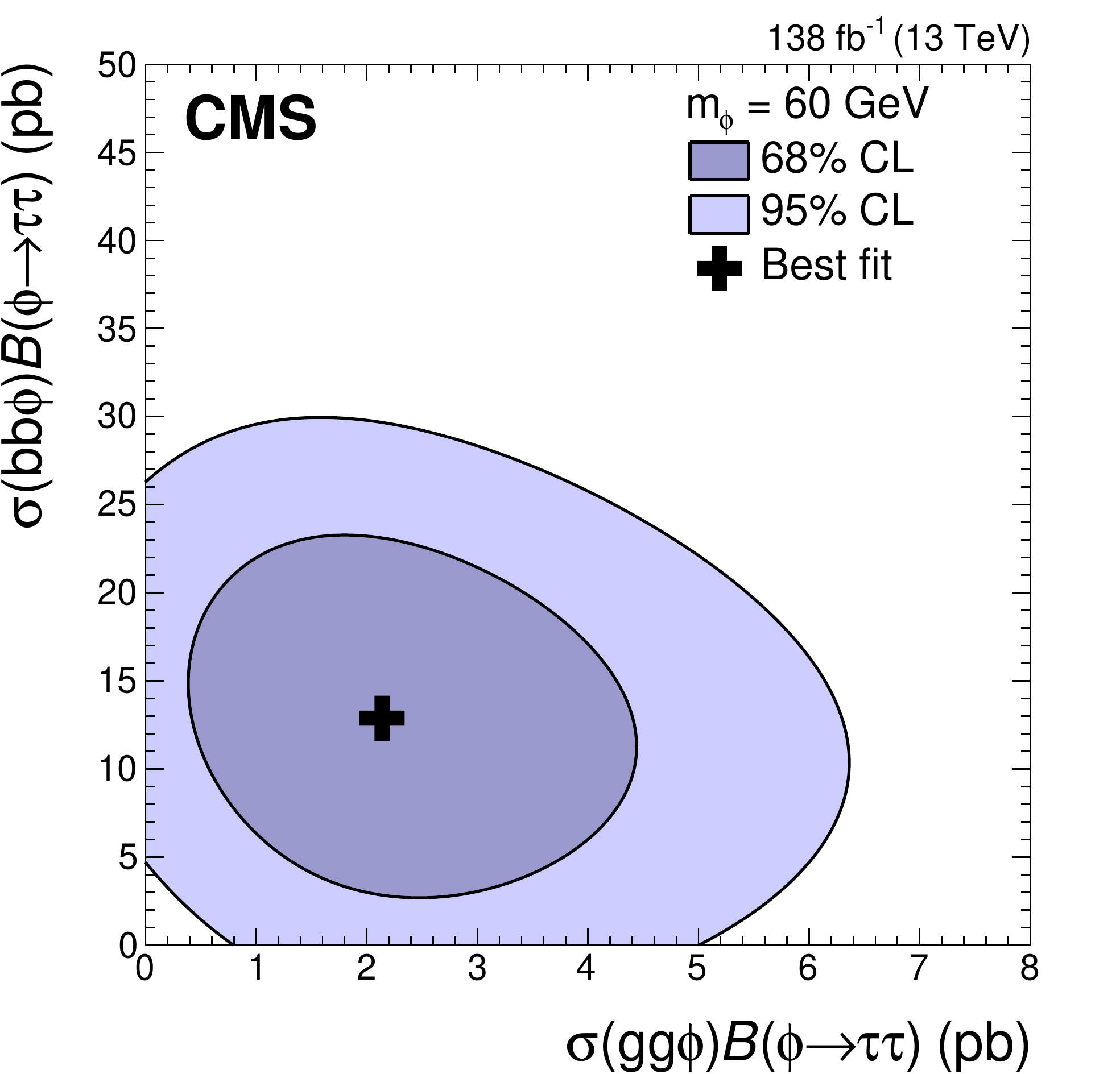}
  \includegraphics[width=0.3\textwidth]{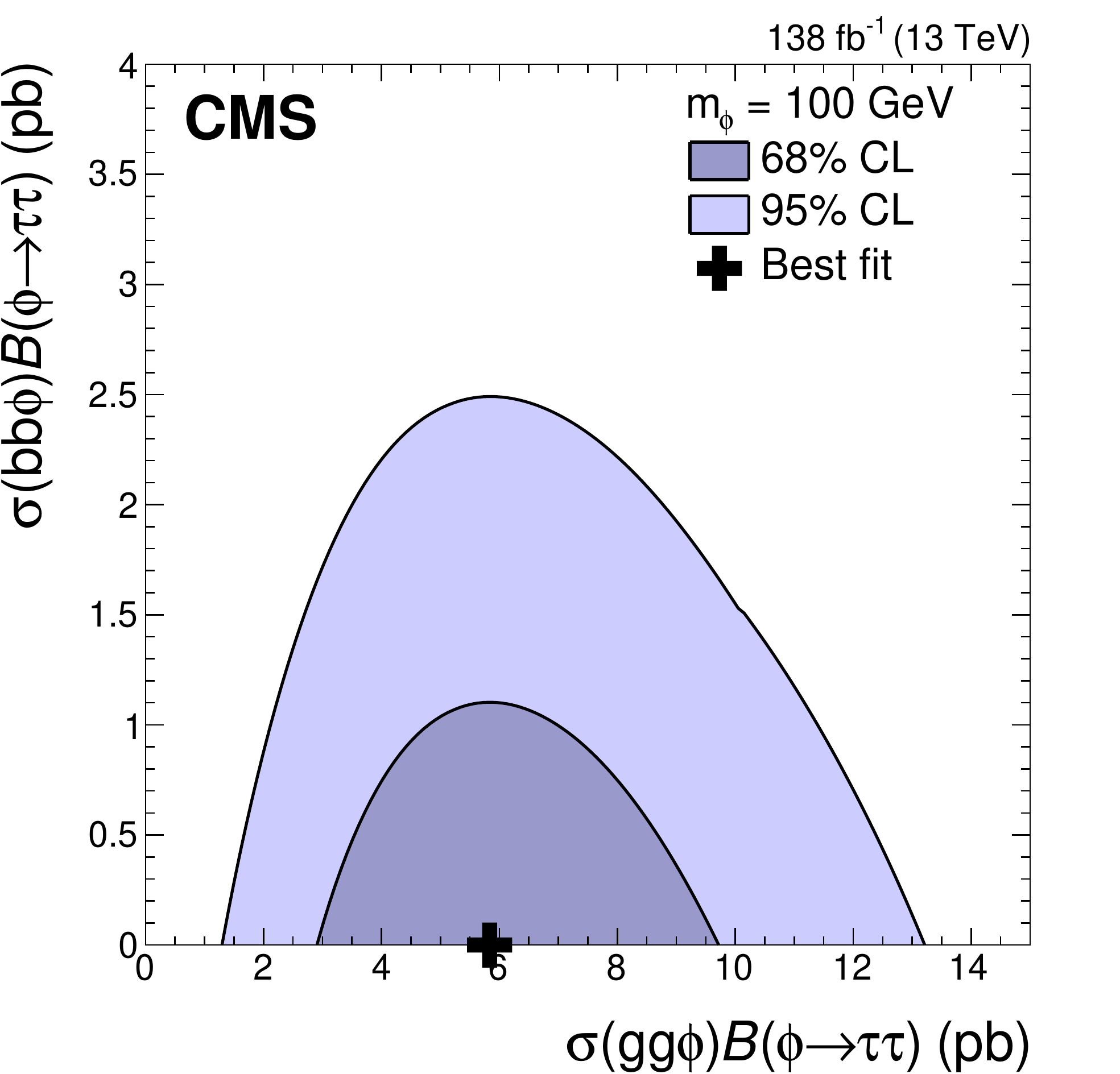}
  \includegraphics[width=0.3\textwidth]{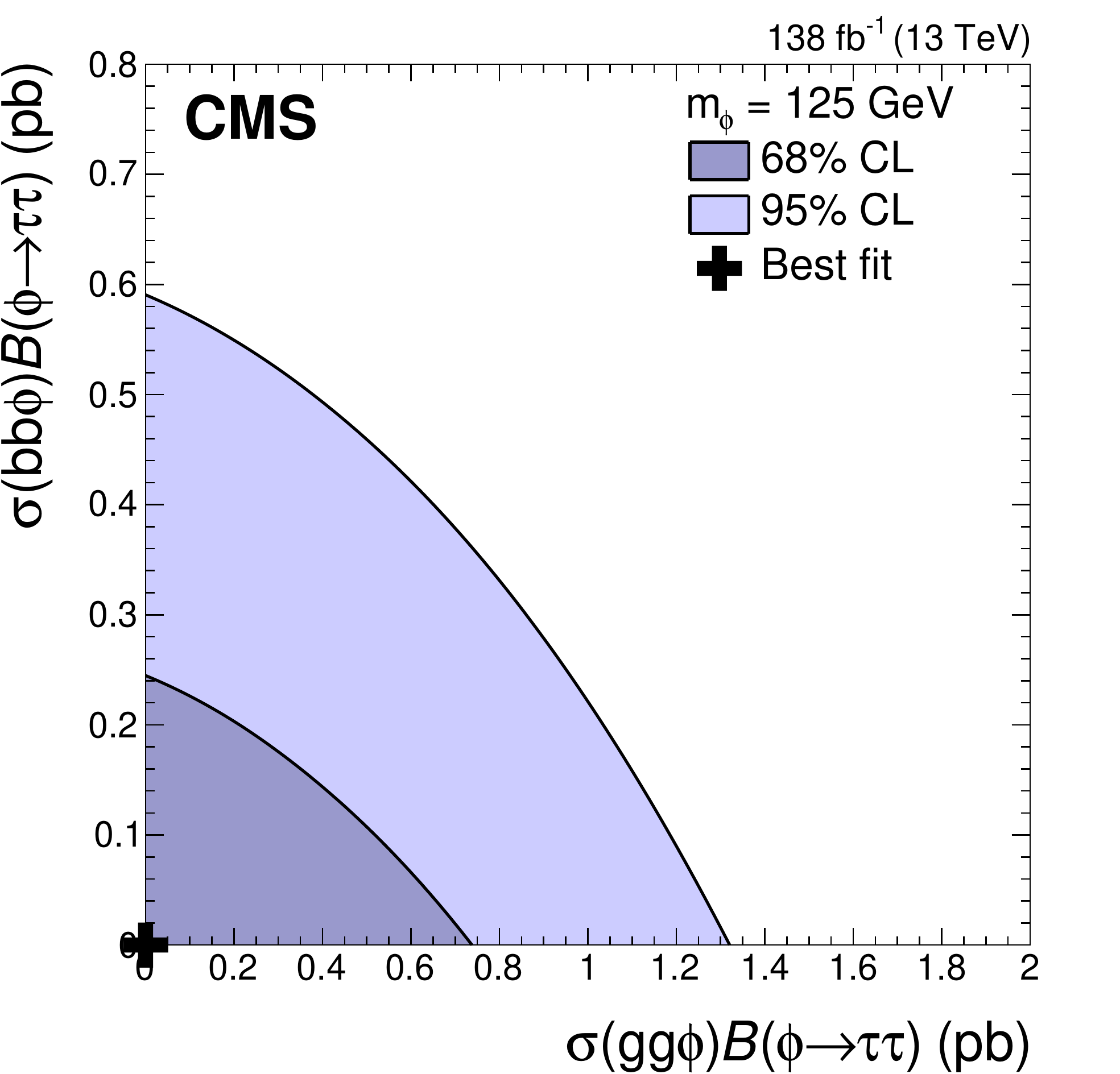} \
  \includegraphics[width=0.3\textwidth]{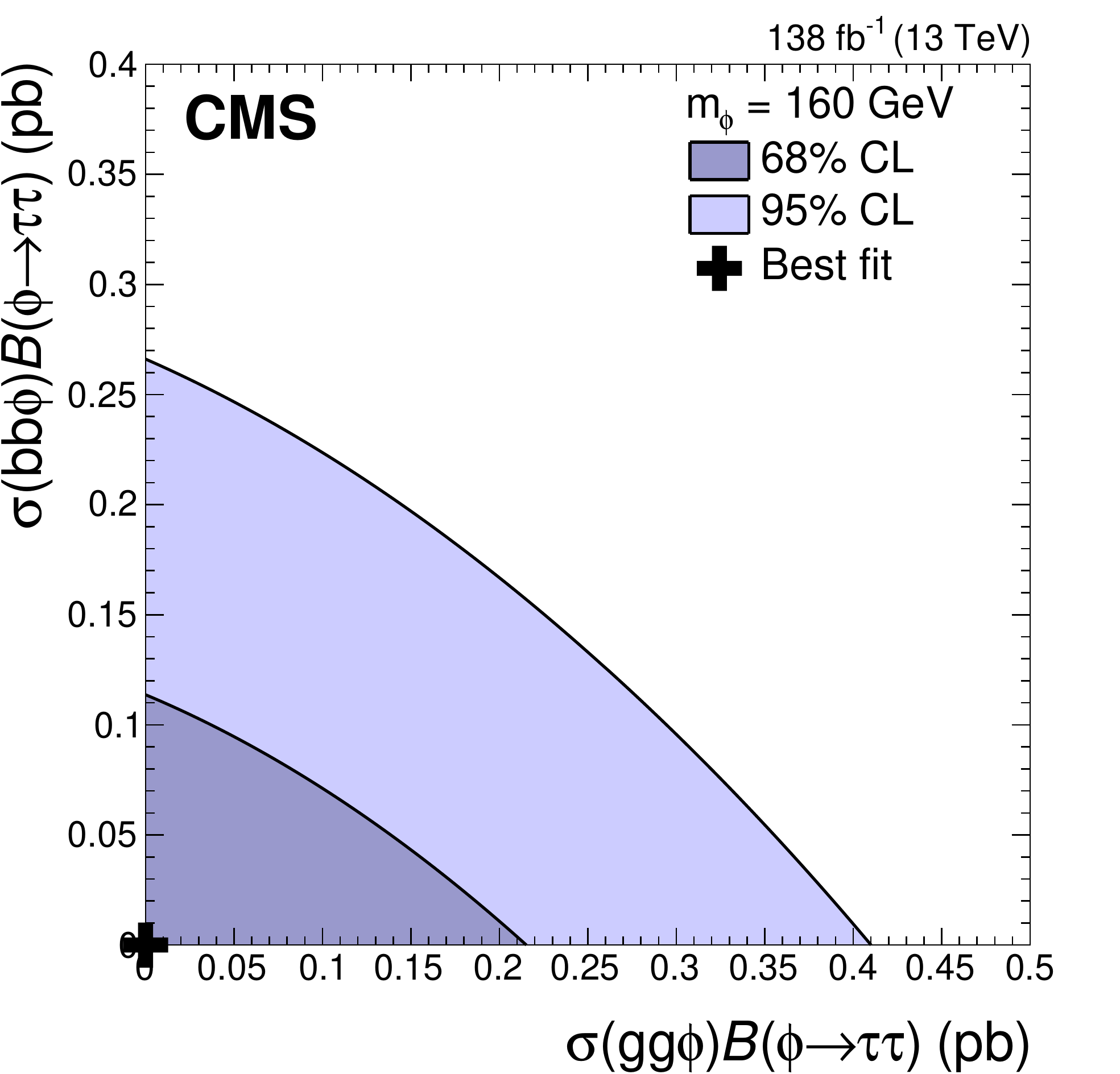}
  \includegraphics[width=0.3\textwidth]{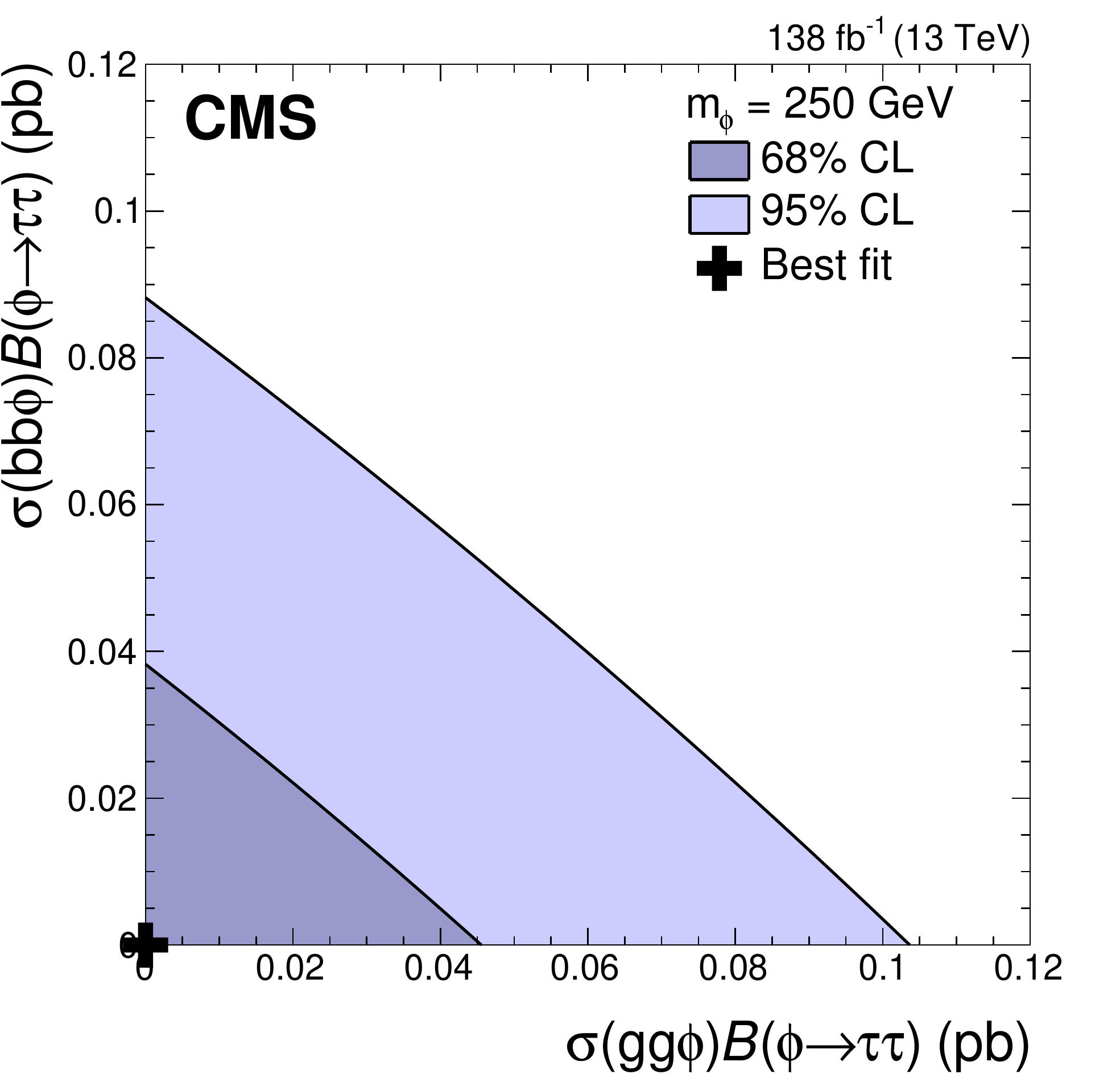}
  \includegraphics[width=0.3\textwidth]{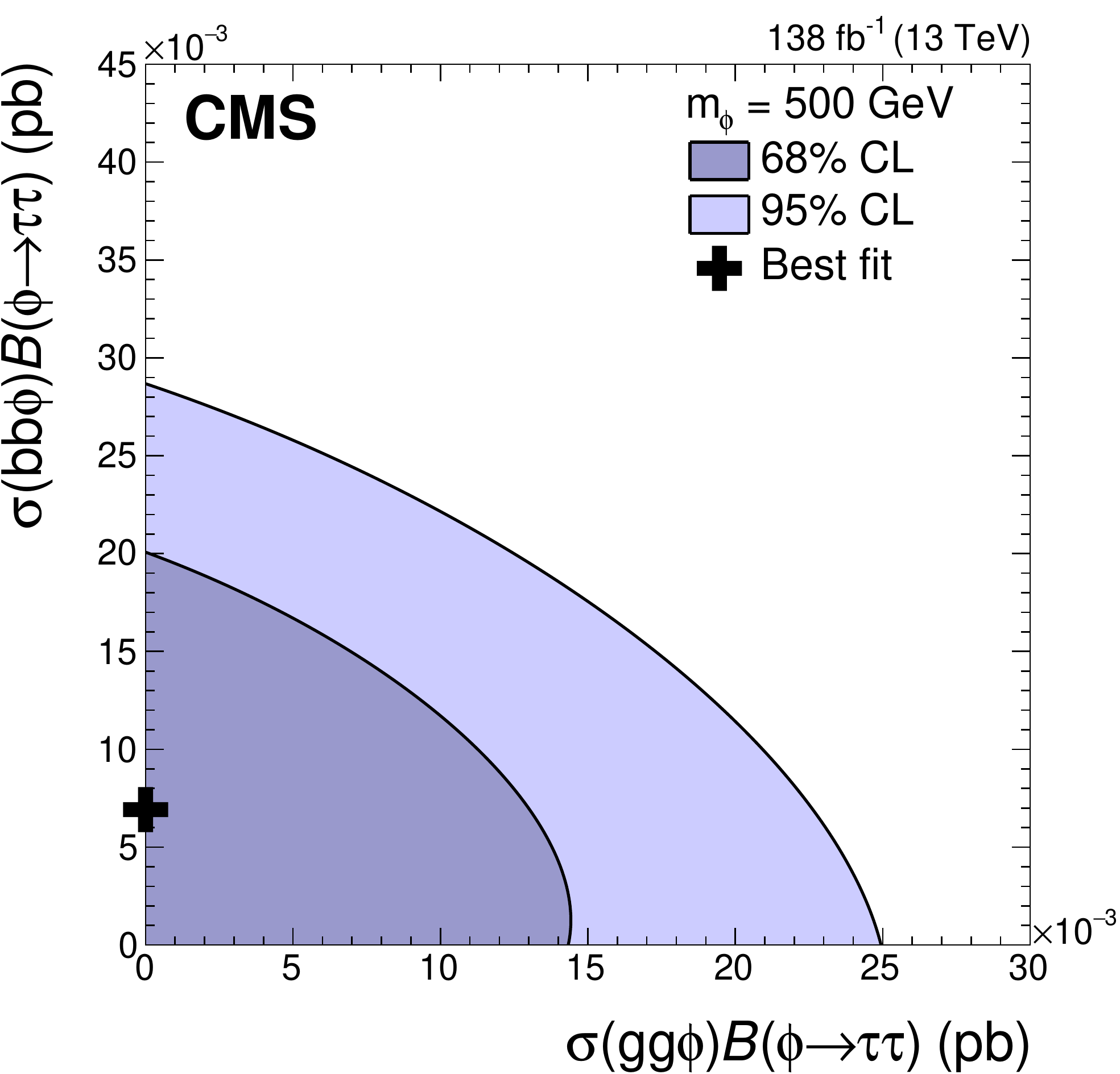} \
  \includegraphics[width=0.3\textwidth]{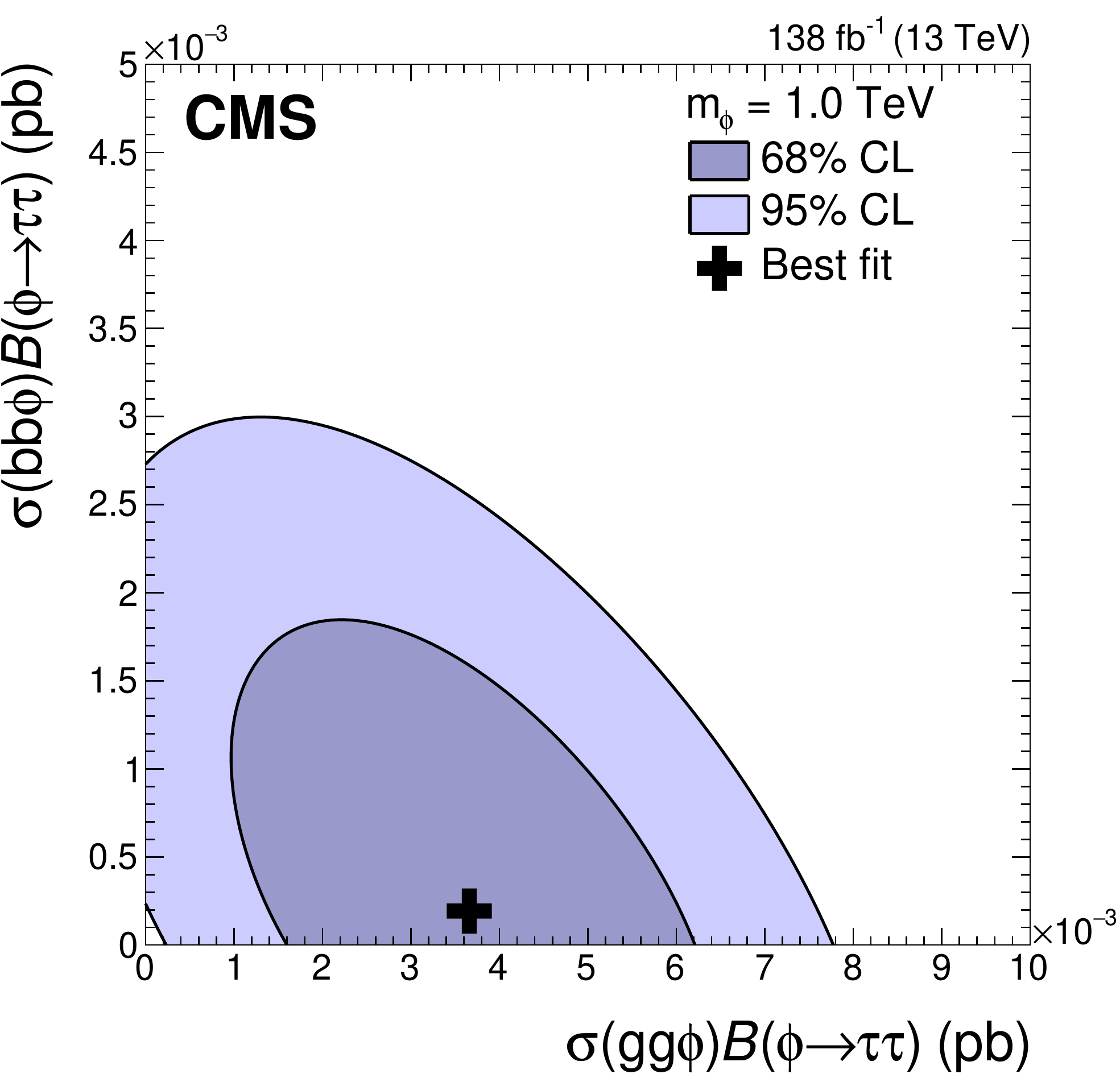}
  \includegraphics[width=0.3\textwidth]{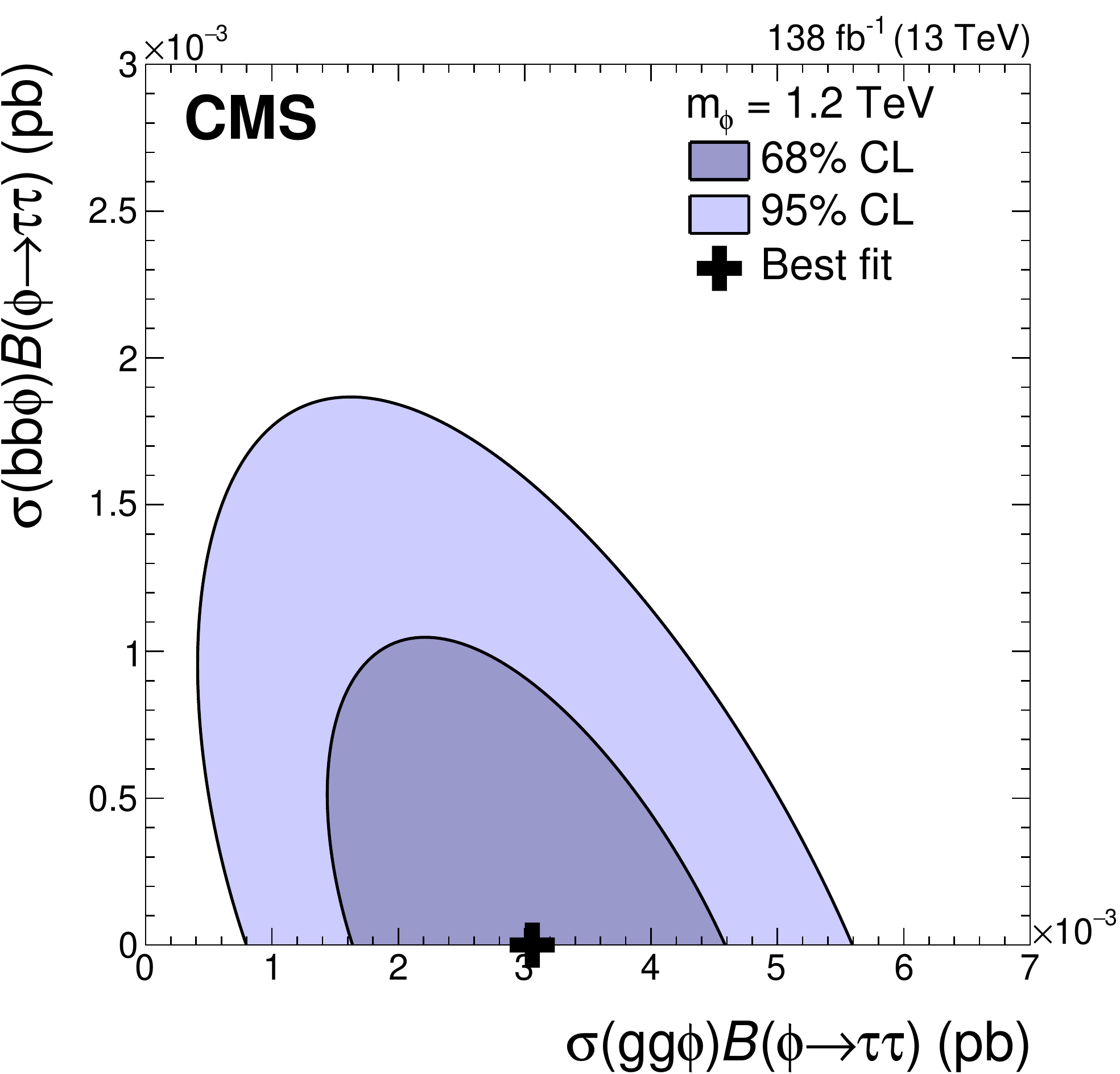}
  \includegraphics[width=0.3\textwidth]{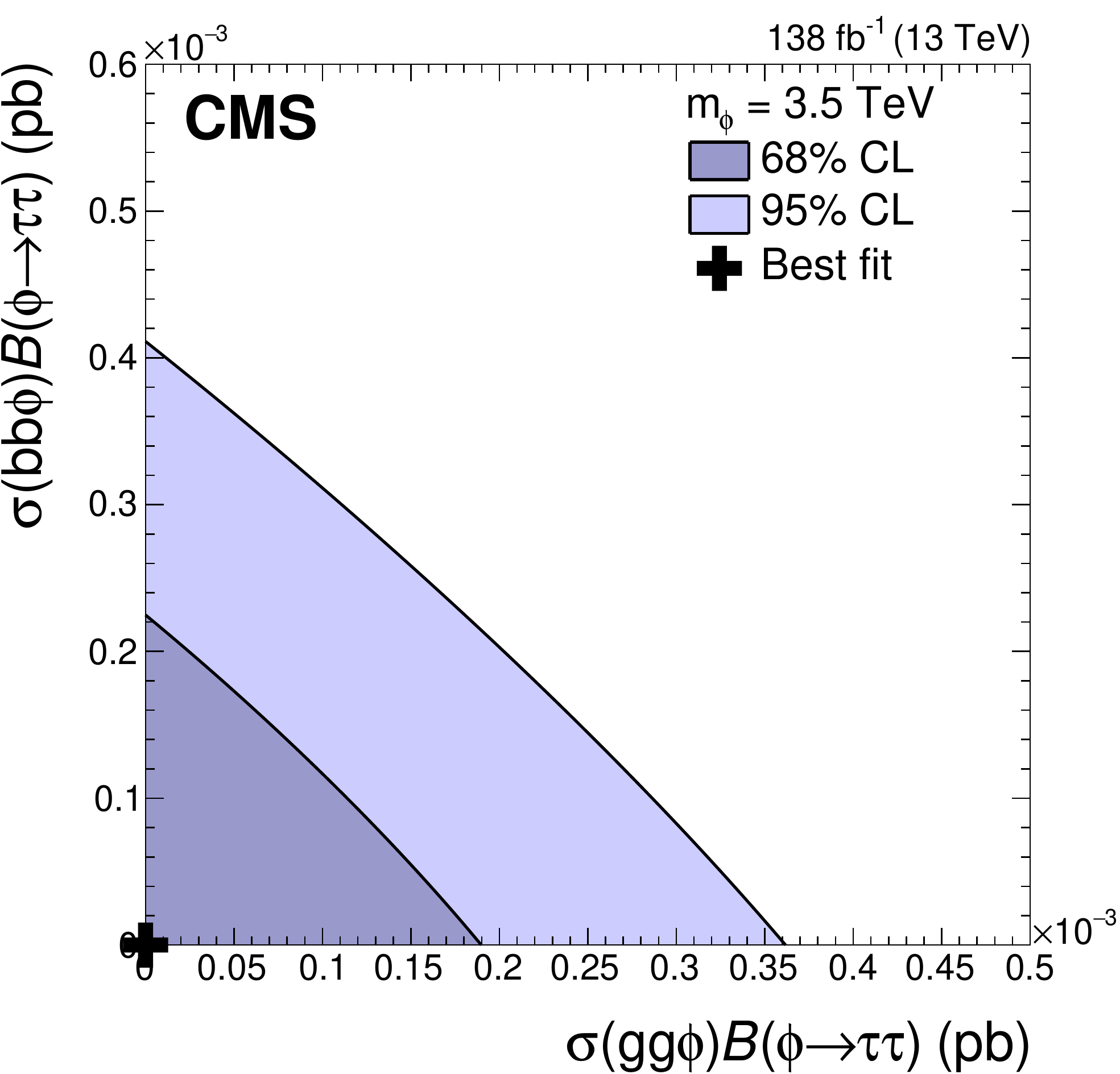}
  \caption{
    Maximum likelihood estimates, and 68\% and 95\% \CL contours obtained from scans 
    of the signal likelihood for the model-independent \Pphi search. The scans are 
    shown for selected values of \mphi between 60\GeV and 3.5\TeV. In each figure 
    the SM expectation is $(0, 0)$. 
  }
  \label{fig:likelihood-scans}
\end{figure}

Figure~\ref{fig:results_modelindep_cmb} shows the expected and observed 95\% 
confidence level (\CL) upper limits on the product of the cross sections and 
branching fraction for the decay into \PGt leptons for \ggPhi and \bbPhi 
production in a mass range of $60\leq\mphi\leq3500\GeV$. These limits have been 
obtained following the modified frequentist approach described in 
Refs.~\cite{Junk:1999kv,Read:2002hq}. When setting the limit in one production 
mode the POI of the other production mode is profiled. The limits are shown with 
a separation into the low-mass ($\mphi<250\GeV$) and high-mass ($\mphi\geq 250\GeV$) 
regions of the search. 

The expected limits in the absence of a signal span four orders of magnitude 
between ${\approx}10\unit{pb}$ (at $\mphi=60\GeV$) and ${\approx}0.3\unit{fb}$ 
(at $\mphi=3.5\TeV$) for both production modes, with a falling slope for increasing 
values of \mphi. In general, the observation falls within the central 95\% interval 
of the expectation. For the low-mass search, the largest deviation from the 
expectation is observed for \ggPhi production at $\mphi=100\GeV$ with a local 
(global) $p$-value equivalent to 3.1 (2.7) standard deviations (s.d.). To turn 
the local into a global $p$-value, a number $N_{\text{trial}}$ of pseudo-data 
from the input distributions of the background model to the maximum likelihood 
fit is created. For each mass hypothesis in consideration, a fit of the signal 
model to these pseudo-data is performed and the fraction of cases, where the 
outcome of these fits with the maximal significance exceeds the observed significance, 
with respect to $N_{\text{trial}}$ is determined. Finally, the local $p$-value 
is reduced by this fraction. The best fit value of the product of the cross section 
with the branching fraction for the decay into \PGt leptons is $\sigma_{\ggPhi}\,
\mathcal{B}(\Pphi\to\ditau)=(5.8\pm {}^{2.5}_{2.0})\unit{pb}$. The excess at 
$\mphi=100\GeV$ exhibits a $p$-value of 50\% (58\%) for the compatibility across 
\ditau final states (data-taking years). Within the resolution of \mtt this 
coincides with a similar excess observed in a previous search for low-mass 
resonances by the CMS Collaboration in the $\PGg\PGg$ final state, where the 
smallest local $p$-value corresponds to a significance of 2.8 s.d.\ for a mass of 
95.3\GeV~\cite{CMS:2018cyk}. The local (global) significance for the \ditau search 
evaluated at $\mphi=95\GeV$ is 2.6 (2.3) s.d.\ and the best fit value of the 
product of the cross section with the branching fraction for the decay into \PGt 
leptons is $\sigma_{\ggPhi}\,\mathcal{B}(\Pphi\to\ditau)=(7.8\pm {}^{3.9}_{3.1})
\unit{pb}$. For the high-mass search, the largest deviation from the expectation 
is observed for \ggPhi production at $\mphi=1.2\TeV$ with a local (global) 
$p$-value equivalent to 2.8 (2.2) s.d., where the best fit value of the product 
of the cross section with the branching fraction for the decay into \PGt leptons 
is $\sigma_{\ggPhi}\,\mathcal{B}(\Pphi\to\ditau)=(3.1\pm {}^{1.0}_{1.1})\unit{fb}$. 
The excess at $\mphi=1.2\TeV$ exhibits a $p$-value of 11\% (63\%) for the compatibility 
across \ditau final states (data-taking years). For \bbPhi production, no deviation 
from the expectation beyond the level of 2 s.d.\ is observed. 
Figure~\ref{fig:likelihood-scans} shows the same results in the form of maximum 
likelihood estimates with 68\% and 95\% \CL contours obtained from scans of the signal 
likelihood along the \ggPhi and \bbPhi cross sections, for selected values of \mphi 
between 60\GeV and 3.5\TeV. 

\subsection{Search for vector leptoquarks}
\label{sec:results_VLQ}

The inputs for the search for vector leptoquarks are the binned template 
distributions of \mTtot in the categories shown in Fig.~\ref{fig:categories} 
resulting in 51 input distributions for signal extraction, for the years 2016--2018. 
Based on these inputs a signal is searched for in the range of $1<\mU<5\TeV$. 

Due to the destructive interference with the \ZTT process discussed in 
Section~\ref{sec:simulation}, depending on \mTtot, a signal from \Uone $t$-channel 
exchange may result in an enhancement or a reduction of the yields as expected 
from the SM. A typical example of this effect for a signal with $\mU=1\TeV$, 
$\gU=1.5$, for the VLQ BM 1 scenario is shown in Fig.~\ref{fig:signal-templates} 
(right). There, a reduction in yield (with respect to the SM) is expected for 
$\mTtot\lesssim250\GeV$ and an enhancement for $250\lesssim\mTtot\lesssim1000
\GeV$. In principle both effects contribute to the sensitivity of the analysis 
to the signal. However, the region of $\mTtot\lesssim250\GeV$, which features 
the deficit, suffers from large backgrounds, reducing the contribution of this 
effect to the overall sensitivity. Studies confirm that the sensitivity of the 
analysis to the signal is driven by the high \mTtot region, which consistently 
features an enhancement of the yields expected from the SM. Since the interference 
reduces this enhancement it also reduces the sensitivity of the analysis, compared 
to a signal without interference. 

\begin{figure}[b]
  \centering
  \includegraphics[width=0.48\textwidth]{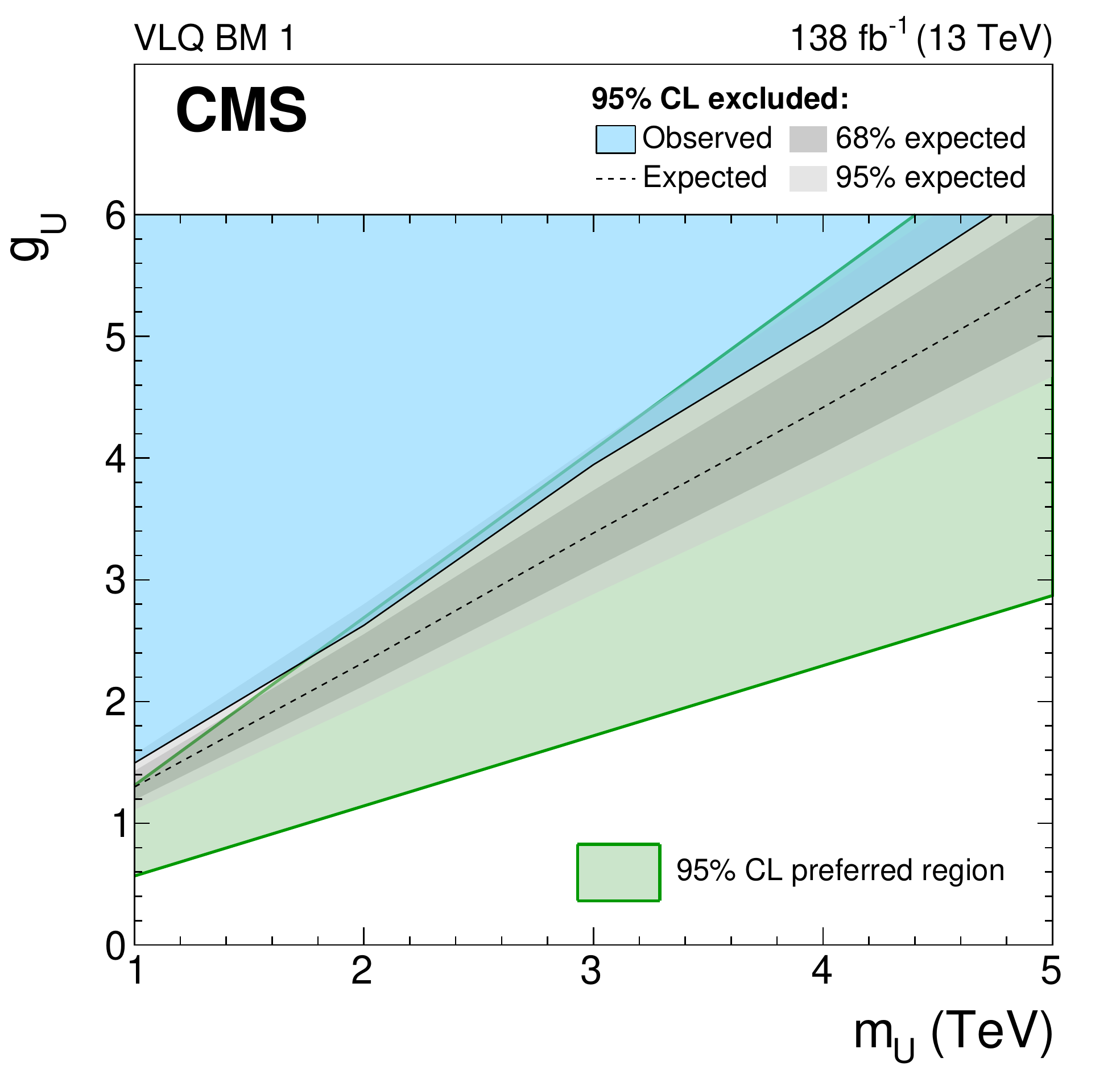}
  \includegraphics[width=0.48\textwidth]{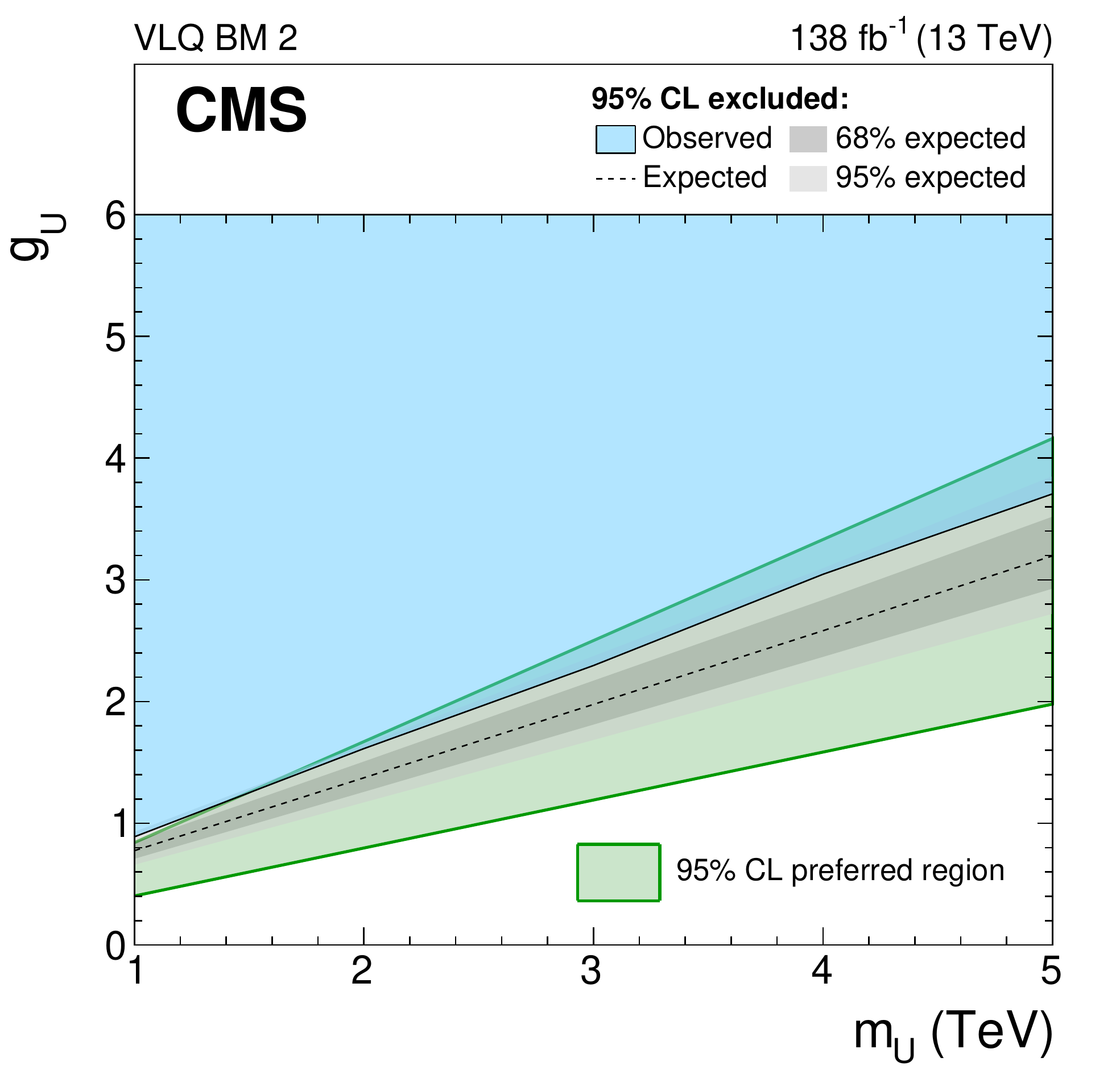}
  \caption{
    Expected and observed 95\% \CL upper limits on \gU in the VLQ BM (left) 1 and 
    (right) 2 scenarios, in a mass range of $1<\mU<5\TeV$. The expected 
    median of the exclusion limit in the absence of signal is shown by the dashed 
    line. The dark and bright grey bands indicate the central 68\% and 95\% 
    intervals of the expected exclusion limit. The observed excluded parameter 
    space is indicated by the coloured blue area. For both scenarios, the 95\% 
    confidence interval for the preferred region from the global fit presented 
    in Ref.~\cite{Cornella:2021sby} is also shown by the green shaded area.
  }
  \label{fig:vlq_exclusion-contours}
\end{figure}

No statistically significant signal is observed and 95\% \CL upper limits on \gU 
are derived for the VLQ BM 1 and 2 scenarios, as shown in 
Fig.~\ref{fig:vlq_exclusion-contours}, again following the modified frequentist 
approach as for the previously discussed search. The expected sensitivity of the 
analysis drops for increasing values of \mU following a linear progression with 
values from $\gU=1.3$ (0.8) to 5.6 (3.2) for the VLQ BM 1 (2) scenario. The observed 
limits fall within the central 95\% intervals for the expected limits in the absence 
of signal. The expected limits are also within the 95\% confidence interval of 
the best fit results reported by Ref.~\cite{Cornella:2021sby}, indicating that 
the search is sensitive to a portion of the parameter space that can explain the 
\PQb physics anomalies.

\subsection{MSSM interpretation of the data}

For the interpretation of the data in MSSM benchmark scenarios, the signal is 
based on the binned distributions of \mTtot in the categories shown in 
Fig.~\ref{fig:categories}, complemented by distributions of the NN output 
function used for the stage-0 simplified template cross section measurement of 
Ref.~\cite{CMS:2022kdi}, as discussed in Section~\ref{sec:event-categories}, 
resulting in 129 input distributions for signal extraction. 

In the MSSM, the signal constitutes a multiresonance structure with contributions 
from \Ph, \PH, and \PA bosons. For the scenarios chosen for this paper \Ph is 
associated with \Phobs. Any MSSM prediction has to match the observed properties 
of \Phobs, in particular its mass, cross sections for various production modes, 
and branching fraction for the decay into \PGt leptons. For the benchmark scenarios 
summarized in Ref.~\cite{Bagnaschi:2791954}, all model parameters have been chosen 
such that \mh is compatible with the observed \Phobs mass of 
125.38\GeV~\cite{Sirunyan:2020xwk}, within an uncertainty of ${\pm}3\GeV$ in most 
of the provided parameter space. The uncertainty of $\pm3\GeV$ in the prediction 
of \mh is supposed to reflect the unknown effect of higher-order corrections, as 
discussed in Ref.~\cite{Slavich:2020zjv}. The value of \mh is allowed to vary 
within these boundaries, according to a flat distribution. For the interpretation 
this is taken into account by simulating the \Ph signal at the observed \Phobs 
mass. For \Ph production, the modes via \ggh, \PQb associated production 
(\bbh), VBF, and \Vh production are included, and all cross sections and the 
branching fraction for the decay into \PGt leptons are scaled according to the 
MSSM predictions. To remove any dependencies of these predictions on the exact 
value of \mh, they are scaled to the expectation for $\mh=125.38\GeV$, following 
the prescription of Ref.~\cite{Bagnaschi:2791954}. For \PA and \PH production, 
gluon fusion (\ggA, \ggH) and \PQb associated production (\bbA, \bbH) are included. 

All kinematic distributions are modelled within the accuracies discussed in 
Section~\ref{sec:simulation}. In particular, the \PH (\PA) boson \pt spectra in 
\ggH (\ggA) production are modelled as a function of \tanb for each tested value 
of \mA, resulting in a softer progression for increasing values of \tanb. In the 
``no \PQb tag'' categories for $\mtt>250\GeV$ the \Ph signal is expected to be 
negligible so it is dropped from the signal templates. A summary of the 
association of signals to the templates used for signal extraction is given in 
Table~\ref{tab:mssm-signal-association}. To interpolate the simulated mass points 
to the exact predicted values of \mH, a linear template morphing algorithm, as 
described in Ref.~\cite{Read:1999kh}, is used. 

\begin{table}[b]
  \centering
  \topcaption{
    Contribution of MSSM signals to the \mTtot and NN output function template 
    distributions used for signal extraction for the interpretation of the data 
    in MSSM benchmark scenarios.
  }
  \begin{tabular}{llcc}
    & & \multicolumn{2}{c}{Signal processes} \\
    \multicolumn{2}{l}{Categories} & \ggh, \bbh, VBF, \Vh & \ggH/\ggA, \bbH/\bbA \\
    \hline
    No \PQb tag & $\mtt<250\GeV$    & $\checkmark$ & \checkmark \\
    No \PQb tag & $\mtt>250\GeV$ & \NA          & $\checkmark$ \\
    \multicolumn{2}{l}{\PQb tag} & $\checkmark$ & $\checkmark$ \\
    \multicolumn{2}{l}{Control regions} & $\checkmark$ & $\NA$ \\
  \end{tabular}
  \label{tab:mssm-signal-association}
\end{table}

The \mA-\tanb plane is scanned and for each tested point in (\mA, \tanb), the 
\CLs ~\cite{Read:2002hq} value is calculated. Those points where \CLs falls below 
5\% define the 95\% \CL exclusion contour for the benchmark scenario under 
consideration. The underlying test compares the MSSM hypothesis, with signal 
contributions for \Ph ($S_{\Ph}$), \PH ($S_{\PH}$), and \PA ($S_{\PA}$), with 
the SM hypothesis ($S_{\text{SM}}$), with only one signal contribution related 
to \Phobs. The test versus the SM hypothesis is justified by the properties of 
\Phobs being in agreement with the SM expectation within the experimental 
accuracy of current measurements. For the hypothesis test the likelihood of 
Eq.~(\ref{eq:likelihood}) is expressed in the form
\begin{linenomath}
  \begin{equation}
    \mathcal{L}\left(\{k_{i}\},\mu\right) =\prod
    \limits_{i}\mathcal{P}\Bigl(k_{i}|\mu\Bigl((S_{\Ph}-S_{\text{SM}})+S_{\PH}
    +S_{\PA}\Bigr)+S_{\text{SM}}+
    \sum\limits_{b}B_{b}\Bigr),
    \label{eq:likelihood-mssm}
  \end{equation}
\end{linenomath}
where for brevity the dependence on the nuisance parameters $\{\theta_{j}\}$ has 
been omitted. Equation~(\ref{eq:likelihood-mssm}) represents a nested likelihood 
model from which the MSSM hypothesis (with $\mu=1$) evolves through continuous 
transformation from the SM hypothesis (with $\mu=0$). We note that the only 
physically meaningful hypotheses in Eq.~(\ref{eq:likelihood-mssm}) correspond to 
$\mu=0$ and 1. On the other hand, in the large number limit this construction 
allows the application of the asymptotic formulas given in Ref.~\cite{Cowan:2010js}, 
as analytic estimates of the sampling distributions for the MSSM and SM hypotheses, 
when using the profile likelihood ratio given in Eq.~(\ref{eq:profile-likelihood-ratio}) 
as the test statistic. We have verified the validity of the large number limit 
for masses of $\mA>1\TeV$ with the help of ensemble tests. Since we are using 
the same template distributions for $S_{\text{SM}}$ and $S_{\Ph}$ the transition 
from $\mu=0$ to 1 corresponds to a normalization change of the signal contribution 
related to \Phobs, only.

\begin{figure}[t]
  \centering
  \includegraphics[width=0.48\textwidth]{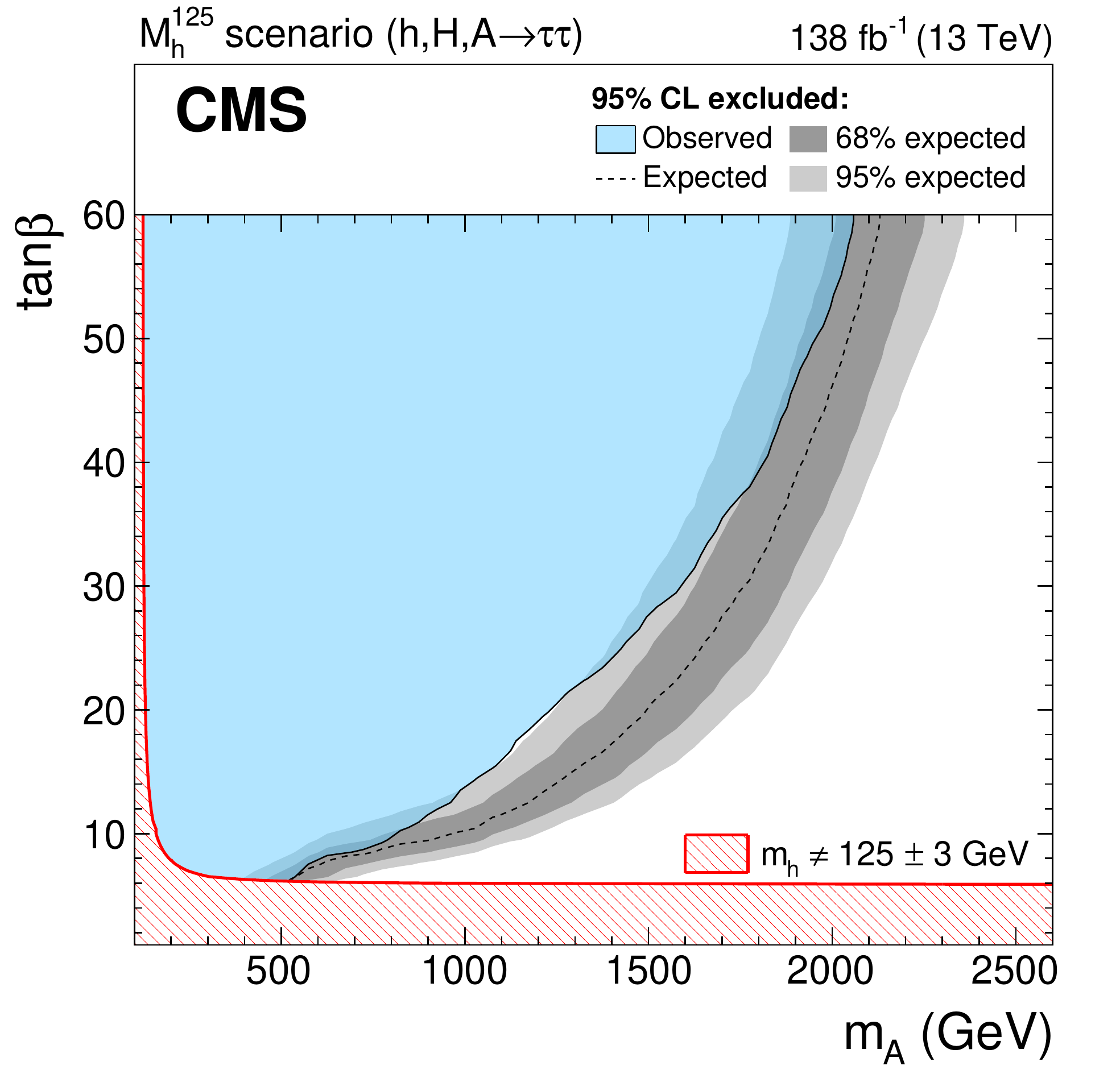}
  \includegraphics[width=0.48\textwidth]{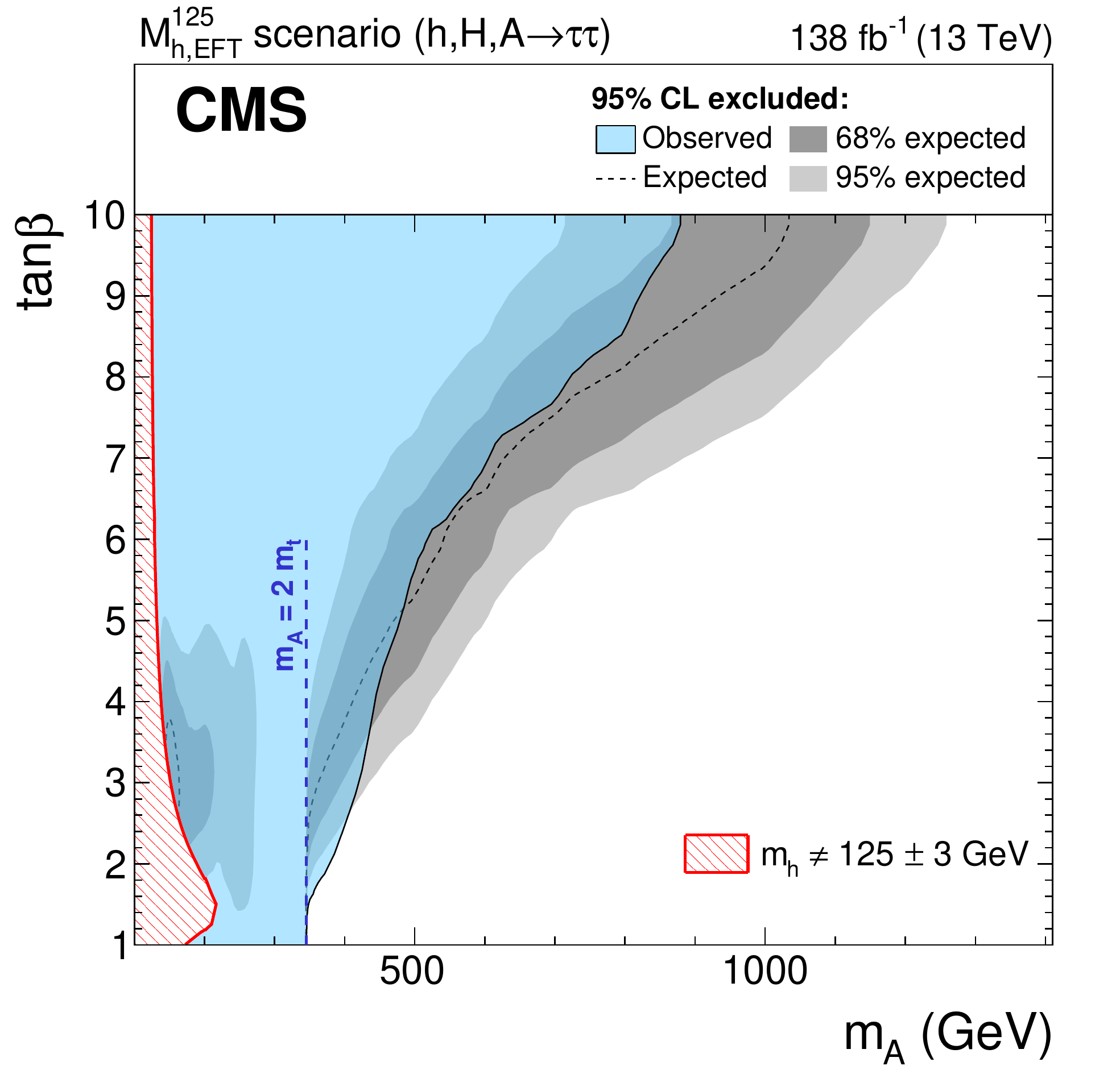}
  \caption{
    Expected and observed 95\% \CL exclusion contours in the MSSM (left) \mhBMPone 
    and (right) \mhBMPtwo scenarios. The expected median in the absence of a signal 
    is shown as a dashed black line. The dark and bright grey bands indicate the 
    central 68\% and 95\% intervals of the expected exclusion. The observed exclusion 
    contour is indicated by the coloured blue area. For both scenarios, those parts 
    of the parameter space where \mh deviates by more then ${\pm}3\GeV$ from the 
    mass of \Phobs are indicated by a red hatched area. For the \mhBMPtwo scenario, 
    the dashed blue line indicates the threshold at $\mA=2\mt$ whereby the $\PA\to
    \ttbar$ decay starts to influence the \ATT branching fraction. The \HTT 
    branching fraction is influenced more gradually close to this threshold since 
    \PA and \PH are not completely degenerate in mass.  
  }
  \label{fig:exclusion-contours}
\end{figure}

Figure~\ref{fig:exclusion-contours} shows the exclusion contours in the 
\mA-\tanb plane for two representative benchmark scenarios of the MSSM, 
\mhBMPone~\cite{Bahl:2018zmf} and \mhBMPtwo~\cite{Bahl:2019ago}. The 
red hatched areas indicate the regions where the compatibility of \mh with the 
observed \Phobs mass could not be achieved within the previously discussed ${\pm}
3\GeV$ boundary. For low values of \tanb, higher scales for the additional SUSY 
particle masses (referred to collectively as ``\mSUSY'') are required to accommodate 
a mass of $\mh\approx125\GeV$. In the \mhBMPone scenario, where \mSUSY is fixed, the 
prediction of \mh falls below 122\GeV. In the \mhBMPtwo scenario, \mSUSY is adjusted 
to values that meet the required prediction for \mh in each point in (\mA, \tanb) 
individually. The growing logarithmic corrections associated with the large values 
of \mSUSY are resummed using an effective field theory approach. The \mhBMPtwo 
scenario can thus be viewed as a continuation of the \mhBMPone scenario for 
$\tanb\lesssim10$. In this case the red hatched area at very low values of \mA in 
Fig.~\ref{fig:exclusion-contours} (right) indicates the parameter space where the 
values required for \mSUSY exceed the GUT scale. For both scenarios the Higgs boson 
masses, mixing angle $\alpha$, and effective Yukawa couplings have been calculated 
with the code \textsc{FeynHiggs}~\cite{Heinemeyer:1998yj,Heinemeyer:1998np,
Degrassi:2002fi,Frank:2006yh,Hahn:2013ria,Bahl:2016brp,Bahl:2017aev,Bahl:2018qog}. 
Branching fractions for the decay into \PGt leptons and other final states have 
been obtained from a combination of the codes \textsc{FeynHiggs} (and 
\textsc{HDECAY}~\cite{Djouadi:1997yw,Djouadi:2018xqq}) for the \mhBMPtwo 
(\mhBMPone) scenario, as described in Ref.~\cite{Bagnaschi:2791954} following the 
prescriptions given in Refs.~\cite{LHCHiggsCrossSectionWorkingGroup:2013rie,
deFlorian:2016spz,Denner:2011mq}. 

Inclusive cross sections for the production via \ggPhi have been calculated 
using the program \textsc{SusHi} 1.7.0~\cite{Harlander:2012pb,Harlander:2016hcx}, 
including NLO corrections in \alpS for the \PQt- and \PQb-quark contributions to 
the cross section~\cite{Spira:1995rr,Harlander:2005rq}, NNLO corrections in \alpS 
in the heavy \PQt quark limit, for the \PQt quark contribution~\cite{Harlander:2002wh,
Anastasiou:2002yz,Ravindran:2003um,Harlander:2002vv,Anastasiou:2002wq}, and 
next-to-NNLO contributions in \alpS for \Ph production~\cite{Anastasiou:2014lda,
Anastasiou:2015yha,Anastasiou:2016cez}. Electroweak corrections mediated by 
light-flavour quarks are included at two-loop accuracy reweighting the SM results 
of Refs.~\cite{Aglietti:2004nj,Bonciani:2010ms}. Contributions from squarks and 
gluinos are included at NLO precision in \alpS following 
Refs.~\cite{Degrassi:2010eu,Degrassi:2011vq,Degrassi:2012vt}. The \tanb-enhanced 
SUSY contributions to the Higgs-\PQb couplings have been resummed using the 
one-loop $\Delta_{\mathrm{b}}$ terms from Ref.~\cite{Hofer:2009xb} as provided by 
\textsc{FeynHiggs}. Uncertainties in these $\Delta_{\mathrm{b}}$ terms which range 
${\approx}10\%$ are not included in the overall uncertainties in the predictions as 
they are subdominant with respect to the other theoretical uncertainties.  

For \bbH production, cross sections have been calculated for the SM Higgs boson as a function 
of its mass, based on soft-collinear effective theory~\cite{Bonvini:2015pxa,
Bonvini:2016fgf} which combines the merits of both the 4FS~\cite{Dittmaier:2003ej,
Dawson:2003kb} and 5FS~\cite{Harlander:2003ai,Duhr:2019kwi} calculations. These 
cross sections coincide with the results of the so-called ``fixed order plus 
next-to-leading log'' approach of Refs.~\cite{Forte:2015hba,Forte:2016sja}. The 
pure \PQt- and loop-induced $\PQt\PQb$-interference contributions are separately 
reweighted with effective Higgs couplings, using an effective mixing angle $\alpha$, 
and including the resummation of \tanb-enhanced SUSY contributions as in the \ggPhi 
case. The same SM cross sections are also used to obtain the reweighted cross 
section for \bbA production. A more detailed discussion is given in 
Ref.~\cite{Bagnaschi:2791954}. All Higgs boson masses, effective mixing angles 
$\alpha$, Yukawa couplings, branching fractions, cross sections, and their 
uncertainties, which are included for the exclusion contours, are obtained from 
Ref.~\cite{MSSM_benchmark}.  

In the figure, the exclusion sensitivity, estimated from the expected median in 
the absence of a signal, is indicated by the dashed black line. We note that the 
central 68\% and 95\% intervals, also given for the exclusion sensitivity, should 
not be misinterpreted as an uncertainty in the analysis, but they rather reflect 
the variation of the expected signal yield in the probed parameter space of the 
chosen benchmark scenarios. For the \mhBMPtwo scenario the sensitivity sharply 
drops at $\mA=2\,\mt$, caused by a drop of the branching fractions for the decay 
of \PA and \PH into $\PGt$ leptons where the \PA and \PH decays into two on-shell 
\PQt quarks become kinematically accessible. The distinct boundary is related to 
the fact that in \textsc{FeynHiggs}, which is used for the calculation of all 
branching fractions for this benchmark scenario, only the decay into on-shell 
\ttbar pairs is implemented. The parameter space of each benchmark scenario that 
is excluded at 95\% \CL by the data is indicated by the coloured blue area. 

Both scenarios are excluded at 95\% \CL for $\mA\lesssim350\GeV$. The local 
excess observed at 1.2\TeV causes the deviation of the observed exclusion from 
the expectation. For $\mA\lesssim250\GeV$, most of the \ggH/\ggA events do not 
enter the ``no \PQb tag'' categories due to the $\mtt>250\GeV$ requirement, 
although these events still contribute to the signal yields in the NN categories. 
In this parameter space the sensitivity to the MSSM is driven by the measurements 
of the \Phobs production rates, while the sensitivity to the \PH and \PA enters 
mainly via the \bbPhi signal in the ``\PQb tag'' categories, especially for 
increasing values of \tanb. 

\section{Summary}
\label{sec:summary}

Three searches have been presented for signatures of physics beyond the standard 
model (SM) in \ditau final states in proton-proton collisions at the LHC, 
using a data sample collected with the CMS detector at $\sqrt{s} = 13\TeV$, 
corresponding to an integrated luminosity of 138\fbinv. Upper limits at 95\% 
confidence level (CL) have been set on the products of the branching fraction for 
the decay into \PGt leptons and the cross sections for the production of a resonance 
\Pphi in addition to the observed Higgs boson via gluon fusion (\ggPhi) or in 
association with \PQb quarks, ranging from \order(10\unit{pb}) for a mass of 
$60\GeV$ to 0.3\unit{fb} for a mass of $3.5\TeV$ each. The data reveal two excesses 
for \ggPhi production with local $p$-values equivalent to about three standard 
deviations at $\mphi=0.1$ and 1.2\TeV. Within the resolution of the reconstructed 
invariant mass of the \ditau system, the excess at 100\GeV coincides with a similar 
excess observed in a previous search for low-mass resonances by the CMS Collaboration 
in the $\PGg\PGg$ final state at a mass of ${\approx}95\GeV$. In a search for 
$t$-channel exchange of a vector leptoquark \Uone, 95\% CL upper limits are set 
on the \Uone coupling to quarks and \PGt leptons ranging from 1 for a mass of 
1\TeV to 6 for a mass of 5\TeV, depending on the scenario. The search is sensitive 
to and excludes a portion of the parameter space that can explain the \PQb physics 
anomalies. In the interpretation of the \mhBMPone and \mhBMPtwo minimal 
supersymmetric SM benchmark scenarios, additional Higgs bosons with masses below 
350\GeV are excluded at 95\% CL.

\begin{acknowledgments}
We congratulate our colleagues in the CERN accelerator departments for the excellent performance of the LHC and thank the technical and administrative staffs at CERN and at other CMS institutes for their contributions to the success of the CMS effort. In addition, we gratefully acknowledge the computing centres and personnel of the Worldwide LHC Computing Grid and other centres for delivering so effectively the computing infrastructure essential to our analyses. Finally, we acknowledge the enduring support for the construction and operation of the LHC, the CMS detector, and the supporting computing infrastructure provided by the following funding agencies: BMBWF and FWF (Austria); FNRS and FWO (Belgium); CNPq, CAPES, FAPERJ, FAPERGS, and FAPESP (Brazil); MES and BNSF (Bulgaria); CERN; CAS, MoST, and NSFC (China); MINCIENCIAS (Colombia); MSES and CSF (Croatia); RIF (Cyprus); SENESCYT (Ecuador); MoER, ERC PUT and ERDF (Estonia); Academy of Finland, MEC, and HIP (Finland); CEA and CNRS/IN2P3 (France); BMBF, DFG, and HGF (Germany); GSRI (Greece); NKFIH (Hungary); DAE and DST (India); IPM (Iran); SFI (Ireland); INFN (Italy); MSIP and NRF (Republic of Korea); MES (Latvia); LAS (Lithuania); MOE and UM (Malaysia); BUAP, CINVESTAV, CONACYT, LNS, SEP, and UASLP-FAI (Mexico); MOS (Montenegro); MBIE (New Zealand); PAEC (Pakistan); MES and NSC (Poland); FCT (Portugal); MESTD (Serbia); MCIN/AEI and PCTI (Spain); MOSTR (Sri Lanka); Swiss Funding Agencies (Switzerland); MST (Taipei); MHESI and NSTDA (Thailand); TUBITAK and TENMAK (Turkey); NASU (Ukraine); STFC (United Kingdom); DOE and NSF (USA).

\hyphenation{Rachada-pisek} Individuals have received support from the Marie-Curie programme and the European Research Council and Horizon 2020 Grant, contract Nos.\ 675440, 724704, 752730, 758316, 765710, 824093, 884104, and COST Action CA16108 (European Union); the Leventis Foundation; the Alfred P.\ Sloan Foundation; the Alexander von Humboldt Foundation; the Belgian Federal Science Policy Office; the Fonds pour la Formation \`a la Recherche dans l'Industrie et dans l'Agriculture (FRIA-Belgium); the Agentschap voor Innovatie door Wetenschap en Technologie (IWT-Belgium); the F.R.S.-FNRS and FWO (Belgium) under the ``Excellence of Science -- EOS" -- be.h project n.\ 30820817; the Beijing Municipal Science \& Technology Commission, No. Z191100007219010; the Ministry of Education, Youth and Sports (MEYS) of the Czech Republic; the Hellenic Foundation for Research and Innovation (HFRI), Project Number 2288 (Greece); the Deutsche Forschungsgemeinschaft (DFG), under Germany's Excellence Strategy -- EXC 2121 ``Quantum Universe" -- 390833306, and under project number 400140256 - GRK2497; the Hungarian Academy of Sciences, the New National Excellence Program - \'UNKP, the NKFIH research grants K 124845, K 124850, K 128713, K 128786, K 129058, K 131991, K 133046, K 138136, K 143460, K 143477, 2020-2.2.1-ED-2021-00181, and TKP2021-NKTA-64 (Hungary); the Council of Science and Industrial Research, India; the Latvian Council of Science; the Ministry of Education and Science, project no. 2022/WK/14, and the National Science Center, contracts Opus 2021/41/B/ST2/01369 and 2021/43/B/ST2/01552 (Poland); the Funda\c{c}\~ao para a Ci\^encia e a Tecnologia, grant CEECIND/01334/2018 (Portugal); the National Priorities Research Program by Qatar National Research Fund; MCIN/AEI/10.13039/501100011033, ERDF ``a way of making Europe", and the Programa Estatal de Fomento de la Investigaci{\'o}n Cient{\'i}fica y T{\'e}cnica de Excelencia Mar\'{\i}a de Maeztu, grant MDM-2017-0765 and Programa Severo Ochoa del Principado de Asturias (Spain); the Chulalongkorn Academic into Its 2nd Century Project Advancement Project, and the National Science, Research and Innovation Fund via the Program Management Unit for Human Resources \& Institutional Development, Research and Innovation, grant B05F650021 (Thailand); the Kavli Foundation; the Nvidia Corporation; the SuperMicro Corporation; the Welch Foundation, contract C-1845; and the Weston Havens Foundation (USA).
\end{acknowledgments}

\bibliography{auto_generated} 
\cleardoublepage \appendix\section{The CMS Collaboration \label{app:collab}}\begin{sloppypar}\hyphenpenalty=5000\widowpenalty=500\clubpenalty=5000\input{HIG-21-001-public-authorlist.tex}\end{sloppypar}
\end{document}

%% file: HIG-21-001-public-authorlist.tex
\cmsinstitute{Yerevan Physics Institute, Yerevan, Armenia}
{\tolerance=6000
A.~Tumasyan\cmsAuthorMark{1}\cmsorcid{0009-0000-0684-6742}
\par}
\cmsinstitute{Institut f\"{u}r Hochenergiephysik, Vienna, Austria}
{\tolerance=6000
W.~Adam\cmsorcid{0000-0001-9099-4341}, J.W.~Andrejkovic, T.~Bergauer\cmsorcid{0000-0002-5786-0293}, S.~Chatterjee\cmsorcid{0000-0003-2660-0349}, K.~Damanakis\cmsorcid{0000-0001-5389-2872}, M.~Dragicevic\cmsorcid{0000-0003-1967-6783}, A.~Escalante~Del~Valle\cmsorcid{0000-0002-9702-6359}, P.S.~Hussain\cmsorcid{0000-0002-4825-5278}, M.~Jeitler\cmsAuthorMark{2}\cmsorcid{0000-0002-5141-9560}, N.~Krammer\cmsorcid{0000-0002-0548-0985}, L.~Lechner\cmsorcid{0000-0002-3065-1141}, D.~Liko\cmsorcid{0000-0002-3380-473X}, I.~Mikulec\cmsorcid{0000-0003-0385-2746}, P.~Paulitsch, F.M.~Pitters, J.~Schieck\cmsAuthorMark{2}\cmsorcid{0000-0002-1058-8093}, R.~Sch\"{o}fbeck\cmsorcid{0000-0002-2332-8784}, D.~Schwarz\cmsorcid{0000-0002-3821-7331}, S.~Templ\cmsorcid{0000-0003-3137-5692}, W.~Waltenberger\cmsorcid{0000-0002-6215-7228}, C.-E.~Wulz\cmsAuthorMark{2}\cmsorcid{0000-0001-9226-5812}
% ...
\par}
\vskip\cmsinstskip
\dag:~Deceased\\
$^{1}$Also at Yerevan State University, Yerevan, Armenia\\
$^{2}$Also at TU Wien, Vienna, Austria\\
% ...
